\documentclass[%
 reprint,
 amsmath,amssymb,
 aps,
]{revtex4-2}

\usepackage{siunitx}
\usepackage{graphicx}
\usepackage{dcolumn}
\usepackage{bm}
\usepackage{braket}
\usepackage{multirow,tabularx}

\begin{document}

\preprint{APS/123-QED}

\title{The Arm Qubit: A Superconducting Qubit Co-Designed for Coherence and Coupling}

\author{Jeremy B. Kline$^{1,2}$}
\author{Alec Yen$^{1,2}$}
\author{Stanley Chen$^{1,2}$}
\author{Kevin P. O'Brien$^{1,2}$}
 \email{kpobrien@mit.edu}
\affiliation{$^{1}$Department of Electrical Engineering and Computer Science, Massachusetts Institute of Technology, Cambridge, MA 02139, USA,}
\affiliation{$^{2}$Research Laboratory of Electronics, Massachusetts Institute of Technology, Cambridge, MA 02139, USA}

\begin{abstract}
We present a superconducting qubit which consists of two strongly coupled modes: one for data storage and one for coupling, allowing faster, higher-fidelity entangling gates and readout.
The use of a dedicated coupling mode allows nonlinear couplings of several hundred MHz between the data mode and other elements, with minimal linear coupling to the data mode. 
Including decoherence, simulations show that this architecture enables microwave-only CZ gates with an infidelity of \SI{8.6e-5}{} in \SI{17}{ns} and always-on ZZ interaction less than \SI{0.4}{kHz}.
Numerical simulations also show readout with state assignment error of \SI{1e-4}{} in \SI{27}{ns} (assuming quantum efficiency $\eta=0.5$), Purcell-limited lifetime of \SI{167}{ms} without a Purcell filter, and a mechanism to suppress shot-noise dephasing ($1/\Gamma_{\phi}=15.8$ ms).
Single-qubit gate infidelities are below \SI{1e-5}{} including decoherence. 
These beyond experimental state-of-the-art gate and readout fidelities rely only on capacitive coupling between arm qubits, making the arm qubit a promising scalable building block for fault-tolerant quantum computers.
\end{abstract}

\maketitle

\section{\label{sec:introduction}Introduction}
Higher fidelity logic operations can significantly reduce the overhead of quantum error correction, accelerating the timeline to useful, fault-tolerant quantum computation~\cite{google_error-correction_2025, krinner_realizing_2022, putterman_hardware-efficient_2025}. 
The fidelity of these operations is ultimately bounded by the ratio of the operation time to the coherence time of the qubits. 
Single-qubit gate speeds are limited by the qubit anharmonicity, and can therefore be improved by using a highly nonlinear qubit potential such as that of fluxonium qubits, which, relative to transmon qubits, have shown higher single-qubit gate fidelities~\cite{rower_suppressing_2024}. 
The speed of operations between modes, such as readout and two-qubit gates however, scales with the strength of coupling between the modes~\cite{blais_circuit_2021, krantz_quantum_engineer}.

The simplest approach to coupling modes is to use linear capacitive or inductive couplings, but this tends to hybridize the coupled modes. 
In the case of two-qubit gates, this hybridization of the coupled modes results in an always-on ZZ coupling between the two qubits~\cite{sung_realization_2021, kandala_cnot-fixed-frequency_2021}.
Thus, two-qubit gates are typically complicated by the requirement to engineer cancellation or suppression of this ZZ interaction~\cite{sung_realization_2021, ding_high-fidelity_2023}. 
In the case of readout, mode hybridization results in measurement-induced state transitions and Purcell decay~\cite{sank_measurement-induced_2016,dumas_measurement-induced_2024}. 
These effects ultimately limit the strength of the nonlinear coupling between the qubit and resonator by forcing operation in the dispersive regime.
However, recent developments in qubit readout have highlighted the ability of nonlinear coupling elements to circumvent these undesirable effects and improve readout speed ~\cite{chapple_balanced_2025, ye_ultrafast_2024, wang_999-fidelity_2024}, suggesting that nonlinear coupling can help circumvent the tradeoff between increasing coupling strengths and preserving isolation of computational states.

Motivated by the advantages nonlinear coupling has been shown to provide in the context of qubit readout, we suggest here a qubit architecture consisting of two nonlinearly coupled modes, co-designed for high-fidelity gates and readout while preserving coherence.
One mode is fluxonium-like, with large anharmonicity and relatively low frequency (1-2 GHz), and can store data with $T_{2}^E> \SI{380}{\micro s}$ (assuming a dielectric quality factor $Q=\SI{3.5e6}{}$).
The other mode is transmon-like, with higher frequency ($>\SI{7}{GHz}$), and can be used to facilitate interactions between data modes (for entangling gates) or between a data mode and a resonator (for readout).
We call this qubit the ``arm qubit'' because the higher frequency mode can be thought of as an ``arm'' of the data mode, reaching out to interact with other components of the quantum processor.

Previous work using multi-mode qubits to engineer favorable coupling properties for readout and gates has focused on weakly anharmonic transmon-like modes~\cite{gambetta_tunable-coupling-qubit_2011, zhang_suppression_2017, hazra_benchmarking_2025, diniz_ultrafast_2013, roy_jrm_2017, pfeiffer_pmon_2024, finck_tcq-gates_2021, dassonneville_transmon-molecule_2020, dassonneville_transmon-molecule-bifurcation_2023, roy_generalized-jrm_2018, salunkhe_quantromon_2025}.
Other multi-mode qubits, some with strong anharmonicity, have been designed for noise protection~\cite{brooks_0pi_2013, smith_cos2phi_2020, larsen_cos2phi_2020, hays_harmonium_2025, kalashnikov_bifluxon_2020, schrade_cos2phi_2022} and erasure detection~\cite{levine_transmon-dual-rail_2024, huang_dual-rail-entanglement_2025, chou_dual-rail-measurement_2024, mehta_bias-preserving_2025} rather than for coupling.
In contrast, the arm qubit is a multi-mode qubit designed specifically for coupling to other modes, but consisting of a strongly anharmonic, fluxonium-like data mode and a weakly anharmonic, transmon-like arm mode.
We show in simulation that this design enables faster, higher-fidelity operation than previous approaches.
The arm mode can facilitate cross-Kerr interactions of several hundred MHz between the data mode and auxiliary modes while suppressing ZZ interactions between the data modes to $< 0.4$ kHz.  
This enables beyond-state-of-the-art CZ gates with an infidelity of \SI{8.6e-5}{} in \SI{17}{ns}, including qubit decoherence.
The large anharmonicity of the data mode enables single-qubit gates with infidelity below \SI{1e-5}{}.
Our simulations also indicate that this scheme allows significantly faster readout than typical dispersive readout, with state assignment error of \SI{1e-4}{} and quantum nondemolition (QND) infidelity of \SI{2.75e-3} in \SI{27}{ns}, assuming quantum efficiency $\eta=0.5$.
Use of the arm mode as an intermediary protects the data mode from Purcell decay, enabling a Purcell-limited lifetime of  \SI{167}{ms}, while also providing a tuning knob with which to suppress shot noise dephasing when not performing readout ($T_\phi =$ \SI{15.8}{ms}).

\section{The arm qubit}
\begin{figure*}
\includegraphics[width=\textwidth]{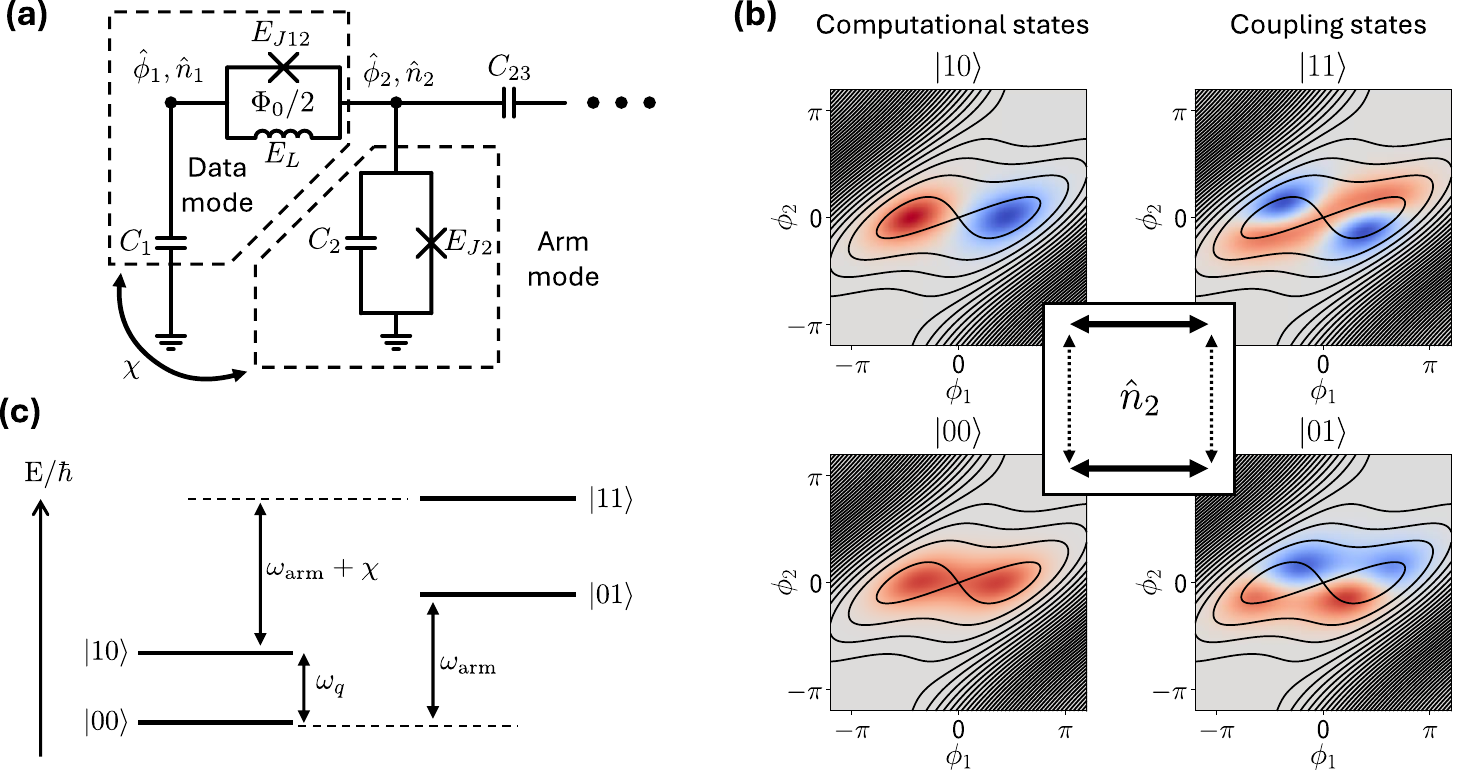}
\caption{\label{fig:Fig1} (a) Circuit diagram of the arm qubit. The data mode energy is primarily stored in the capacitance $C_1$ and the inductances $E_L$ and $E_{J12}$, while the arm mode is primarily stored in $C_2$ and $E_{J2}$. A DC flux bias of half a flux quantum, $\Phi_0/2$, threads the data mode loop. (b) Wave functions of the arm qubit. Because the two minima of the double-well potential lie exactly along $\hat{\phi}_2 = 0$, the computational states primarily represent oscillation in the $\hat{\phi}_1$ direction. This means that any capacitive coupling to the $\hat{n}_2$ degree of freedom is approximately decoupled from the computational states and instead only affects the coupling states.  (c) Energy level diagram for the arm qubit. Low lying levels $\ket{00}, \ket{10}$ form the computational subspace, while states $\ket{01}, \ket{11}$ provide a mechanism for coupling the qubit to other modes, whether for entangling gates or readout.}
\end{figure*}

The circuit diagram for the arm qubit is shown in Fig.~\ref{fig:Fig1}a, and can be divided into a data mode ($C_1, E_{J12}, E_L$) and an arm mode ($C_2, E_{J2}$).
The motivation for dividing the circuit this way can be seen by considering the potential energy of the circuit,
\begin{equation}\label{main_potential}
U = -E_{J12}\cos(\phi_1-\phi_2) + \frac{E_L}{2}(\phi_1 - \phi_2 + \tilde\phi)^2 -E_{J2}\cos(\phi_2).
\end{equation}
By choosing $E_{J12} > E_L$ and biasing the loop at half a flux quantum ($\tilde\phi =\pi$), this potential will have a double well in the $\phi_1 - \phi_2$ direction, much like a fluxonium qubit. 
Crucially, the $E_{J2}\cos(\phi_2)$ term forces the minima of these double wells to lie \textit{exactly} along $\phi_2 = 0$.
This causes the two lowest energy eigenstates, which we will use as the computational states, to primarily represent oscillation in the $\phi_1$ direction, despite the obvious coupling between the $\phi_1$ and $\phi_2$ degrees of freedom in this Hamiltonian.
Flux basis wavefunctions of the arm qubit are plotted on top of the potential in Fig.~\ref{fig:Fig1}b for a representative choice of circuit parameters.

Because the computational states lie primarily in the $\phi_1$ direction, any capacitive coupling to the $\hat{n}_2$ degree of freedom will couple much more strongly to the arm mode transition ($\ket{i0}\leftrightarrow\ket{i1}$) than the data mode transition ($\ket{0i}\leftrightarrow\ket{1i}$). 
More formally, we have

\begin{equation}\label{eq:arm_n_ratio}
\lvert\bra{00}\hat{n}_2\ket{01}\rvert, \lvert\bra{10}\hat{n}_2\ket{11}\rvert \gg\lvert\bra{00}\hat{n}_2\ket{10}\rvert
\end{equation}
This means the states of any mode coupled by $C_{23}$ can strongly couple to and even hybridize with the coupling states $\ket{01}$ and $\ket{11}$ while only minimally hybridizing with the computational states $\ket{00}$ and $\ket{10}$.

Despite this isolation from direct coupling, we can still use the $C_{23}$ capacitance to interact with the data mode via the large nonlinear coupling between the data mode and the arm mode. 
As illustrated by the energy level diagram in Fig.~\ref{fig:Fig1}c, the data mode state shifts the arm mode transition (which the $\hat{n}_2$ operator is sensitive to) by a cross-Kerr interaction which in practice can be greater than a gigahertz. One can then imagine the arm mode conditionally ``pushing'' the energy levels of an auxiliary mode (level repulsion), thus resulting in an effective cross-Kerr interaction between the data mode and the auxiliary mode.
In this paper we show how this effect can be used for both entangling gates and qubit readout.

Intuitively, one way to understand the cross-Kerr interaction between the data mode and the arm mode is to see the the junction and inductor between nodes 1 and 2 as a quarton-like nonlinear coupling element~\cite{ye_engineering_2021} which, despite having a double-welled rather than quartic potential, still provides a strong $\hat{\phi}^2_1\hat{\phi}^2_2$ term.
Since the data mode is primarily an oscillation in $\phi_1$, and the arm mode is primarily an oscillation in $\phi_2$, this nonlinear coupling between the $\phi_1$ and $\phi_2$ degrees of freedom causes a cross-Kerr interaction between the two modes.

\section{Two-qubit Gates}

\begin{figure}
\includegraphics[width=0.5\textwidth]{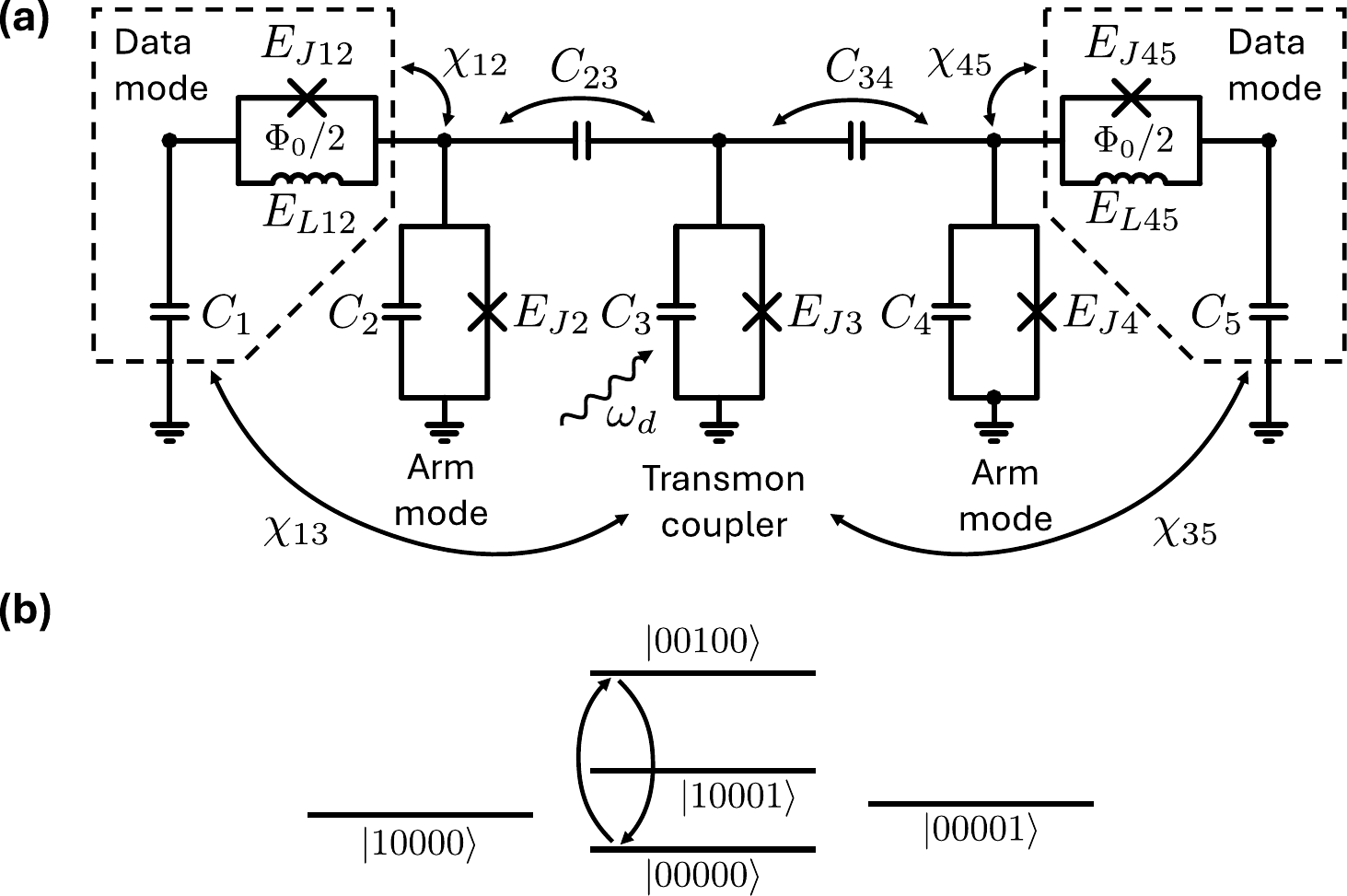}
\caption{\label{fig:gate_circuit} (a) Circuit diagram for two-qubit gates. The arm modes of each arm qubit are capacitively coupled to a center transmon, leading to effective cross-Kerr interactions $\chi_{13}, \chi_{35}$ between the data modes and the transmon. A conditional Rabi oscillation on the transmon results in a $\pi$ phase accumulation when the data modes are in the $\ket{00}$ state. (b) Energy levels in the two-qubit gate circuit. The gate consists of driving a $2\pi$ oscillation between the $\ket{00000}$ and $\ket{00100}$ states (ordering the modes from left to right). When either of the data modes is excited (the $\ket{10000}$, $\ket{00001}$, or $\ket{10001}$ states), the drive is detuned from the transmon frequency.}
\end{figure}

The circuit diagram for a gate between two arm qubits is shown in Fig.~\ref{fig:gate_circuit}a.
The arm modes of each qubit are capacitively coupled with a transmon coupler, and level repulsion between the arm modes and the transmon results in effective cross-Kerr interactions between the data modes and the transmon.
Since the transmon frequency then depends on the state of the data modes, we can drive a conditional Rabi oscillation between $\ket{00000}$ and $\ket{00100}$ (ordering the 5 modes of the circuit from left to right), as shown in Fig.~\ref{fig:gate_circuit}b.
This oscillation imparts a $\pi$ phase on the $\ket{00000}$ state, thereby performing a CZ gate similar to those in~\cite{ding_high-fidelity_2023, rosenfeld_fluxonium-resonator_2024}.

\subsection{Coherent error}
The circuit parameters we use to simulate two-qubit gates are shown in Table~\ref{tab:gate_parameters}, and yield cross-Kerr interactions over 350 MHz between the data modes and the center transmon.
For these parameters, the ratio of charge operator matrix elements described in Eq.~\ref{eq:arm_n_ratio} is $\lvert\bra{00}\hat{n}_2\ket{01}\rvert/\lvert\bra{00}\hat{n}_2\ket{10}\rvert=8.5$ and 8.1 for the first and second qubit respectively. This isolation of the data modes enables a low static ZZ coupling between them of only \SI{0.32}{kHz}.

Throughout this paper we use a hierarchical diagonalization scheme, detailed in Appendix~\ref{app:hierarchical_diagonalization}, to solve for circuit eigenstates before performing time domain simulations. We truncate to the lowest 576 eigenstates of the circuit and then simulate the gate with QuTiP\cite{lambert_qutip5_2024}, computing the fidelity as \cite{pedersen_fidelity_2007}
\begin{equation}\label{eq:gate_fidelity}
    F = \frac{\mathrm{Tr}(\hat{T}^\dagger \hat{T}) + |\mathrm{Tr}(\hat{U}_\mathrm{CZ}^\dagger \hat{T})|^2}{20}
\end{equation}
where we construct the simulation propagator $\hat T$ in the computational basis by repeating the simulation with each of the four computational basis states as the initial state, projecting onto the computational basis (so $\hat T$ is not necessarily unitary), and finally applying virtual Z rotations on the data modes.
$\hat{U}_\mathrm{CZ}$ represents an ideal CZ gate.
We implement the charge drive on the transmon with a simple cosine pulse envelope, numerically optimizing the amplitude of the cosine envelope to maximize gate fidelity. In Fig.~\ref{fig:gate_error}a we plot the simulated coherent error of the gate as a function of gate time and see that at 14 ns it is already below $5\times10^{-5}$.

\begin{table}
    \centering
    \begin{tabular}{cccc}
        \hline
        \hline
        \multicolumn{1}{l}{$E_{J12}/2\pi$} &
        \multicolumn{1}{c|}{38.5 GHz} & 
        \multicolumn{1}{l}{$C_1$} &
        \multicolumn{1}{r}{20.8 fF} \\
        
        \multicolumn{1}{l}{$E_{L12}/2\pi$} & 
        \multicolumn{1}{c|}{26.2 GHz} &
        \multicolumn{1}{l}{$C_2$} &
        \multicolumn{1}{r}{33.1 fF} \\
        
        \multicolumn{1}{l}{$E_{J45}/2\pi$} &
        \multicolumn{1}{c|}{38.4 GHz} &
        \multicolumn{1}{l}{$C_3$} &
        \multicolumn{1}{r}{29.1 fF} \\
        
        \multicolumn{1}{l}{$E_{L45}/2\pi$} &
        \multicolumn{1}{c|}{26.2 GHz} &
        \multicolumn{1}{l}{$C_4$} &
        \multicolumn{1}{r}{33.0 fF} \\
        
        \multicolumn{1}{l}{$E_{J2}/2\pi$} &
        \multicolumn{1}{c|}{19.8 GHz} &
        \multicolumn{1}{l}{$C_5$} &
        \multicolumn{1}{r}{19.4 fF} \\
        
        \multicolumn{1}{l}{$E_{J3}/2\pi$} &
        \multicolumn{1}{c|}{20.5 GHz} &
        \multicolumn{1}{l}{$C_{23}$} &
        \multicolumn{1}{r}{4.7 fF} \\
        
        \multicolumn{1}{l}{$E_{J4}/2\pi$} &
        \multicolumn{1}{c|}{17.6 GHz} &
        \multicolumn{1}{l}{$C_{34}$} &
        \multicolumn{1}{r}{4.8 fF} \\
        \hline
        \multicolumn{2}{c}{Data mode $\omega_1/2\pi$} &
        \multicolumn{2}{r}{1.6 GHz} \\
        
        \multicolumn{2}{c}{Arm mode $\omega_2/2\pi$} &
        \multicolumn{2}{r}{7.21 GHz} \\ 
        
        \multicolumn{2}{c}{Transmon coupler $\omega_3/2\pi$} &
        \multicolumn{2}{r}{8.79 GHz} \\
        
        \multicolumn{2}{c}{Arm mode $\omega_4/2\pi$} &
        \multicolumn{2}{r}{7.02 GHz} \\
        
        \multicolumn{2}{c}{Data mode $\omega_5/2\pi$} &
        \multicolumn{2}{r}{1.70 GHz} \\
        
        \multicolumn{2}{c}{Data anharmonicity $\alpha_1/2\pi$} & 
        \multicolumn{2}{r}{3.13 GHz} \\

        \multicolumn{2}{c}{Data anharmonicity $\alpha_2/2\pi$} &
        \multicolumn{2}{r}{3.08 GHz} \\
        \hline
        \hline
    \end{tabular}
    \caption{Parameters used in the two-qubit gate simulation. Additional parasitic capacitance is included as described in Appendix~\ref{app:hierarchical_diagonalization}.}
    \label{tab:gate_parameters}
\end{table}

\subsection{Decoherence contribution to gate error}
The benefits of fast gates could be lost if the proposed architecture resulted in lower qubit lifetimes.
We follow standard procedures to compute the dominant sources of decoherence (see Appendix~\ref{app:decoherence} for details) and show that these qubits are primarily limited by dielectric loss and should have similar lifetimes to current state-of-the-art superconducting qubits, with $T_{2}^E>380$ $\mu$s.
The estimated decoherence rate contributions are summarized in Table~\ref{tab:decoherence_table}. 

\begin{table}
    \centering
    \begin{tabular}{c|c|c|c|c}
        & Mechanism & Data mode 1 & Data mode 2 & Transmon\\
        \hline
         \multirow{3}{*}{$T_1$} & Dielectric loss & 421 $\mu$s & 394 $\mu$s & 65 $\mu$s\\
         & Flux noise & 1.21 ms & 1.30 ms & - \\
         & Total& 313 $\mu$s & 303 $\mu$s & 65 $\mu$s\\
         \hline
         \multirow{3}{*}{$T_\phi$} & Flux noise (echo) & 1.64 ms & 1.79 ms & - \\
         & Charge noise & 4.76 ms & 2.46 ms & 43 us \\
         & Total & 1.22 ms & 1.04 $\mu$s & 43 $\mu$s \\
         \hline
         $T_{2}^E$ & Total & 413 $\mu$s & 382 $\mu$s & 33 $\mu$s\\
    \end{tabular}
\caption{Estimated coherence times for the qubits in the two-qubit gate circuit (Fig.~\ref{fig:gate_circuit}a) assuming the parameter values in Table~\ref{tab:gate_parameters}. The data mode coherence times are primarily limited by the dielectric loss rate ($\Gamma\approx\omega/Q$), where we assume a dielectric quality factor ($Q$) of \SI{3.5e6}{}.}
    \label{tab:decoherence_table}
\end{table}

To estimate the impact of these coherence times on the gate fidelity, we perform a Lindblad master equation simulation, using as collapse operators $\sqrt{1/T_1}\ket{0}\bra{1}$ for relaxation errors and $\sqrt{1/T_\phi}\ket{1}\bra{1}$ for dephasing.
We include these collapse operators for each of the data modes and for the transmon coupler, constructing them within the space spanned by the five primarily populated states $\ket{00000}, \ket{00001}, \ket{10000}, \ket{10001},$ and $\ket{00100}$.
To reduce the computational requirements of the master equation simulation, we first perform a lossless Schrödinger equation simulation with a 7 ns pulse width and then truncate the Hilbert space to include only states which were populated to at least \SI{1e-6}{} at any time during the simulation.

We then perform quantum process tomography by repeating the master equation simulation for a set of initial states spanning the space of density matrices and reconstructing the operators $\hat{E}_i$ which define the map on the density matrix within the computational subspace as \cite{nielsen_quantum-computation_2010}
\begin{equation}
    \rho \rightarrow \sum_{i}\hat{E}_i\rho\hat{E}^\dagger_i.
\end{equation}
This allows us to express the average gate fidelity as \cite{pedersen_fidelity_2007}
\begin{equation}\label{eq:kraus_fidelity}
    F = \frac{1}{d(d+1)}\sum_i\bigg(\mathrm{Tr}\big(\hat{E}_i^\dagger \hat{E}_i\big) + |\mathrm{Tr}(\hat{U}_\mathrm{CZ}^\dagger \hat{E}_i)|^2\bigg),
\end{equation}
where $d=4$ is the dimension of the computational subspace. Due to leakage to noncomputational states, the $\hat{E}_i$ operators are not normalized ($\sum_i\hat{E}^\dagger_i\hat{E}_i\neq I$) and so the $\mathrm{Tr}\big(\hat{E}_i^\dagger \hat{E}_i\big)$ term contributes meaningfully to the infidelity. Using the decoherence times from Table~\ref{tab:decoherence_table}, which assume a dielectric quality $Q=3.5\times 10^6$, we find a gate infidelity of \SI{8.6e-5}{} in 17 ns in the presence of dissipation.
Since dielectric loss is the strongest contributor to qubit decoherence, we also compute gate fidelity assuming a pessimistic dielectric quality factor $Q=1.5\times10^{6}$ as well as a state-of-the-art $Q=8\times10^{6}$~\cite{wang_towards-practical_2022, bland_2d-transmons_2025}.
For the pessimistic $Q$, we achieve an infidelity of \SI{1.5e-4} in \SI{14}{ns}, while for the state-of-the-art $Q$ we reach an infidelity of \SI{5.8e-5}{} in \SI{17}{ns}.

\subsection{Robustness to parameter variation}
\begin{figure}
\includegraphics[width=0.5\textwidth]{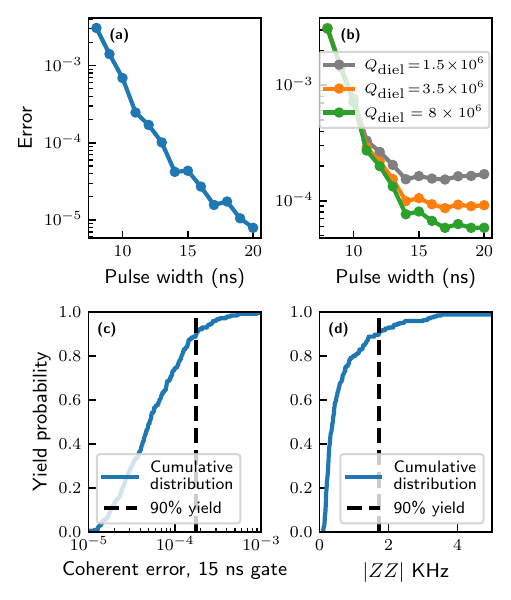}
\caption{\label{fig:gate_error}(a) Simulated CZ gate coherent error as a function of pulse width. (b) Simulated CZ gate error including qubit decoherence, for varying dielectric quality factor $Q_\textrm{diel}$. (c-d) Monte Carlo simulation results assuming a 5\% standard deviation on all circuit parameters. (c) The cumulative distribution of gate fidelities, with the black dashed line denoting the value which we expect to outperform 90\% of the time. (d) The always-on ZZ interaction between the data modes, again with the black dashed line denoting the 90\% yield value.}
\end{figure}

To check whether this gate scheme is robust to fabrication variance, we perform Monte Carlo simulation by generating 240 sets of parameters with a 5\% relative standard deviation on all junction energies and capacitances.
We then assume that the arm modes and transmon coupler can have their junctions replaced with superconducting quantum interference devices (SQUIDs), and therefore be tuned to compensate for fabrication variance.
For each noisy parameter set, we use the DIRECT global optimization algorithm \cite{jones_direct_1993, gablonsky_direct-L_2001} to optimize $E_{J2}, E_{J3}$, and $E_{J4}$. 
In this optimization loop, we set the gate time to be 15 ns, choose the drive amplitude such that the pulse envelope area is $2\pi/|\bra{00000}\hat{n}_2\ket{00100}|$, and for simplicity compute fidelity only starting from the initial state $\ket{++}$ rather than computing the full propagator.
After choosing optimal $E_{J2}, E_{J3}, E_{J4}$ values, we then numerically optimize the drive amplitude for each parameter set, yielding the cumulative distribution of coherent gate error in Fig.~\ref{fig:gate_error}C, in which we see that there is a 90\% probability to achieve less than $1.8\times10^{-4}$ coherent error with a 15 ns pulse.
In this latter optimization of the drive amplitude, we use the full propagator to compute fidelity as in Eq.~\ref{eq:gate_fidelity}.

We also plot the cumulative distribution of static ZZ coupling in Fig.~\ref{fig:gate_error}d, where we see that there is a 90\% probability for the static ZZ to be under \SI{1.8}{kHz}.

\section{Single-qubit gates}
Using the same circuit and parameters as in Fig.~\ref{fig:gate_circuit}a and Table~\ref{tab:gate_parameters}, we simulate single-qubit gate performance.
As in fluxonium qubits, the double-well potential seen by the data modes suppresses their charge matrix elements compared to their flux matrix elements~\cite{rower_suppressing_2024}.
We therefore drive single-qubit gates through the flux bias of the arm qubit, resulting in a time-dependent addition to the Hamiltonian of the form
\begin{equation}
    H_{\textrm{drive}} = d(t)(\hat{\phi}_i-\hat{\phi}_j)
\end{equation}
where $d(t)$ is a real time-dependent coefficient, and $(i,j)=(1,2)$ for the first data mode and $(4,5)$ for the second.
We use a cosine envelope for the drive, with amplitude and frequency numerically optimized to implement an $X/2$ gate.

As in the two-qubit gate simulation, we perform a master equation simulation with the decoherence times in Table~\ref{tab:decoherence_table}.
The Hilbert space is truncated by keeping only states which are populated to at least $1\times 10^{-7}$ at any time during a lossless Schrödinger equation simulation with a \SI{5}{ns} gate pulse.
We then perform quantum process tomography in the single-qubit space to characterize the map on the density matrix, computing the fidelity as in Eq.~(\ref{eq:kraus_fidelity}), with $d=2$ and $U_\textrm{CZ}$ replaced by the $X/2$ unitary.
We again assume a dielectric quality factor of $Q=3.5\times 10^6$ and find that both qubits can achieve single-qubit gate infidelity below \SI{1e-5}{} with a 5 ns gate pulse.

The high single-qubit gate fidelities are explained by the large ~3 GHz anharmonicity of the data modes.
As in fluxonium, this large anharmonicity enables very fast single-qubit gates.
However, the qubit frequencies are around 1.5 GHz, unlike fluxonium which is often only a few hundred MHz.
This is an advantage because the Larmor period of fluxonium qubits can cause a breakdown in the rotating wave approximation and limit the speed of single-qubit gates~\cite{rower_suppressing_2024}.

\section{Readout}
\begin{figure}
\includegraphics[width=0.5\textwidth]{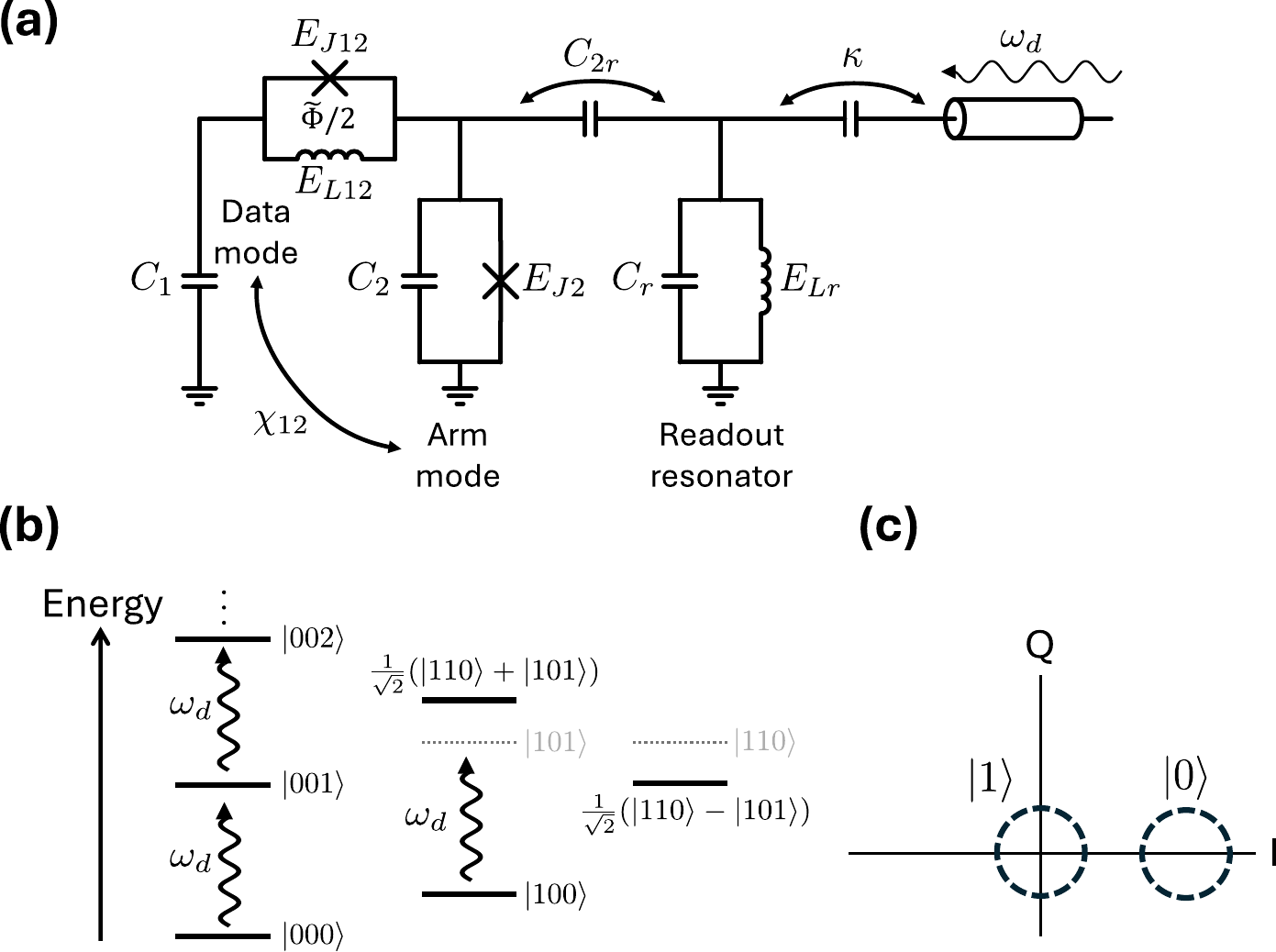}
\caption{\label{fig:readout_circuit} (a) Circuit for readout of an arm qubit. The arm mode is capacitively coupled to a standard coplanar waveguide readout resonator. (b) Energy level diagram. The states are written as $\ket{\textrm{data, arm, resonator}}$. When the data mode is in state $\ket{0}$, the arm and resonator are detuned by $\chi_{12}/2\pi>1$ GHz. When the data mode is in state $\ket{1}$, the arm and resonator are on resonance with each other, and this degeneracy is split by their coupling. (c) Amplitude readout scheme. When the data mode is in state $\ket{0}$, the drive is on resonance with the readout resonator and populates it with a coherent state. When the data mode is in state $\ket{1}$, the drive is detuned from the resonator and the resonator remains mostly unpopulated.}
\end{figure}
The signal-to-noise ratio of standard dispersive readout improves with both the qubit-resonator dispersive shift (which allows for increased resonator loss $\kappa$) and the photon number $\bar n$. However, increasing the dispersive shift causes increased thermal shot noise dephasing~\cite{clerk_shot-noise_2007}, and increasing $\bar n$ tends to lead to measurement-induced state transitions~\cite{sank_measurement-induced_2016, dumas_measurement-induced_2024}.
Shot noise dephasing is often a significant contribution to decoherence rates, especially for fluxonium qubits, and leakage to noncomputational states is particularly damaging in the context of error correction~\cite{google_error-correction_2025}. In the following, we describe how the arm mode of the arm qubit allows us to improve on state-of-the-art dispersive readout speed without sacrificing coherence time or QND fidelity.

The readout circuit, shown in Fig.~\ref{fig:readout_circuit}a, consists of a standard coplanar waveguide readout resonator capacitively coupled to the arm mode.
Due to the inherent isolation between the data mode and any mode capacitively coupled to the arm mode, it is not necessary operate in the dispersive regime.
Instead, we choose our mode frequencies such that, when the data mode is in the ground state, the arm mode is detuned below the resonator mode by approximately $\chi_{12}$, where $\chi_{12}$ is the cross-Kerr shift between the arm and data modes.
This means that when the data mode is in the excited state, the arm mode frequency is increased by $\chi_{12}$, which places it on resonance with the readout resonator.
As shown in Fig.~\ref{fig:readout_circuit}b, the degeneracy of these states ($\ket{110}$ and $\ket{101}$) will be split by the capacitive coupling $C_{2r}$, resulting approximately in an even state $\ket{1}\ket{\psi_e}=\frac{1}{\sqrt{2}}\left(\ket{101}+\ket{110}\right)$ and an odd state $\ket{1}\ket{\psi_o}=\frac{1}{\sqrt{2}}\left(\ket{101}-\ket{110}\right)$.

By driving the resonator at frequency $\omega_d = E_{\ket{001}} - E_{\ket{000}}$, we excite it to a coherent state when the data mode is in the ground state.
When the data mode is in the excited state, the drive is instead detuned from the hybridized states of $\ket{1}\ket{\psi_e}$ and $\ket{1}\ket{\psi_o}$ by detunings $\Delta_e=\omega_d-E_{\ket{1}\ket{\psi_e}}+E_{\ket{100}}$ and $\Delta_o=\omega_d-E_{\ket{1}\ket{\psi_o}}+E_{\ket{100}}$, respectively.
As long as the resonator linewidth $\kappa$ is significantly smaller than $\Delta_e$ and $\Delta_o$, the resonator is left nearly unpopulated when the data mode is in the excited state.
This conditional excitation of the resonator is illustrated in the IQ-plane in Fig.~\ref{fig:readout_circuit}c.

\begin{table}
    \centering
    \begin{tabular}{lr|lr}
         \hline
         \hline
         $E_L/2\pi$& \SI{26.2}{GHz} & $C_1$ & \SI{20.4}{fF}\\
         $E_{J12}/2\pi$& \SI{38.5}{GHz} & $C_2$ & \SI{36.2}{fF}\\
         $E_{J2}/2\pi$& \SI{33.6}{GHz} &  $C_r$& \SI{293}{fF}\\
         $E_{Lr}/2\pi$& \SI{223}{GHz} & $C_{2r}$ & \SI{6.5}{fF} \\
         $\omega_q/2\pi$& \SI{1.6}{GHz} & $\chi_{12}/2\pi$& \SI{1.46}{GHz} \\
          $\omega_{arm}/2\pi$& \SI{9.35}{GHz} & $\Delta_e/2\pi$ & \SI{-271}{MHz} \\
         $\omega_r/2\pi$& \SI{10.81}{GHz}& $\Delta_o/2\pi$ & \SI{268}{MHz}\\
         \hline
         \hline
    \end{tabular}
    \caption{Parameters used in the readout simulation. The resonator inductance and capacitance are chosen to correspond approximately to a 50 Ohm coplanar waveguide resonator. The qubit capacitance is chosen such that its charging energy $E_c$ matches that in the two-qubit gate simulation. Additional parasitic capacitance is included, as described in Appendix~\ref{app:hierarchical_diagonalization}. The cross-Kerr between the data and arm modes $\chi_{12}$ is calculated as $E_{\ket{110}}-E_{\ket{100}}-E_{\ket{010}}+E_{\ket{000}}$, where $E_{\ket{110}}$ is calculated by taking the average of $E_{\ket{1}\ket{\psi_e}}$ and $E_{\ket{1}\ket{\psi_o}}$.}
    \label{tab:readout_parameters}.
\end{table}

\begin{figure}
\includegraphics[width=0.5\textwidth]{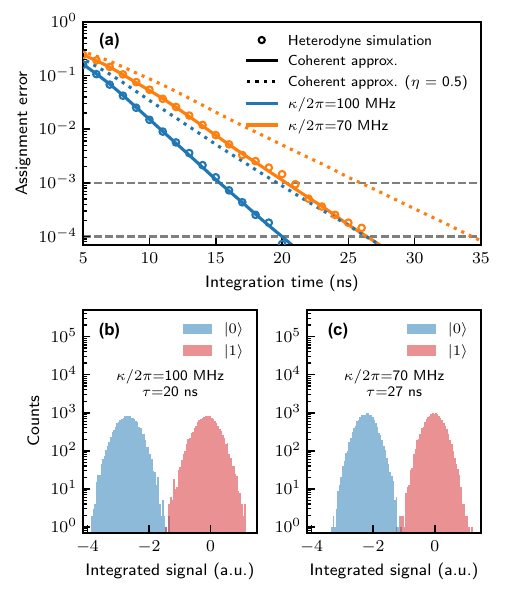}
\caption{\label{fig:readout_error} Readout fidelities and histograms of 13,800 single-shot heterodyne detection simulations of arm qubit readout. Two separate simulations are conducted, one with a 20 ns square pulse and $\kappa/2\pi=100$ MHz, and the other with a 27 ns square pulse and $\kappa/2\pi=70$ MHz. Pulse lengths are chosen such that state assignment errors below \SI{1e-4}{} are achieved. (a) Readout fidelity as a function of time (calculated at 1-ns intervals) for different quantum efficiencies. (b-c) Histograms of the integrated signal for (b) $\kappa/2\pi=100$ MHz and (c) $\kappa/2\pi=70$ MHz, assuming ideal quantum efficiency. }
\end{figure}

\begin{table}
\centering
    \begin{tabular}{l|c c}
        \hline
        \hline
            $\kappa/2\pi$ & \SI{100}{MHz} & \SI{70}{MHz} \\
            \hline
            Readout time, $\eta=1$  & \SI{20}{ns} & \SI{27}{ns} \\
            Readout time, $\eta=0.5$  & \SI{27}{ns} & \SI{35}{ns} \\
            QND infidelity, $\ket{0}$ & \SI{2.0e-3}{} & \SI{1.5e-3}{} \\
            QND infidelity, $\ket{1}$ & \SI{3.5e-3}{} & \SI{0.3e-3}{} \\
            QND infidelity, average & \SI{2.75e-3}{} & \SI{0.9e-3}{} \\
            Average photon number $\bar{n}$, $\ket{0}$ &  3.49 &3.47 \\
            Average photon number $\bar{n}$, $\ket{1}$ & 0.21 & 0.06 \\
            $T_\phi$, shot noise dephasing & \SI{15.8}{ms} & \SI{11.1}{ms}\\
            $T_{1,\textrm{Purcell}}$ & \SI{167}{ms} & \SI{239}{ms}\\
        \hline
        \hline
    \end{tabular}
    \caption{Comparison of simulated readout results for resonator linewidth of $\kappa/2\pi=\SI{100}{MHz}$ and \SI{70}{MHz}. Readout times are chosen to ensure a readout assignment error below \SI{1e-4}{}. QND infidelities are calculated for offset charge on the arm mode of $n_g=0$. Shot noise dephasing is calculated after reducing the arm mode junction to $E_{J2}/2\pi=13$ GHz to suppress the cross-Kerr interaction between the data mode and the resonator mode.}
    \label{tab:kappa-compare}
\end{table}

We simulate this readout scheme using a stochastic Schrödinger equation simulation in QuTiP~\cite{lambert_qutip5_2024} (see further details in Appendix~\ref{app:readout}).
We use the parameters in Table~\ref{tab:readout_parameters}, where we have kept the same data mode parameters ($E_{J12}, E_L, C_1$) as for the gates simulation.
The large detunings from $\omega_d$ of $\Delta_e/2\pi=\SI{-271}{MHz}$ and $\Delta_o/2\pi=\SI{268}{MHz}$ suppress leakage during readout when the data mode is in the excited state, and allow the use of a very large resonator linewidth $\kappa/2\pi=\SI{100}{MHz}$.
The readout drive is set to a square pulse.
From the stochastic Schrödinger equation, we simulate 13,800 integrated heterodyne trajectories (see Fig.~\ref{fig:readout_error}).
For $\eta=1$, state assignment error of \SI{1e-4}{} is achieved for integration times as short as \SI{20}{ns}.
By approximating the integrated signals as Gaussian distributions, we compute the signal to noise ratio~\cite{gambetta_readout_2007, swiadek_readout_2024} and scale this by $\sqrt{\eta}$ to compute the expected readout fidelity for realistic $\eta$.
For $\eta=0.5$, we estimate readout assignment error below \SI{1e-4}{} to be achieved in \SI{27}{ns}.

We evaluate the QND infidelity of this readout scheme using a Lindblad master equation simulation (see Appendix~\ref{app:readout} for further details).
We find that the drive populates the resonator with relatively low photon numbers of $\bar{n}\approx3.49$ for data mode in state $\ket{0}$ and $\bar{n} \approx 0.21$ for data mode in state $\ket{1}$.
This low photon number reduces the sensitivity to measurement-induced state transitions (MIST)~\cite{sank_measurement-induced_2016, dumas_measurement-induced_2024} and allows us to directly compute QND infidelity without relying on techniques such as branch analysis~\cite{dumas_measurement-induced_2024}.
To quantify QND infidelity, we wait \SI{50}{ns} for the resonator to ring down and then compute the overlap with respect to the initial state. 
For offset charge on the arm mode of $n_g=0$, this results in QND infidelities of \SI{2.0e-3}{} and \SI{3.5e-3}{} when the data mode is in the ground and excited state, respectively.
We repeat this simulation for varying offset charge on the arm mode and find that QND infidelity remains below \SI{1e-2}{} (see Appendix~\ref{app:readout}).
We note that leakage can be reduced with smaller $\kappa$ at the expense of increased readout time; for $\kappa/2\pi=\SI{70}{MHz}$, we obtain QND infidelities of \SI{1.5e-3}{} and \SI{0.3e-3}{} for the data mode in the ground and excited states, respectively.
For completeness, we repeat all simulations with $\kappa/2\pi=\SI{70}{MHz}$ and include the results in Table~\ref{tab:kappa-compare}.

This readout scheme allows us to suppress shot noise dephasing simply by replacing the $E_{J2}$ junction with a SQUID and tuning the arm mode away from the resonator when not performing readout.
Simulating this with the parameters in Table~\ref{tab:readout_parameters} shows suppression of the approximately \SI{270}{MHz} qubit-state-dependent frequency shift down to a cross-Kerr of $\chi/2\pi=10$ MHz when tuning $E_{J2}/2\pi$ down to \SI{13}{GHz}.
Combined with the relatively high resonator frequency of \SI{10.81}{GHz} and large $\kappa/2\pi=100$ MHz, this results in an expected thermal photon shot noise dephasing limit of \SI{15.8}{ms}.
We compute the shot noise dephasing rate according to~\cite{clerk_shot-noise_2007,rigetti_shot-noise_2012}
\begin{equation}
    \frac{1}{T_\phi} = \Gamma_\phi = \frac{\kappa}{2}\textrm{Re}\bigg[\sqrt{\bigg(1+\frac{i\chi n_\textrm{th}}{\kappa}\bigg)^2 + \frac{4i\chi n_\textrm{th}}{\kappa}} -1\bigg],
\end{equation}
where $n_\textrm{th}$ is the thermal occupation of the resonator, assuming an effective temperature of 45 mK~\cite{wang_cavity-attenuators_2019}.

We also note that this readout scheme does not require a Purcell filter because the data mode is only minimally mixed with the arm mode, resulting in a Purcell-limited lifetime of \SI{167}{ms}.
The Purcell lifetime is computed as
\begin{equation}
    \frac{1}{T_{1,\textrm{Purcell}}}  = |\bra{000}\hat{d}\ket{100}|^2,
\end{equation}
where the collapse operator $\hat d$ includes the resonator decay rate and is constructed as described in Appendix~\ref{app:readout}.

\section{Discussion}
A key advantage of the presented design is that it simultaneously provides high $E_c$ (low capacitance) in the fluxonium-like data modes and large capacitance in the transmon-like arm modes.
The relatively high $E_c$ of the data modes (compared to the arm modes or to typical transmon qubits) allows them to have large anharmonicity and suppressed flux noise sensitivity, as in a fluxonium qubit.
Meanwhile, the larger capacitance of the arm mode provides a generous capacitance budget with which to scale to a two-dimensional lattice.
With the parameters in Tables \ref{tab:gate_parameters} and \ref{tab:readout_parameters}, we see that coupling to a transmon coupler to perform two-qubit gates and coupling to a readout resonator each require a coupling capacitance around 5~fF.
In a two-dimensional tiling, we expect each arm mode to couple to four transmon couplers qubits plus one readout resonator, summing to a total coupling capacitance of $\sim25$~fF, which makes up only 66\% of the $\sim38$~fF total arm mode capacitance in the gates circuit.

We note also that the proposed amplitude-based readout naturally lends itself to readout using a Josephson photomultiplier~\cite{opremcak_JPM-readout_2018, opremcak_JPM-readout_2021}, which promises several advantages in scalability over the typical heterodyne detection approach, such as removing the need for a quantum-limited amplifier and potentially the associated isolators.

Our simulations indicate that the arm qubit can exhibit faster, higher-fidelity gates and readout compared to current superconducting qubit architectures while maintaining state-of-the-art coherence times.
In the context of quantum error correction, the improved fidelities will reduce the error correction overhead, and the faster operation times could enable faster error correction cycles.
The arm qubit therefore promises to accelerate the timeline towards fault-tolerant quantum computation.

\begin{acknowledgments}
This research was supported in part by the Army Research Office under award no. W911NF-23-1-045.
J.B.K acknowledges support from the MIT CQE-LPS Doc Bedard Fellowship. The authors acknowledge the MIT Supercloud and Lincoln Laboratory Supercomputing center for providing high performance computing resources that have contributed to the research results reported within this paper.
\end{acknowledgments}

\appendix

\section{Hierarchical Diagonalization}\label{app:hierarchical_diagonalization}
Throughout the paper, we employ a hierarchical diagonalization scheme~\cite{kerman_efficient-simulation_2020, chitta_scqubits_2022, groszkowski_scqubits_2021} in which subsystems of a given circuit are diagonalized independently, truncated in their dimension to include only some number of the lowest energy eigenstates, and then one at a time added together, diagonalized, and truncated.

Note that the potential energy for the arm qubit (Eq.~\ref{main_potential} of the main text) is periodic in the $\phi_1+\phi_2$ direction (with $\phi_1-\phi_2$ held constant).
To properly account for this periodicity when numerically diagonalizing the Hamiltonian, it is easiest to perform a change of variables following the method in~\cite{ding_mode-removal_2021}.
In summary, the technique is to define the new flux variables $\vec{\phi}'$ in terms of the old flux variables $\vec{\phi}$ with
\begin{equation}\label{eq:supp_phi_transform}
    \vec{\phi}' = W\vec{\phi}.
\end{equation}
To maintain commutation relations, this then requires redefining the charge operators as
\begin{equation}
    \vec{n}' = (W^T)^{-1}\vec{n},
\end{equation}
which has the effect of also transforming the charging energy matrix to
\begin{equation}\label{eq:supp_ec_transform}
    E_C' = WE_CW^T.
\end{equation}
We choose the matrix $W$ in each case to isolate the periodic degree of freedom, so that it can be initially diagonalized in the charge basis.
We then choose either the Fock basis or the charge basis to initially diagonalize the various modes of the circuit, depending on whether their potential is periodic in flux.

Below, we detail the diagonalization process for each of the circuits analyzed in the main text.

\subsection{CZ gate circuit}
\label{app:gates}
To simulate CZ gate fidelity between two arm qubits, we use the circuit from Fig.~\ref{fig:gate_circuit} A.
The Hamiltonian for this circuit can be written as
\begin{equation}\label{eq:gate_hamiltonian_notransform}
\begin{split}
    H =\ &\frac{E_{L12}}{2}(\hat{\phi}_1 - \hat{\phi}_2)^2 + E_{J12}\cos(\hat{\phi}_1 - \hat{\phi}_2) \\
    &-E_{J2}\cos(\hat{\phi}_2) - E_{J3}\cos(\hat{\phi}_3) - E_{J4}\cos(\hat{\phi}_4) \\
    &\frac{E_{L45}}{2}(\hat{\phi}_4 - \hat{\phi}_5)^2 + E_{J45}\cos(\hat{\phi}_4 - \hat{\phi}_5) \\
    & +\sum_{ij}4E^{-1}_{Cij}\hat{n}_i\hat{n}_j\:,
\end{split}
\end{equation}

where $E^{-1}_{Cij}$ is the charging energy matrix, defined from the capacitance matrix as
$$E_C = \frac{e^2}{2}C^{-1},$$
where $C^{-1}$ is the inverse capacitance matrix of the circuit and $e$ is the electron charge.

When constructing the capacitance matrix, we include a parasitic capacitance of 1 fF between nearest-neighbor nodes, 0.5 fF between next-nearest-neighbor nodes, and 0.1 fF between next-next-nearest-neighbor nodes. We neglect direct capacitance between the two data nodes (nodes 1 and 5) as that would correspond to a next-next-next-nearest neighbor parasitic capacitance.
We additionally include a capacitance in parallel with every Josephson junction, computed by assuming the junctions are fabricated with a critical current density of \SI{1}{\micro A/\micro m^2} and a capacitance per unit area of \SI{67}{fF/\micro m^2}.

We then perform a change of variables in order to isolate the periodicity of the cosine terms in the circuit potential energy. Following equations~\ref{eq:supp_phi_transform}-\ref{eq:supp_ec_transform}, we define the transformation matrix $W$ as
\begin{equation}
    W =
    \begin{pmatrix}
    1 & -1 & 0 & 0 & 0 \\
    0 & 1 & 0 & 0 & 0 \\
    0 & 0 & 1 & 0 & 0 \\
    0 & 0 & 0 & 1 & 0 \\
    0 & 0 & 0 & -1 & 1
    \end{pmatrix}.
\end{equation}
This allows us to rewrite the Hamiltonian in the transformed coordinates as 
\begin{equation}
\begin{split}
    H =\ &\frac{E_{L12}}{2}\hat{\phi}_1^2 + E_{J12}\cos(\hat{\phi}_1) \\
    &-E_{J2}\cos(\hat{\phi}_2) - E_{J3}\cos(\hat{\phi}_3) - E_{J4}\cos(\hat{\phi}_4) \\
    &\frac{E_{L45}}{2}\hat{\phi}_5^2 + E_{J45}\cos(\hat{\phi}_5) \\
    & \sum_{ij}4E^{'-1}_{Cij}\hat{n}_i\hat{n}_j\:
\end{split}
\end{equation}
Because each of the five modes of the circuit is primarily associated with a single node, we can separate out a ``bare'' Hamiltonian for each of these modes.
Then, because of the strong coupling within each of the arm qubits, it is computationally convenient to add the coupling between each data mode and its associated arm mode.
We next add the coupling between the left arm qubit and the transmon coupler and, finally, the coupling between the left three modes and the right two modes.
This process can be understood as dividing the full circuit Hamiltonian into
\begin{equation}
    \begin{split}
        H =\ & H_1 + H_2 + H_3 + H_4 + H_5\\
        & +H_{1,2} + H_{4,5} + H_{12,3} + H_{123,45}
    \end{split}
\end{equation}
where 
$$H_1 = \frac{E_{L12}}{2}\hat{\phi}_1^2 + E_{J12}\cos(\hat{\phi}_1) + 4E_{C11}^{-1}\hat{n}_1^2$$
$$H_2 = - E_{J2}\cos(\hat{\phi}_2) + 4E_{C22}^{-1}\hat{n}_2^2$$
$$H_3 = - E_{J3}\cos(\hat{\phi}_3) + 4E_{C33}^{-1}\hat{n}_3^2$$
$$H_4 = - E_{J4}\cos(\hat{\phi}_4) + 4E_{C44}^{-1}\hat{n}_4^2$$
$$H_5 = \frac{E_{L45}}{2}\hat{\phi}_5^2 + E_{J45}\cos(\hat{\phi}_5) + 4E_{C55}^{-1}\hat{n}_5^2$$
are the bare Hamiltonians seen by each mode. 
In the transformed coordinates, all couplings are capacitive, and the coupling Hamiltonians are given by
$$H_{1,2} = 8E_{C12}^{-1}\hat{n}_1\hat{n}_2$$
$$H_{4,5} = 8E_{C45}^{-1}\hat{n}_4\hat{n}_5$$
$$H_{12,3} = 8E_{C13}^{-1}\hat{n}_1\hat{n}_3 + 8E_{C23}^{-1}\hat{n}_2\hat{n}_3$$
$$H_{123, 45} = \sum_{i,j}8E_{Cij}^{-1}\hat{n}_i\hat{n}_j \quad\textrm{for}\ i\in(1,2,3),\ j\in(4,5)$$

To diagonalize the system, we begin by diagonalizing each of the bare modes.
For modes 1 and 5, which are not periodic in flux, we initially discretize the Hamiltonian in the Fock basis, with $\phi_{zpf}$ of each mode chosen to be $(2E_C)^{\frac{1}{4}}$.
For modes 2, 3, and 4, which are periodic in flux, we initially discretize the Hamiltonian in the charge basis.
After diagonalizing each of the bare modes, we truncate and keep a reduced number of eigenstates.

In the basis defined by diagonalizing the bare modes, we then diagonalize each arm qubit by adding in the appropriate coupling Hamiltonian.
For example, to diagonalize the leftmost arm qubit, we diagonalize the matrix $H_1 + H_2 + H_{1,2}$.
We then again truncate the dimension of each arm qubit Hamiltonian, and combine them with the transmon coupler by first by adding the transmon coupler to the left arm qubit and diagonalizing the matrix
$$H_1 + H_2 +H_3 + H_{1,2} + H_{12,3},$$
again truncating, and then finally diagonalizing the full Hamiltonian. 

\subsection{Readout circuit}
To simulate readout of an arm qubit, we use the circuit of Fig.~\ref{fig:readout_circuit}a. 
The Hamiltonian for this circuit is given by
\begin{equation}
\begin{aligned}
    H =\ &\frac{E_{L12}}{2}(\hat{\phi}_1 - \hat{\phi}_2)^2 + E_{J12}\cos(\hat{\phi}_1 - \hat{\phi}_2) \\
    &-E_{J2}\cos(\hat{\phi}_2) + \frac{E_{Lr}}{2}\hat{\phi}_r^2 \\
    &+\sum_{i,j \in \{1,2,r\}} 4 E_{C_{ij}}^{-1} \hat{n}_i \hat{n}_j\:.\\
\end{aligned}
\end{equation}

We follow many of the same assumptions as in the gates simulation. When constructing the capacitance matrix, a 1 fF parasitic capacitance is included between nearest-neighbor nodes and 0.1 fF parasitic capacitance between the next-nearest-neighbor nodes. 
Furthermore, a junction capacitance is included in parallel with every Josephson junction by assuming a critical current density of \SI{1}{\micro A/\micro m^2} and a capacitance per unit area of \SI{67}{fF/\micro m^2}
Finally, we again perform a change of variables to isolate the periodicity of the cosine terms in the circuit's potential energy, where our transformation matrix W is defined as
\begin{equation}
    W =
    \begin{pmatrix}
    1 & -1 & 0  \\
    0 & 1 & 0 \\
    0 & 0 & 1  \\
    \end{pmatrix}.
\end{equation}
Our Hamiltonian is rewritten in transformed coordinates as
\begin{equation}
\begin{aligned}
    H =\ &\frac{E_{L12}}{2}(\hat{\phi}_1)^2 + E_{J12}\cos(\hat{\phi}_1) \\
    &-E_{J2}\cos(\hat{\phi}_2) + \frac{E_{Lr}}{2}\hat{\phi}_r^2 \\
    &+\sum_{i,j \in \{1,2,r\}} 4 E_{C_{ij}}^{\prime -1} \hat{n}_i \hat{n}_j\:.
\end{aligned}
\end{equation}
As in the gates simulation, we separate out a ``bare'' Hamiltonian for each of the three modes, and then combine the modes one at a time.
The full circuit Hamiltonian is thus divided as
\begin{equation}
    H = H_1 + H_2 + H_r + H_{1,2} + H_{12,r}\:,
\end{equation}
where
\begin{equation}
    H_1 = \frac{E_{L12}}{2}\hat{\phi}_1^2 + E_{J12}\cos(\hat{\phi}_1) + 4E_{C11}^{-1}\hat{n}_1^2\:,
\end{equation}
\begin{equation}
    H_2 = - E_{J2}\cos(\hat{\phi}_2) + 4E_{C22}^{-1}\hat{n}_2^2\:,
\end{equation}
\begin{equation}
    H_r = \frac{E_{Lr}}{2}\hat{\phi}_r^2 + 4E_{Crr}^{-1}\hat{n}_r^2\:,
\end{equation}
\begin{equation}
    H_{1,2} = 8E_{C12}^{-1}\hat{n}_1\hat{n}_2\:,
\end{equation}
and
\begin{equation}
    H_{12,r} = 8E_{C1r}^{-1}\hat{n}_1\hat{n}_r + 8E_{C2r}^{-1}\hat{n}_2\hat{n}_r\:.
\end{equation}
We then follow the same iterative diagonalization procedure described in Appendix~\ref{app:gates}.

\section{Readout Simulation Methods}\label{app:readout}
For time domain simulation of the readout scheme, we follow a procedure similar to \cite{ye_ultrafast_2024}. We use the Linblad-form master equation for the density matrix $\rho$:

\begin{equation}
\dot{\rho}=-i\left[\hat{H}_0 + \hat{H}_d(t), \rho\right]+ \kappa \mathcal{D}[\hat{d}] \rho \:.
\label{eq:lindblad_me}
\end{equation}

The resonator drive is modeled as the time-dependent operator $\hat{H}_d(t) = \varepsilon(t)\hat{n}_r$ and the dissipator as $\mathcal{D}(\hat{d}) \rho ={\hat d} \rho {\hat d}^{\dagger}-\frac{1}{2}({\hat d}^{\dagger} {\hat d} \rho + \rho {\hat d}^{\dagger} {\hat d})$.
The collapse operator $\hat d$ is given by \cite{beaudoin_dissipation_2011}

\begin{equation}
\sqrt{\kappa} \hat{d} = \sum_{i, j>i} \sqrt{\kappa_{\text{eff},ij}} \big | \bra{e_j}\hat{\xi}\ket{e}_i \big| \ket{e_i} \bra{e_j},
\label{eq:kappa}
\end{equation}
where $\hat \xi$ is a normalized resonator mode charge operator $\hat{\xi} = \frac{\hat{n}_{r}}{\bra{000}\hat{n}_r\ket{001}}$, and the effective decay rate $\kappa_{\text{eff},ij}$ is given by~\cite{blais_circuit_2021}
\begin{equation}
    \kappa_{\text{eff},ij} = \kappa\big(\frac{\omega_{ij}}{\omega_r}\big)^2.
\end{equation}
Here, we have assumed a zero temperature bath where transitions only occur from high to low energy eigenstates, i.e. $\ket{e_j} \rightarrow \ket{e_i}$ with $j>i$.
Dissipations with rates below 1 MHz were discarded as much slower than the simulation timescale. 

To speed up the master equation simulation, we truncate the diagonalized Hamiltonian to include only the states that are significantly populated during the duration of the readout pulse.
We perform this truncation in an iterative manner by first truncating to an initial subset of states that are expected to be populated.
The truncated system is then evolved in time and any states whose populations never exceed $10^{-5}$ are discarded.
We then expand the simulation state space by considering the matrix elements of the drive and collapse operators and adding any states which have a matrix element greater than $10^{-2}$ to one of the simulation states.
We then repeat the simulation, again discarding states whose populations never exceed $10^{-5}$.
This process is repeated until the set of simulation states no longer changes, indicating convergence. 

\begin{figure}
\includegraphics[width=0.5\textwidth]{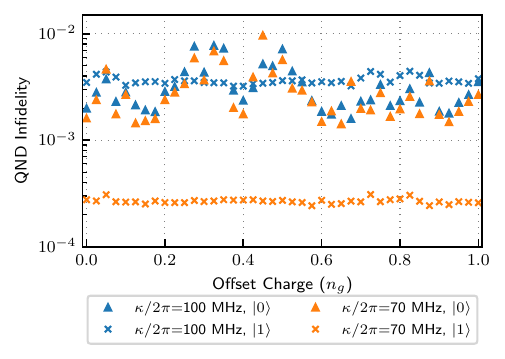}
\caption{\label{fig:offset_charge} QND infidelity as a function of gate charge $n_g$. Triangles represent QND infidelity when the data mode is in state $\ket{0}$, and crosses represent QND infidelity when the data mode is in state $\ket{1}$.}
\end{figure}

We evaluate the QND infidelity for different gate charge $n_g$ as shown in Fig.~\ref{fig:offset_charge}, and observe that the QND infidelity does not exceed $10^{-2}$.
Note that the QNDness for qubit in state $\ket{1}$ is insensitive to offset charge due to the very low photon population in the resonator.

To sample individual readout trajectories, we use QuTiP~\cite{lambert_qutip5_2024} to solve the stochastic Schrödinger equation.
We use the same collapse operator and set of truncated states as in the master equation simulation.
The measured signal is the real and imaginary parts of the collapse operator expectation value, which we integrate with weighting determined by the average contrast (between the signal for qubit in state $\ket{0}$ and state $\ket{1}$) at each timestep.

\section{Decoherence channels}\label{app:decoherence}

\subsection{Dielectric loss}
We model dielectric loss as voltage noise where the voltage spectral density for each capacitor is given by \cite{schoelkopf_qubits-as-spectrometers_2003,nguyen_high-coherence-fluxonium_2019, hays_harmonium_2025}
\begin{equation}
    S_{V}(\omega, C) = \frac{\hbar}{CQ}\big|(\coth(\frac{\hbar\omega}{2k_BT}) + 1)\big|,
\end{equation}
We then note that the capacitive energy terms in a circuit Hamiltonian can be rewritten as
\begin{equation}
    H_{cap} = \sum_{i\neq j}\frac{1}{2}C_{ij}(\hat{V}_i - \hat{V}_j)^2 + \sum_i\frac{1}{2}C_i\hat{V}_i^2\,,
\end{equation}
where $C_{ij}$ is the capacitance between nodes $i$ and $j$, $C_i$ is the capacitance from node $i$ to ground, and the $\hat{V}_i$ are node voltage operators defined from the charge operators as 
\begin{equation}
    \mathbf{\hat V} = 2e\mathbf{C^{-1}}\mathbf{\hat{n}},
\end{equation}
where $\mathbf{C^{-1}}$ is the inverse capacitance matrix.
We can now add an additional noisy voltage $\tilde{v}$ in series with any capacitance to obtain a term of the form
\begin{equation}
    \frac{1}{2}C_{ij}(\hat{V}_i - \hat{V}_j+ \tilde{v})^2.
\end{equation}
This shows that the relevant operators for Fermi's golden rule are of the form $C_{ij}(\hat{V}_i - \hat{V}_j)$ and $C_{i}\hat{V}_i$.
For each capacitor to ground, we can then evaluate a transition rate
\begin{equation}
    \Gamma_{C_i} = \frac{1}{\hbar^2}|\bra{1}C_{i}\hat{V}_i\ket{0}|^2(S_{V}(\omega, C_i) + S_{V}(-\omega, C_i)),
\end{equation}
where we included both up (heating) and down (relaxation) transition rates by evaluating the spectral density at $\pm\omega$.
For capacitors between two circuit nodes, we can similarly evaluate
\begin{equation}
    \Gamma_{C_{ij}} = \frac{1}{\hbar^2}|\bra{1}C_{ij}(\hat{V}_i - \hat{V}_j)\ket{0}|^2(S_{V}(\omega, C_{ij}) + S_{V}(-\omega, C_{ij}))
\end{equation}
Summing the dielectric loss rates for every capacitance in the circuit for two-qubit gates (Fig.~\ref{fig:gate_error}a, Table~\ref{tab:gate_parameters}) and assuming a dielectric quality factor $Q=3.5\times10^6$ and temperature $T = 20$ mK, we find that the dielectric loss T1 is typical for qubits at these relatively low frequencies, corresponding to T1 limits of 421 $\mu$s and 394 $\mu$s for the left and right data modes of Fig.~\ref{fig:gate_error}a) respectively.

\subsection{Flux noise}
Noise in the flux threading the arm qubit loop can cause both dephasing ($T_\phi$) and bit flip ($T_1$) errors.
We compute dephasing errors by generating a 1-hour flux noise time series $\tilde\phi(t)$ with power spectral density given by
\begin{equation}
    S_\Phi(\omega) = A_\Phi^2\bigg(\frac{2\pi\times1\textrm{Hz}}{\omega}\bigg),
\end{equation}
with $A_\Phi^2 = (1\mu\Phi_0)^2/\textrm{Hz}$.
We then diagonalize the circuit Hamiltonian for a range of $\tilde\phi(t)$ values and interpolate to convert $\tilde\phi(t)$ into a fluctuating qubit frequency $\omega(t)$.
Integrating short segments of $\omega(t)$ with a sign flip halfway through allows us to simulate repeated echo measurements, which we average to obtain echo dephasing times of 1.64 ms and 1.79 ms for the left and right data modes of Fig.~\ref{fig:gate_error}a) respectively.

We compute $T_1$ errors according to Fermi's golden rule \cite{krantz_quantum_engineer} 
\begin{equation}\label{eq:flux_noise}
\Gamma_{1, \Phi}=\left|\bra{0}\frac{\partial H}{\partial \Phi}\ket{1}\right|^2 S_{\Phi}\left(\omega_q\right).
\end{equation}
We use $\frac{\partial H}{\partial \Phi} = \frac{2\pi}{\Phi_0}E_{L12}(\hat\phi_1 - \hat\phi_2)$ for qubit 1 (similar for qubit 2), corresponding to writing the flux bias on the inductor in the untransformed varables of equation~\ref{eq:gate_hamiltonian_notransform}, which results in an irrotational Hamiltonian as described in \cite{you_circuit-quantization-flux_2019}. This imposes $T_1$ limits of 1.21 ms for the left data mode of the Fig.~\ref{fig:gate_error}a circuit and 1.3 ms for the right data mode. 

\subsection{Charge noise}
The transmon-like arm modes of the arm qubits have a relatively low $E_J/E_C\approx 40$.
This means that the data mode frequencies depend weakly on the offset gate charge of the arm modes.
We estimate the charge noise dephasing time of the data modes by including the offset charge in the CZ gate circuit Hamiltonian, such that the capacitive energy term becomes
\begin{equation}
    \sum_{ij}4E^{-1}_{Cij}(\hat{n}_i- n_{g,i})(\hat{n}_j - n_{g, j}),
\end{equation}
where the offset charge $n_{g,i}$ is always zero for the data modes.
By diagonalizing the Hamiltonian with varying offset charge, we find that the data mode frequency dependence on the neighboring arm mode offset charge is approximately sinusoidal, following
\begin{equation}
    \omega_q \approx \omega_0 + A\cos(2\pi n_g),
\end{equation}
for some average frequency $\omega_0$ and amplitude of charge-dependent frequency shift $A$.
This cosine dependence is consistent with the arm mode having the charge dispersion of a typical transmon~\cite{koch_charge-insensitive_2007}.
We then pessimistically compute the dephasing rate $\Gamma_\phi$ as the average deviation from the mean frequency
\begin{equation}
    \Gamma_\phi = \int_0^1dn_g|A\cos(2\pi n_g)| = \frac{A}{\pi}.
\end{equation}
This estimate corresponds to assuming that the offset charge is randomly set to a static value.
If instead the offset charge fluctuates, as it typically does in experiment, this would reduce the dephasing rate seen by the qubit.

\subsection{Quasiparticles}
Quasiparticles tunneling across a Josephson junction can cause $T_1$ errors.
These can be modeled as~\cite{nguyen_fluxonium_blueprint}

\begin{equation}
    \Gamma_\mathrm{qp} =  \left|\bra{0}\sin \frac{\hat{\phi}}{2}\ket{1}\right|^2 \frac{8 E_{J}}{\pi \hbar} x_{\mathrm{qp}} \sqrt{\frac{2 \Delta}{\hbar \omega_{q}}}\;,
\end{equation}
where $\hat{\phi}$ is the phase across the junction, $E_J$ is the energy of the junction, $x_{\mathrm{qp}}$ is the quasiparticle density, and $\Delta$ is the superconducting gap of the superconductor used.
In fluxonium qubits biased at $\phi_{ext} = \pi$, this matrix element goes to zero and quasiparticle losses are limited to the junction array~\cite{pop_coherent-suppression_2014}.
The same is true of the arm qubit data modes — the matrix element for $\left|\bra{0}\sin \frac{\hat{\phi}}{2}\ket{1}\right|^2$ across the $E_{J12}$ junction in an arm qubit is zero when the qubit is biased to half a flux quantum.
We therefore expect that the primary $T_1$ contribution from quasiparticles occurs in the junction array of the arm qubit, following a similar expression~\cite{nguyen_fluxonium_blueprint}
\begin{equation}\label{eq:array_qp}
    \Gamma_\mathrm{qp} = \left|\bra{0}\frac{\hat{\phi}}{2}\ket{1}\right|^2 \frac{8 E_{L}}{\pi \hbar} x_{\mathrm{qp}} \sqrt{\frac{2 \Delta}{\hbar \omega_{q}}}\;,
\end{equation}
where $E_L$ is the inductive energy of the junction array.
Experimental measurements of fluxonium coherence times, however, suggest that the relevant quasiparticle density $x_\mathrm{qp}$ of the array is significantly lower than the $10^{-9} - 10^{-7}$ values typically measured near junctions~\cite{somoroff_millisecond_2023}.
To our knowledge, the contribution to loss from quasiparticles in an array is too small to have been accurately measured yet, as fluxonium qubits tend to be dominated by dielectric loss instead~\cite{somoroff_millisecond_2023}.
Additionally, it has been shown~\cite{mcewen_resisting_2024} that gap engineering can significantly reduce quasiparticle-induced relaxation.
We therefore neglect quasiparticle dissipation in the coherence times of these circuits quoted in the main text.

It is possible to use data such as that of~\cite{somoroff_millisecond_2023} to pessimistically upper bound the quasiparticle density in the array to $x_\mathrm{qp}\approx2\times10^{-9}$ by assuming that \textit{all} of the qubit relaxation is caused by quasiparticles.
Substituting this value into equation~\ref{eq:array_qp} and using the $E_L$ values from table~\ref{tab:gate_parameters}, we get quasiparticle-induced relaxation rates  $1/\Gamma_\mathrm{qp} =$ 540 $\mu$s and 560 $\mu$s for the left and right qubits respectively.

\bibliography{apssamp}

\begin{thebibliography}{64}%
\makeatletter
\providecommand \@ifxundefined [1]{%
 \@ifx{#1\undefined}
}%
\providecommand \@ifnum [1]{%
 \ifnum #1\expandafter \@firstoftwo
 \else \expandafter \@secondoftwo
 \fi
}%
\providecommand \@ifx [1]{%
 \ifx #1\expandafter \@firstoftwo
 \else \expandafter \@secondoftwo
 \fi
}%
\providecommand \natexlab [1]{#1}%
\providecommand \enquote  [1]{``#1''}%
\providecommand \bibnamefont  [1]{#1}%
\providecommand \bibfnamefont [1]{#1}%
\providecommand \citenamefont [1]{#1}%
\providecommand \href@noop [0]{\@secondoftwo}%
\providecommand \href [0]{\begingroup \@sanitize@url \@href}%
\providecommand \@href[1]{\@@startlink{#1}\@@href}%
\providecommand \@@href[1]{\endgroup#1\@@endlink}%
\providecommand \@sanitize@url [0]{\catcode `\\12\catcode `\$12\catcode `\&12\catcode `\#12\catcode `\^12\catcode `\_12\catcode `\%12\relax}%
\providecommand \@@startlink[1]{}%
\providecommand \@@endlink[0]{}%
\providecommand \url  [0]{\begingroup\@sanitize@url \@url }%
\providecommand \@url [1]{\endgroup\@href {#1}{\urlprefix }}%
\providecommand \urlprefix  [0]{URL }%
\providecommand \Eprint [0]{\href }%
\providecommand \doibase [0]{https://doi.org/}%
\providecommand \selectlanguage [0]{\@gobble}%
\providecommand \bibinfo  [0]{\@secondoftwo}%
\providecommand \bibfield  [0]{\@secondoftwo}%
\providecommand \translation [1]{[#1]}%
\providecommand \BibitemOpen [0]{}%
\providecommand \bibitemStop [0]{}%
\providecommand \bibitemNoStop [0]{.\EOS\space}%
\providecommand \EOS [0]{\spacefactor3000\relax}%
\providecommand \BibitemShut  [1]{\csname bibitem#1\endcsname}%
\let\auto@bib@innerbib\@empty
\bibitem [{\citenamefont {{Google Quantum AI and Collaborators}}(2025)}]{google_error-correction_2025}%
  \BibitemOpen
  \bibfield  {author} {\bibinfo {author} {\bibnamefont {{Google Quantum AI and Collaborators}}},\ }\bibfield  {title} {\bibinfo {title} {Quantum error correction below the surface code threshold},\ }\href {https://doi.org/10.1038/s41586-024-08449-y} {\bibfield  {journal} {\bibinfo  {journal} {Nature}\ }\textbf {\bibinfo {volume} {638}},\ \bibinfo {pages} {920} (\bibinfo {year} {2025})}\BibitemShut {NoStop}%
\bibitem [{\citenamefont {Krinner}\ \emph {et~al.}(2022)\citenamefont {Krinner}, \citenamefont {Lacroix}, \citenamefont {Remm}, \citenamefont {Di~Paolo}, \citenamefont {Genois}, \citenamefont {Leroux}, \citenamefont {Hellings}, \citenamefont {Lazar}, \citenamefont {Swiadek}, \citenamefont {Herrmann}, \citenamefont {Norris}, \citenamefont {Andersen}, \citenamefont {M{\"u}ller}, \citenamefont {Blais}, \citenamefont {Eichler},\ and\ \citenamefont {Wallraff}}]{krinner_realizing_2022}%
  \BibitemOpen
  \bibfield  {author} {\bibinfo {author} {\bibfnamefont {S.}~\bibnamefont {Krinner}}, \bibinfo {author} {\bibfnamefont {N.}~\bibnamefont {Lacroix}}, \bibinfo {author} {\bibfnamefont {A.}~\bibnamefont {Remm}}, \bibinfo {author} {\bibfnamefont {A.}~\bibnamefont {Di~Paolo}}, \bibinfo {author} {\bibfnamefont {E.}~\bibnamefont {Genois}}, \bibinfo {author} {\bibfnamefont {C.}~\bibnamefont {Leroux}}, \bibinfo {author} {\bibfnamefont {C.}~\bibnamefont {Hellings}}, \bibinfo {author} {\bibfnamefont {S.}~\bibnamefont {Lazar}}, \bibinfo {author} {\bibfnamefont {F.}~\bibnamefont {Swiadek}}, \bibinfo {author} {\bibfnamefont {J.}~\bibnamefont {Herrmann}}, \bibinfo {author} {\bibfnamefont {G.~J.}\ \bibnamefont {Norris}}, \bibinfo {author} {\bibfnamefont {C.~K.}\ \bibnamefont {Andersen}}, \bibinfo {author} {\bibfnamefont {M.}~\bibnamefont {M{\"u}ller}}, \bibinfo {author} {\bibfnamefont {A.}~\bibnamefont {Blais}}, \bibinfo {author} {\bibfnamefont {C.}~\bibnamefont {Eichler}},\ and\ \bibinfo {author} {\bibfnamefont
  {A.}~\bibnamefont {Wallraff}},\ }\bibfield  {title} {\bibinfo {title} {Realizing repeated quantum error correction in a distance-three surface code},\ }\href {https://doi.org/10.1038/s41586-022-04566-8} {\bibfield  {journal} {\bibinfo  {journal} {Nature}\ }\textbf {\bibinfo {volume} {605}},\ \bibinfo {pages} {669} (\bibinfo {year} {2022})}\BibitemShut {NoStop}%
\bibitem [{\citenamefont {Putterman}\ \emph {et~al.}(2025)\citenamefont {Putterman}, \citenamefont {Noh}, \citenamefont {Hann}, \citenamefont {MacCabe}, \citenamefont {Aghaeimeibodi}, \citenamefont {Patel}, \citenamefont {Lee}, \citenamefont {Jones}, \citenamefont {Moradinejad}, \citenamefont {Rodriguez}, \citenamefont {Mahuli}, \citenamefont {Rose}, \citenamefont {Owens}, \citenamefont {Levine}, \citenamefont {Rosenfeld}, \citenamefont {Reinhold}, \citenamefont {Moncelsi}, \citenamefont {Alcid}, \citenamefont {Alidoust}, \citenamefont {{Arrangoiz-Arriola}}, \citenamefont {Barnett}, \citenamefont {Bienias}, \citenamefont {Carson}, \citenamefont {Chen}, \citenamefont {Chen}, \citenamefont {Chinkezian}, \citenamefont {Chisholm}, \citenamefont {Chou}, \citenamefont {Clerk}, \citenamefont {Clifford}, \citenamefont {Cosmic}, \citenamefont {Curiel}, \citenamefont {Davis}, \citenamefont {DeLorenzo}, \citenamefont {D'Ewart}, \citenamefont {Diky}, \citenamefont {D'Souza}, \citenamefont {Dumitrescu}, \citenamefont
  {Eisenmann}, \citenamefont {Elkhouly}, \citenamefont {Evenbly}, \citenamefont {Fang}, \citenamefont {Fang}, \citenamefont {Fling}, \citenamefont {Fon}, \citenamefont {Garcia}, \citenamefont {Gorshkov}, \citenamefont {Grant}, \citenamefont {Gray}, \citenamefont {Grimberg}, \citenamefont {Grimsmo}, \citenamefont {Haim}, \citenamefont {Hand}, \citenamefont {He}, \citenamefont {Hernandez}, \citenamefont {Hover}, \citenamefont {Hung}, \citenamefont {Hunt}, \citenamefont {Iverson}, \citenamefont {Jarrige}, \citenamefont {Jaskula}, \citenamefont {Jiang}, \citenamefont {Kalaee}, \citenamefont {Karabalin}, \citenamefont {Karalekas}, \citenamefont {Keller}, \citenamefont {Khalajhedayati}, \citenamefont {Kubica}, \citenamefont {Lee}, \citenamefont {Leroux}, \citenamefont {Lieu}, \citenamefont {Ly}, \citenamefont {Madrigal}, \citenamefont {Marcaud}, \citenamefont {McCabe}, \citenamefont {Miles}, \citenamefont {Milsted}, \citenamefont {Minguzzi}, \citenamefont {Mishra}, \citenamefont {Mukherjee}, \citenamefont
  {Naghiloo}, \citenamefont {Oblepias}, \citenamefont {Ortuno}, \citenamefont {Pagdilao}, \citenamefont {Pancotti}, \citenamefont {Panduro}, \citenamefont {Paquette}, \citenamefont {Park}, \citenamefont {Peairs}, \citenamefont {Perello}, \citenamefont {Peterson}, \citenamefont {Ponte}, \citenamefont {Preskill}, \citenamefont {Qiao}, \citenamefont {Refael}, \citenamefont {Resnick}, \citenamefont {Retzker}, \citenamefont {Reyna}, \citenamefont {Runyan}, \citenamefont {Ryan}, \citenamefont {Sahmoud}, \citenamefont {Sanchez}, \citenamefont {Sanil}, \citenamefont {Sankar}, \citenamefont {Sato}, \citenamefont {Scaffidi}, \citenamefont {Siavoshi}, \citenamefont {Sivarajah}, \citenamefont {Skogland}, \citenamefont {Su}, \citenamefont {Swenson}, \citenamefont {Teo}, \citenamefont {Tomada}, \citenamefont {Torlai}, \citenamefont {Wollack}, \citenamefont {Ye}, \citenamefont {Zerrudo}, \citenamefont {Zhang}, \citenamefont {Brand{\~a}o}, \citenamefont {Matheny},\ and\ \citenamefont
  {Painter}}]{putterman_hardware-efficient_2025}%
  \BibitemOpen
  \bibfield  {author} {\bibinfo {author} {\bibfnamefont {H.}~\bibnamefont {Putterman}}, \bibinfo {author} {\bibfnamefont {K.}~\bibnamefont {Noh}}, \bibinfo {author} {\bibfnamefont {C.~T.}\ \bibnamefont {Hann}}, \bibinfo {author} {\bibfnamefont {G.~S.}\ \bibnamefont {MacCabe}}, \bibinfo {author} {\bibfnamefont {S.}~\bibnamefont {Aghaeimeibodi}}, \bibinfo {author} {\bibfnamefont {R.~N.}\ \bibnamefont {Patel}}, \bibinfo {author} {\bibfnamefont {M.}~\bibnamefont {Lee}}, \bibinfo {author} {\bibfnamefont {W.~M.}\ \bibnamefont {Jones}}, \bibinfo {author} {\bibfnamefont {H.}~\bibnamefont {Moradinejad}}, \bibinfo {author} {\bibfnamefont {R.}~\bibnamefont {Rodriguez}}, \bibinfo {author} {\bibfnamefont {N.}~\bibnamefont {Mahuli}}, \bibinfo {author} {\bibfnamefont {J.}~\bibnamefont {Rose}}, \bibinfo {author} {\bibfnamefont {J.~C.}\ \bibnamefont {Owens}}, \bibinfo {author} {\bibfnamefont {H.}~\bibnamefont {Levine}}, \bibinfo {author} {\bibfnamefont {E.}~\bibnamefont {Rosenfeld}}, \bibinfo {author} {\bibfnamefont
  {P.}~\bibnamefont {Reinhold}}, \bibinfo {author} {\bibfnamefont {L.}~\bibnamefont {Moncelsi}}, \bibinfo {author} {\bibfnamefont {J.~A.}\ \bibnamefont {Alcid}}, \bibinfo {author} {\bibfnamefont {N.}~\bibnamefont {Alidoust}}, \bibinfo {author} {\bibfnamefont {P.}~\bibnamefont {{Arrangoiz-Arriola}}}, \bibinfo {author} {\bibfnamefont {J.}~\bibnamefont {Barnett}}, \bibinfo {author} {\bibfnamefont {P.}~\bibnamefont {Bienias}}, \bibinfo {author} {\bibfnamefont {H.~A.}\ \bibnamefont {Carson}}, \bibinfo {author} {\bibfnamefont {C.}~\bibnamefont {Chen}}, \bibinfo {author} {\bibfnamefont {L.}~\bibnamefont {Chen}}, \bibinfo {author} {\bibfnamefont {H.}~\bibnamefont {Chinkezian}}, \bibinfo {author} {\bibfnamefont {E.~M.}\ \bibnamefont {Chisholm}}, \bibinfo {author} {\bibfnamefont {M.-H.}\ \bibnamefont {Chou}}, \bibinfo {author} {\bibfnamefont {A.}~\bibnamefont {Clerk}}, \bibinfo {author} {\bibfnamefont {A.}~\bibnamefont {Clifford}}, \bibinfo {author} {\bibfnamefont {R.}~\bibnamefont {Cosmic}}, \bibinfo {author}
  {\bibfnamefont {A.~V.}\ \bibnamefont {Curiel}}, \bibinfo {author} {\bibfnamefont {E.}~\bibnamefont {Davis}}, \bibinfo {author} {\bibfnamefont {L.}~\bibnamefont {DeLorenzo}}, \bibinfo {author} {\bibfnamefont {J.~M.}\ \bibnamefont {D'Ewart}}, \bibinfo {author} {\bibfnamefont {A.}~\bibnamefont {Diky}}, \bibinfo {author} {\bibfnamefont {N.}~\bibnamefont {D'Souza}}, \bibinfo {author} {\bibfnamefont {P.~T.}\ \bibnamefont {Dumitrescu}}, \bibinfo {author} {\bibfnamefont {S.}~\bibnamefont {Eisenmann}}, \bibinfo {author} {\bibfnamefont {E.}~\bibnamefont {Elkhouly}}, \bibinfo {author} {\bibfnamefont {G.}~\bibnamefont {Evenbly}}, \bibinfo {author} {\bibfnamefont {M.~T.}\ \bibnamefont {Fang}}, \bibinfo {author} {\bibfnamefont {Y.}~\bibnamefont {Fang}}, \bibinfo {author} {\bibfnamefont {M.~J.}\ \bibnamefont {Fling}}, \bibinfo {author} {\bibfnamefont {W.}~\bibnamefont {Fon}}, \bibinfo {author} {\bibfnamefont {G.}~\bibnamefont {Garcia}}, \bibinfo {author} {\bibfnamefont {A.~V.}\ \bibnamefont {Gorshkov}}, \bibinfo {author}
  {\bibfnamefont {J.~A.}\ \bibnamefont {Grant}}, \bibinfo {author} {\bibfnamefont {M.~J.}\ \bibnamefont {Gray}}, \bibinfo {author} {\bibfnamefont {S.}~\bibnamefont {Grimberg}}, \bibinfo {author} {\bibfnamefont {A.~L.}\ \bibnamefont {Grimsmo}}, \bibinfo {author} {\bibfnamefont {A.}~\bibnamefont {Haim}}, \bibinfo {author} {\bibfnamefont {J.}~\bibnamefont {Hand}}, \bibinfo {author} {\bibfnamefont {Y.}~\bibnamefont {He}}, \bibinfo {author} {\bibfnamefont {M.}~\bibnamefont {Hernandez}}, \bibinfo {author} {\bibfnamefont {D.}~\bibnamefont {Hover}}, \bibinfo {author} {\bibfnamefont {J.~S.~C.}\ \bibnamefont {Hung}}, \bibinfo {author} {\bibfnamefont {M.}~\bibnamefont {Hunt}}, \bibinfo {author} {\bibfnamefont {J.}~\bibnamefont {Iverson}}, \bibinfo {author} {\bibfnamefont {I.}~\bibnamefont {Jarrige}}, \bibinfo {author} {\bibfnamefont {J.-C.}\ \bibnamefont {Jaskula}}, \bibinfo {author} {\bibfnamefont {L.}~\bibnamefont {Jiang}}, \bibinfo {author} {\bibfnamefont {M.}~\bibnamefont {Kalaee}}, \bibinfo {author} {\bibfnamefont
  {R.}~\bibnamefont {Karabalin}}, \bibinfo {author} {\bibfnamefont {P.~J.}\ \bibnamefont {Karalekas}}, \bibinfo {author} {\bibfnamefont {A.~J.}\ \bibnamefont {Keller}}, \bibinfo {author} {\bibfnamefont {A.}~\bibnamefont {Khalajhedayati}}, \bibinfo {author} {\bibfnamefont {A.}~\bibnamefont {Kubica}}, \bibinfo {author} {\bibfnamefont {H.}~\bibnamefont {Lee}}, \bibinfo {author} {\bibfnamefont {C.}~\bibnamefont {Leroux}}, \bibinfo {author} {\bibfnamefont {S.}~\bibnamefont {Lieu}}, \bibinfo {author} {\bibfnamefont {V.}~\bibnamefont {Ly}}, \bibinfo {author} {\bibfnamefont {K.~V.}\ \bibnamefont {Madrigal}}, \bibinfo {author} {\bibfnamefont {G.}~\bibnamefont {Marcaud}}, \bibinfo {author} {\bibfnamefont {G.}~\bibnamefont {McCabe}}, \bibinfo {author} {\bibfnamefont {C.}~\bibnamefont {Miles}}, \bibinfo {author} {\bibfnamefont {A.}~\bibnamefont {Milsted}}, \bibinfo {author} {\bibfnamefont {J.}~\bibnamefont {Minguzzi}}, \bibinfo {author} {\bibfnamefont {A.}~\bibnamefont {Mishra}}, \bibinfo {author} {\bibfnamefont
  {B.}~\bibnamefont {Mukherjee}}, \bibinfo {author} {\bibfnamefont {M.}~\bibnamefont {Naghiloo}}, \bibinfo {author} {\bibfnamefont {E.}~\bibnamefont {Oblepias}}, \bibinfo {author} {\bibfnamefont {G.}~\bibnamefont {Ortuno}}, \bibinfo {author} {\bibfnamefont {J.}~\bibnamefont {Pagdilao}}, \bibinfo {author} {\bibfnamefont {N.}~\bibnamefont {Pancotti}}, \bibinfo {author} {\bibfnamefont {A.}~\bibnamefont {Panduro}}, \bibinfo {author} {\bibfnamefont {J.~P.}\ \bibnamefont {Paquette}}, \bibinfo {author} {\bibfnamefont {M.}~\bibnamefont {Park}}, \bibinfo {author} {\bibfnamefont {G.~A.}\ \bibnamefont {Peairs}}, \bibinfo {author} {\bibfnamefont {D.}~\bibnamefont {Perello}}, \bibinfo {author} {\bibfnamefont {E.~C.}\ \bibnamefont {Peterson}}, \bibinfo {author} {\bibfnamefont {S.}~\bibnamefont {Ponte}}, \bibinfo {author} {\bibfnamefont {J.}~\bibnamefont {Preskill}}, \bibinfo {author} {\bibfnamefont {J.}~\bibnamefont {Qiao}}, \bibinfo {author} {\bibfnamefont {G.}~\bibnamefont {Refael}}, \bibinfo {author} {\bibfnamefont
  {R.}~\bibnamefont {Resnick}}, \bibinfo {author} {\bibfnamefont {A.}~\bibnamefont {Retzker}}, \bibinfo {author} {\bibfnamefont {O.~A.}\ \bibnamefont {Reyna}}, \bibinfo {author} {\bibfnamefont {M.}~\bibnamefont {Runyan}}, \bibinfo {author} {\bibfnamefont {C.~A.}\ \bibnamefont {Ryan}}, \bibinfo {author} {\bibfnamefont {A.}~\bibnamefont {Sahmoud}}, \bibinfo {author} {\bibfnamefont {E.}~\bibnamefont {Sanchez}}, \bibinfo {author} {\bibfnamefont {R.}~\bibnamefont {Sanil}}, \bibinfo {author} {\bibfnamefont {K.}~\bibnamefont {Sankar}}, \bibinfo {author} {\bibfnamefont {Y.}~\bibnamefont {Sato}}, \bibinfo {author} {\bibfnamefont {T.}~\bibnamefont {Scaffidi}}, \bibinfo {author} {\bibfnamefont {S.}~\bibnamefont {Siavoshi}}, \bibinfo {author} {\bibfnamefont {P.}~\bibnamefont {Sivarajah}}, \bibinfo {author} {\bibfnamefont {T.}~\bibnamefont {Skogland}}, \bibinfo {author} {\bibfnamefont {C.-J.}\ \bibnamefont {Su}}, \bibinfo {author} {\bibfnamefont {L.~J.}\ \bibnamefont {Swenson}}, \bibinfo {author} {\bibfnamefont {S.~M.}\
  \bibnamefont {Teo}}, \bibinfo {author} {\bibfnamefont {A.}~\bibnamefont {Tomada}}, \bibinfo {author} {\bibfnamefont {G.}~\bibnamefont {Torlai}}, \bibinfo {author} {\bibfnamefont {E.~A.}\ \bibnamefont {Wollack}}, \bibinfo {author} {\bibfnamefont {Y.}~\bibnamefont {Ye}}, \bibinfo {author} {\bibfnamefont {J.~A.}\ \bibnamefont {Zerrudo}}, \bibinfo {author} {\bibfnamefont {K.}~\bibnamefont {Zhang}}, \bibinfo {author} {\bibfnamefont {F.~G. S.~L.}\ \bibnamefont {Brand{\~a}o}}, \bibinfo {author} {\bibfnamefont {M.~H.}\ \bibnamefont {Matheny}},\ and\ \bibinfo {author} {\bibfnamefont {O.}~\bibnamefont {Painter}},\ }\bibfield  {title} {\bibinfo {title} {Hardware-efficient quantum error correction via concatenated bosonic qubits},\ }\href {https://doi.org/10.1038/s41586-025-08642-7} {\bibfield  {journal} {\bibinfo  {journal} {Nature}\ }\textbf {\bibinfo {volume} {638}},\ \bibinfo {pages} {927} (\bibinfo {year} {2025})}\BibitemShut {NoStop}%
\bibitem [{\citenamefont {Rower}\ \emph {et~al.}(2024)\citenamefont {Rower}, \citenamefont {Ding}, \citenamefont {Zhang}, \citenamefont {Hays}, \citenamefont {An}, \citenamefont {Harrington}, \citenamefont {Rosen}, \citenamefont {Gertler}, \citenamefont {Hazard}, \citenamefont {Niedzielski}, \citenamefont {Schwartz}, \citenamefont {Gustavsson}, \citenamefont {Serniak}, \citenamefont {Grover},\ and\ \citenamefont {Oliver}}]{rower_suppressing_2024}%
  \BibitemOpen
  \bibfield  {author} {\bibinfo {author} {\bibfnamefont {D.~A.}\ \bibnamefont {Rower}}, \bibinfo {author} {\bibfnamefont {L.}~\bibnamefont {Ding}}, \bibinfo {author} {\bibfnamefont {H.}~\bibnamefont {Zhang}}, \bibinfo {author} {\bibfnamefont {M.}~\bibnamefont {Hays}}, \bibinfo {author} {\bibfnamefont {J.}~\bibnamefont {An}}, \bibinfo {author} {\bibfnamefont {P.~M.}\ \bibnamefont {Harrington}}, \bibinfo {author} {\bibfnamefont {I.~T.}\ \bibnamefont {Rosen}}, \bibinfo {author} {\bibfnamefont {J.~M.}\ \bibnamefont {Gertler}}, \bibinfo {author} {\bibfnamefont {T.~M.}\ \bibnamefont {Hazard}}, \bibinfo {author} {\bibfnamefont {B.~M.}\ \bibnamefont {Niedzielski}}, \bibinfo {author} {\bibfnamefont {M.~E.}\ \bibnamefont {Schwartz}}, \bibinfo {author} {\bibfnamefont {S.}~\bibnamefont {Gustavsson}}, \bibinfo {author} {\bibfnamefont {K.}~\bibnamefont {Serniak}}, \bibinfo {author} {\bibfnamefont {J.~A.}\ \bibnamefont {Grover}},\ and\ \bibinfo {author} {\bibfnamefont {W.~D.}\ \bibnamefont {Oliver}},\ }\bibfield  {title}
  {\bibinfo {title} {Suppressing {{Counter-Rotating Errors}} for {{Fast Single-Qubit Gates}} with {{Fluxonium}}},\ }\href {https://doi.org/10.1103/PRXQuantum.5.040342} {\bibfield  {journal} {\bibinfo  {journal} {PRX Quantum}\ }\textbf {\bibinfo {volume} {5}},\ \bibinfo {pages} {040342} (\bibinfo {year} {2024})}\BibitemShut {NoStop}%
\bibitem [{\citenamefont {Blais}\ \emph {et~al.}(2021)\citenamefont {Blais}, \citenamefont {Grimsmo}, \citenamefont {Girvin},\ and\ \citenamefont {Wallraff}}]{blais_circuit_2021}%
  \BibitemOpen
  \bibfield  {author} {\bibinfo {author} {\bibfnamefont {A.}~\bibnamefont {Blais}}, \bibinfo {author} {\bibfnamefont {A.~L.}\ \bibnamefont {Grimsmo}}, \bibinfo {author} {\bibfnamefont {S.~M.}\ \bibnamefont {Girvin}},\ and\ \bibinfo {author} {\bibfnamefont {A.}~\bibnamefont {Wallraff}},\ }\bibfield  {title} {\bibinfo {title} {Circuit {{Quantum Electrodynamics}}},\ }\href {https://doi.org/10.1103/RevModPhys.93.025005} {\bibfield  {journal} {\bibinfo  {journal} {Reviews of Modern Physics}\ }\textbf {\bibinfo {volume} {93}},\ \bibinfo {pages} {025005} (\bibinfo {year} {2021})}\BibitemShut {NoStop}%
\bibitem [{\citenamefont {Krantz}\ \emph {et~al.}(2019)\citenamefont {Krantz}, \citenamefont {Kjaergaard}, \citenamefont {Yan}, \citenamefont {Orlando}, \citenamefont {Gustavsson},\ and\ \citenamefont {Oliver}}]{krantz_quantum_engineer}%
  \BibitemOpen
  \bibfield  {author} {\bibinfo {author} {\bibfnamefont {P.}~\bibnamefont {Krantz}}, \bibinfo {author} {\bibfnamefont {M.}~\bibnamefont {Kjaergaard}}, \bibinfo {author} {\bibfnamefont {F.}~\bibnamefont {Yan}}, \bibinfo {author} {\bibfnamefont {T.~P.}\ \bibnamefont {Orlando}}, \bibinfo {author} {\bibfnamefont {S.}~\bibnamefont {Gustavsson}},\ and\ \bibinfo {author} {\bibfnamefont {W.~D.}\ \bibnamefont {Oliver}},\ }\bibfield  {title} {\bibinfo {title} {A quantum engineer's guide to superconducting qubits},\ }\href@noop {} {\bibfield  {journal} {\bibinfo  {journal} {Applied physics reviews}\ }\textbf {\bibinfo {volume} {6}} (\bibinfo {year} {2019})}\BibitemShut {NoStop}%
\bibitem [{\citenamefont {Sung}\ \emph {et~al.}(2021)\citenamefont {Sung}, \citenamefont {Ding}, \citenamefont {Braum{\"u}ller}, \citenamefont {Veps{\"a}l{\"a}inen}, \citenamefont {Kannan}, \citenamefont {Kjaergaard}, \citenamefont {Greene}, \citenamefont {Samach}, \citenamefont {McNally}, \citenamefont {Kim}, \citenamefont {Melville}, \citenamefont {Niedzielski}, \citenamefont {Schwartz}, \citenamefont {Yoder}, \citenamefont {Orlando}, \citenamefont {Gustavsson},\ and\ \citenamefont {Oliver}}]{sung_realization_2021}%
  \BibitemOpen
  \bibfield  {author} {\bibinfo {author} {\bibfnamefont {Y.}~\bibnamefont {Sung}}, \bibinfo {author} {\bibfnamefont {L.}~\bibnamefont {Ding}}, \bibinfo {author} {\bibfnamefont {J.}~\bibnamefont {Braum{\"u}ller}}, \bibinfo {author} {\bibfnamefont {A.}~\bibnamefont {Veps{\"a}l{\"a}inen}}, \bibinfo {author} {\bibfnamefont {B.}~\bibnamefont {Kannan}}, \bibinfo {author} {\bibfnamefont {M.}~\bibnamefont {Kjaergaard}}, \bibinfo {author} {\bibfnamefont {A.}~\bibnamefont {Greene}}, \bibinfo {author} {\bibfnamefont {G.~O.}\ \bibnamefont {Samach}}, \bibinfo {author} {\bibfnamefont {C.}~\bibnamefont {McNally}}, \bibinfo {author} {\bibfnamefont {D.}~\bibnamefont {Kim}}, \bibinfo {author} {\bibfnamefont {A.}~\bibnamefont {Melville}}, \bibinfo {author} {\bibfnamefont {B.~M.}\ \bibnamefont {Niedzielski}}, \bibinfo {author} {\bibfnamefont {M.~E.}\ \bibnamefont {Schwartz}}, \bibinfo {author} {\bibfnamefont {J.~L.}\ \bibnamefont {Yoder}}, \bibinfo {author} {\bibfnamefont {T.~P.}\ \bibnamefont {Orlando}}, \bibinfo {author}
  {\bibfnamefont {S.}~\bibnamefont {Gustavsson}},\ and\ \bibinfo {author} {\bibfnamefont {W.~D.}\ \bibnamefont {Oliver}},\ }\bibfield  {title} {\bibinfo {title} {Realization of {{High-Fidelity CZ}} and {{Z Z}} -{{Free iSWAP Gates}} with a {{Tunable Coupler}}},\ }\href {https://doi.org/10.1103/PhysRevX.11.021058} {\bibfield  {journal} {\bibinfo  {journal} {Physical Review X}\ }\textbf {\bibinfo {volume} {11}},\ \bibinfo {pages} {021058} (\bibinfo {year} {2021})}\BibitemShut {NoStop}%
\bibitem [{\citenamefont {Kandala}\ \emph {et~al.}(2021)\citenamefont {Kandala}, \citenamefont {Wei}, \citenamefont {Srinivasan}, \citenamefont {Magesan}, \citenamefont {Carnevale}, \citenamefont {Keefe}, \citenamefont {Klaus}, \citenamefont {Dial},\ and\ \citenamefont {McKay}}]{kandala_cnot-fixed-frequency_2021}%
  \BibitemOpen
  \bibfield  {author} {\bibinfo {author} {\bibfnamefont {A.}~\bibnamefont {Kandala}}, \bibinfo {author} {\bibfnamefont {K.~X.}\ \bibnamefont {Wei}}, \bibinfo {author} {\bibfnamefont {S.}~\bibnamefont {Srinivasan}}, \bibinfo {author} {\bibfnamefont {E.}~\bibnamefont {Magesan}}, \bibinfo {author} {\bibfnamefont {S.}~\bibnamefont {Carnevale}}, \bibinfo {author} {\bibfnamefont {G.~A.}\ \bibnamefont {Keefe}}, \bibinfo {author} {\bibfnamefont {D.}~\bibnamefont {Klaus}}, \bibinfo {author} {\bibfnamefont {O.}~\bibnamefont {Dial}},\ and\ \bibinfo {author} {\bibfnamefont {D.~C.}\ \bibnamefont {McKay}},\ }\bibfield  {title} {\bibinfo {title} {Demonstration of a high-fidelity cnot gate for fixed-frequency transmons with engineered $zz$ suppression},\ }\href {https://doi.org/10.1103/PhysRevLett.127.130501} {\bibfield  {journal} {\bibinfo  {journal} {Phys. Rev. Lett.}\ }\textbf {\bibinfo {volume} {127}},\ \bibinfo {pages} {130501} (\bibinfo {year} {2021})}\BibitemShut {NoStop}%
\bibitem [{\citenamefont {Ding}\ \emph {et~al.}(2023)\citenamefont {Ding}, \citenamefont {Hays}, \citenamefont {Sung}, \citenamefont {Kannan}, \citenamefont {An}, \citenamefont {Di~Paolo}, \citenamefont {Karamlou}, \citenamefont {Hazard}, \citenamefont {Azar}, \citenamefont {Kim}, \citenamefont {Niedzielski}, \citenamefont {Melville}, \citenamefont {Schwartz}, \citenamefont {Yoder}, \citenamefont {Orlando}, \citenamefont {Gustavsson}, \citenamefont {Grover}, \citenamefont {Serniak},\ and\ \citenamefont {Oliver}}]{ding_high-fidelity_2023}%
  \BibitemOpen
  \bibfield  {author} {\bibinfo {author} {\bibfnamefont {L.}~\bibnamefont {Ding}}, \bibinfo {author} {\bibfnamefont {M.}~\bibnamefont {Hays}}, \bibinfo {author} {\bibfnamefont {Y.}~\bibnamefont {Sung}}, \bibinfo {author} {\bibfnamefont {B.}~\bibnamefont {Kannan}}, \bibinfo {author} {\bibfnamefont {J.}~\bibnamefont {An}}, \bibinfo {author} {\bibfnamefont {A.}~\bibnamefont {Di~Paolo}}, \bibinfo {author} {\bibfnamefont {A.~H.}\ \bibnamefont {Karamlou}}, \bibinfo {author} {\bibfnamefont {T.~M.}\ \bibnamefont {Hazard}}, \bibinfo {author} {\bibfnamefont {K.}~\bibnamefont {Azar}}, \bibinfo {author} {\bibfnamefont {D.~K.}\ \bibnamefont {Kim}}, \bibinfo {author} {\bibfnamefont {B.~M.}\ \bibnamefont {Niedzielski}}, \bibinfo {author} {\bibfnamefont {A.}~\bibnamefont {Melville}}, \bibinfo {author} {\bibfnamefont {M.~E.}\ \bibnamefont {Schwartz}}, \bibinfo {author} {\bibfnamefont {J.~L.}\ \bibnamefont {Yoder}}, \bibinfo {author} {\bibfnamefont {T.~P.}\ \bibnamefont {Orlando}}, \bibinfo {author} {\bibfnamefont
  {S.}~\bibnamefont {Gustavsson}}, \bibinfo {author} {\bibfnamefont {J.~A.}\ \bibnamefont {Grover}}, \bibinfo {author} {\bibfnamefont {K.}~\bibnamefont {Serniak}},\ and\ \bibinfo {author} {\bibfnamefont {W.~D.}\ \bibnamefont {Oliver}},\ }\href@noop {} {\bibinfo {title} {High-{{Fidelity}}, {{Frequency-Flexible Two-Qubit Fluxonium Gates}} with a {{Transmon Coupler}}}} (\bibinfo {year} {2023}),\ \Eprint {https://arxiv.org/abs/2304.06087} {arXiv:2304.06087 [quant-ph]} \BibitemShut {NoStop}%
\bibitem [{\citenamefont {Sank}\ \emph {et~al.}(2016)\citenamefont {Sank}, \citenamefont {Chen}, \citenamefont {Khezri}, \citenamefont {Kelly}, \citenamefont {Barends}, \citenamefont {Campbell}, \citenamefont {Chen}, \citenamefont {Chiaro}, \citenamefont {Dunsworth}, \citenamefont {Fowler}, \citenamefont {Jeffrey}, \citenamefont {Lucero}, \citenamefont {Megrant}, \citenamefont {Mutus}, \citenamefont {Neeley}, \citenamefont {Neill}, \citenamefont {O'Malley}, \citenamefont {Quintana}, \citenamefont {Roushan}, \citenamefont {Vainsencher}, \citenamefont {White}, \citenamefont {Wenner}, \citenamefont {Korotkov},\ and\ \citenamefont {Martinis}}]{sank_measurement-induced_2016}%
  \BibitemOpen
  \bibfield  {author} {\bibinfo {author} {\bibfnamefont {D.}~\bibnamefont {Sank}}, \bibinfo {author} {\bibfnamefont {Z.}~\bibnamefont {Chen}}, \bibinfo {author} {\bibfnamefont {M.}~\bibnamefont {Khezri}}, \bibinfo {author} {\bibfnamefont {J.}~\bibnamefont {Kelly}}, \bibinfo {author} {\bibfnamefont {R.}~\bibnamefont {Barends}}, \bibinfo {author} {\bibfnamefont {B.}~\bibnamefont {Campbell}}, \bibinfo {author} {\bibfnamefont {Y.}~\bibnamefont {Chen}}, \bibinfo {author} {\bibfnamefont {B.}~\bibnamefont {Chiaro}}, \bibinfo {author} {\bibfnamefont {A.}~\bibnamefont {Dunsworth}}, \bibinfo {author} {\bibfnamefont {A.}~\bibnamefont {Fowler}}, \bibinfo {author} {\bibfnamefont {E.}~\bibnamefont {Jeffrey}}, \bibinfo {author} {\bibfnamefont {E.}~\bibnamefont {Lucero}}, \bibinfo {author} {\bibfnamefont {A.}~\bibnamefont {Megrant}}, \bibinfo {author} {\bibfnamefont {J.}~\bibnamefont {Mutus}}, \bibinfo {author} {\bibfnamefont {M.}~\bibnamefont {Neeley}}, \bibinfo {author} {\bibfnamefont {C.}~\bibnamefont {Neill}}, \bibinfo
  {author} {\bibfnamefont {P.~J.~J.}\ \bibnamefont {O'Malley}}, \bibinfo {author} {\bibfnamefont {C.}~\bibnamefont {Quintana}}, \bibinfo {author} {\bibfnamefont {P.}~\bibnamefont {Roushan}}, \bibinfo {author} {\bibfnamefont {A.}~\bibnamefont {Vainsencher}}, \bibinfo {author} {\bibfnamefont {T.}~\bibnamefont {White}}, \bibinfo {author} {\bibfnamefont {J.}~\bibnamefont {Wenner}}, \bibinfo {author} {\bibfnamefont {A.~N.}\ \bibnamefont {Korotkov}},\ and\ \bibinfo {author} {\bibfnamefont {J.~M.}\ \bibnamefont {Martinis}},\ }\bibfield  {title} {\bibinfo {title} {Measurement-{{Induced State Transitions}} in a {{Superconducting Qubit}}: {{Beyond}} the {{Rotating Wave Approximation}}},\ }\href {https://doi.org/10.1103/PhysRevLett.117.190503} {\bibfield  {journal} {\bibinfo  {journal} {Physical Review Letters}\ }\textbf {\bibinfo {volume} {117}},\ \bibinfo {pages} {190503} (\bibinfo {year} {2016})}\BibitemShut {NoStop}%
\bibitem [{\citenamefont {Dumas}\ \emph {et~al.}(2024)\citenamefont {Dumas}, \citenamefont {{Groleau-Par{\'e}}}, \citenamefont {McDonald}, \citenamefont {{Mu{\~n}oz-Arias}}, \citenamefont {Lled{\'o}}, \citenamefont {D'Anjou},\ and\ \citenamefont {Blais}}]{dumas_measurement-induced_2024}%
  \BibitemOpen
  \bibfield  {author} {\bibinfo {author} {\bibfnamefont {M.~F.}\ \bibnamefont {Dumas}}, \bibinfo {author} {\bibfnamefont {B.}~\bibnamefont {{Groleau-Par{\'e}}}}, \bibinfo {author} {\bibfnamefont {A.}~\bibnamefont {McDonald}}, \bibinfo {author} {\bibfnamefont {M.~H.}\ \bibnamefont {{Mu{\~n}oz-Arias}}}, \bibinfo {author} {\bibfnamefont {C.}~\bibnamefont {Lled{\'o}}}, \bibinfo {author} {\bibfnamefont {B.}~\bibnamefont {D'Anjou}},\ and\ \bibinfo {author} {\bibfnamefont {A.}~\bibnamefont {Blais}},\ }\bibfield  {title} {\bibinfo {title} {Measurement-{{Induced Transmon Ionization}}},\ }\href {https://doi.org/10.1103/PhysRevX.14.041023} {\bibfield  {journal} {\bibinfo  {journal} {Physical Review X}\ }\textbf {\bibinfo {volume} {14}},\ \bibinfo {pages} {041023} (\bibinfo {year} {2024})}\BibitemShut {NoStop}%
\bibitem [{\citenamefont {Chapple}\ \emph {et~al.}(2025)\citenamefont {Chapple}, \citenamefont {{Benhayoune-Khadraoui}}, \citenamefont {Richer},\ and\ \citenamefont {Blais}}]{chapple_balanced_2025}%
  \BibitemOpen
  \bibfield  {author} {\bibinfo {author} {\bibfnamefont {A.~A.}\ \bibnamefont {Chapple}}, \bibinfo {author} {\bibfnamefont {O.}~\bibnamefont {{Benhayoune-Khadraoui}}}, \bibinfo {author} {\bibfnamefont {S.}~\bibnamefont {Richer}},\ and\ \bibinfo {author} {\bibfnamefont {A.}~\bibnamefont {Blais}},\ }\href {https://doi.org/10.48550/arXiv.2501.09010} {\bibinfo {title} {Balanced cross-{{Kerr}} coupling for superconducting qubit readout}} (\bibinfo {year} {2025}),\ \Eprint {https://arxiv.org/abs/2501.09010} {arXiv:2501.09010 [quant-ph]} \BibitemShut {NoStop}%
\bibitem [{\citenamefont {Ye}\ \emph {et~al.}(2024)\citenamefont {Ye}, \citenamefont {Kline}, \citenamefont {Chen}, \citenamefont {Yen},\ and\ \citenamefont {O'Brien}}]{ye_ultrafast_2024}%
  \BibitemOpen
  \bibfield  {author} {\bibinfo {author} {\bibfnamefont {Y.}~\bibnamefont {Ye}}, \bibinfo {author} {\bibfnamefont {J.~B.}\ \bibnamefont {Kline}}, \bibinfo {author} {\bibfnamefont {S.}~\bibnamefont {Chen}}, \bibinfo {author} {\bibfnamefont {A.}~\bibnamefont {Yen}},\ and\ \bibinfo {author} {\bibfnamefont {K.~P.}\ \bibnamefont {O'Brien}},\ }\bibfield  {title} {\bibinfo {title} {Ultrafast superconducting qubit readout with the quarton coupler},\ }\href {https://doi.org/10.1126/sciadv.ado9094} {\bibfield  {journal} {\bibinfo  {journal} {Science Advances}\ }\textbf {\bibinfo {volume} {10}},\ \bibinfo {pages} {eado9094} (\bibinfo {year} {2024})}\BibitemShut {NoStop}%
\bibitem [{\citenamefont {Wang}\ \emph {et~al.}(2024)\citenamefont {Wang}, \citenamefont {Liu}, \citenamefont {Chen}, \citenamefont {Du}, \citenamefont {Ying}, \citenamefont {Wang}, \citenamefont {Huo}, \citenamefont {Peng}, \citenamefont {Zhu}, \citenamefont {Chen}, \citenamefont {Lu},\ and\ \citenamefont {Pan}}]{wang_999-fidelity_2024}%
  \BibitemOpen
  \bibfield  {author} {\bibinfo {author} {\bibfnamefont {C.}~\bibnamefont {Wang}}, \bibinfo {author} {\bibfnamefont {F.-M.}\ \bibnamefont {Liu}}, \bibinfo {author} {\bibfnamefont {H.}~\bibnamefont {Chen}}, \bibinfo {author} {\bibfnamefont {Y.-F.}\ \bibnamefont {Du}}, \bibinfo {author} {\bibfnamefont {C.}~\bibnamefont {Ying}}, \bibinfo {author} {\bibfnamefont {J.-W.}\ \bibnamefont {Wang}}, \bibinfo {author} {\bibfnamefont {Y.-H.}\ \bibnamefont {Huo}}, \bibinfo {author} {\bibfnamefont {C.-Z.}\ \bibnamefont {Peng}}, \bibinfo {author} {\bibfnamefont {X.}~\bibnamefont {Zhu}}, \bibinfo {author} {\bibfnamefont {M.-C.}\ \bibnamefont {Chen}}, \bibinfo {author} {\bibfnamefont {C.-Y.}\ \bibnamefont {Lu}},\ and\ \bibinfo {author} {\bibfnamefont {J.-W.}\ \bibnamefont {Pan}},\ }\href {https://arxiv.org/abs/2412.13849} {\bibinfo {title} {99.9\%-fidelity in measuring a superconducting qubit}} (\bibinfo {year} {2024}),\ \Eprint {https://arxiv.org/abs/2412.13849} {arXiv:2412.13849 [quant-ph]} \BibitemShut {NoStop}%
\bibitem [{\citenamefont {Gambetta}\ \emph {et~al.}(2011)\citenamefont {Gambetta}, \citenamefont {Houck},\ and\ \citenamefont {Blais}}]{gambetta_tunable-coupling-qubit_2011}%
  \BibitemOpen
  \bibfield  {author} {\bibinfo {author} {\bibfnamefont {J.~M.}\ \bibnamefont {Gambetta}}, \bibinfo {author} {\bibfnamefont {A.~A.}\ \bibnamefont {Houck}},\ and\ \bibinfo {author} {\bibfnamefont {A.}~\bibnamefont {Blais}},\ }\bibfield  {title} {\bibinfo {title} {Superconducting qubit with purcell protection and tunable coupling},\ }\href {https://doi.org/10.1103/PhysRevLett.106.030502} {\bibfield  {journal} {\bibinfo  {journal} {Phys. Rev. Lett.}\ }\textbf {\bibinfo {volume} {106}},\ \bibinfo {pages} {030502} (\bibinfo {year} {2011})}\BibitemShut {NoStop}%
\bibitem [{\citenamefont {Zhang}\ \emph {et~al.}(2017)\citenamefont {Zhang}, \citenamefont {Liu}, \citenamefont {Raftery},\ and\ \citenamefont {Houck}}]{zhang_suppression_2017}%
  \BibitemOpen
  \bibfield  {author} {\bibinfo {author} {\bibfnamefont {G.}~\bibnamefont {Zhang}}, \bibinfo {author} {\bibfnamefont {Y.}~\bibnamefont {Liu}}, \bibinfo {author} {\bibfnamefont {J.~J.}\ \bibnamefont {Raftery}},\ and\ \bibinfo {author} {\bibfnamefont {A.~A.}\ \bibnamefont {Houck}},\ }\bibfield  {title} {\bibinfo {title} {Suppression of photon shot noise dephasing in a tunable coupling superconducting qubit},\ }\href {https://doi.org/10.1038/s41534-016-0002-2} {\bibfield  {journal} {\bibinfo  {journal} {npj Quantum Information}\ }\textbf {\bibinfo {volume} {3}},\ \bibinfo {pages} {1} (\bibinfo {year} {2017})}\BibitemShut {NoStop}%
\bibitem [{\citenamefont {Hazra}\ \emph {et~al.}(2025)\citenamefont {Hazra}, \citenamefont {Dai}, \citenamefont {Connolly}, \citenamefont {Kurilovich}, \citenamefont {Wang}, \citenamefont {Frunzio},\ and\ \citenamefont {Devoret}}]{hazra_benchmarking_2025}%
  \BibitemOpen
  \bibfield  {author} {\bibinfo {author} {\bibfnamefont {S.}~\bibnamefont {Hazra}}, \bibinfo {author} {\bibfnamefont {W.}~\bibnamefont {Dai}}, \bibinfo {author} {\bibfnamefont {T.}~\bibnamefont {Connolly}}, \bibinfo {author} {\bibfnamefont {P.~D.}\ \bibnamefont {Kurilovich}}, \bibinfo {author} {\bibfnamefont {Z.}~\bibnamefont {Wang}}, \bibinfo {author} {\bibfnamefont {L.}~\bibnamefont {Frunzio}},\ and\ \bibinfo {author} {\bibfnamefont {M.~H.}\ \bibnamefont {Devoret}},\ }\bibfield  {title} {\bibinfo {title} {Benchmarking the readout of a superconducting qubit for repeated measurements},\ }\href {https://doi.org/10.1103/PhysRevLett.134.100601} {\bibfield  {journal} {\bibinfo  {journal} {Phys. Rev. Lett.}\ }\textbf {\bibinfo {volume} {134}},\ \bibinfo {pages} {100601} (\bibinfo {year} {2025})}\BibitemShut {NoStop}%
\bibitem [{\citenamefont {Diniz}\ \emph {et~al.}(2013)\citenamefont {Diniz}, \citenamefont {Dumur}, \citenamefont {Buisson},\ and\ \citenamefont {Auff\`eves}}]{diniz_ultrafast_2013}%
  \BibitemOpen
  \bibfield  {author} {\bibinfo {author} {\bibfnamefont {I.}~\bibnamefont {Diniz}}, \bibinfo {author} {\bibfnamefont {E.}~\bibnamefont {Dumur}}, \bibinfo {author} {\bibfnamefont {O.}~\bibnamefont {Buisson}},\ and\ \bibinfo {author} {\bibfnamefont {A.}~\bibnamefont {Auff\`eves}},\ }\bibfield  {title} {\bibinfo {title} {Ultrafast quantum nondemolition measurements based on a diamond-shaped artificial atom},\ }\href {https://doi.org/10.1103/PhysRevA.87.033837} {\bibfield  {journal} {\bibinfo  {journal} {Phys. Rev. A}\ }\textbf {\bibinfo {volume} {87}},\ \bibinfo {pages} {033837} (\bibinfo {year} {2013})}\BibitemShut {NoStop}%
\bibitem [{\citenamefont {Roy}\ \emph {et~al.}(2017)\citenamefont {Roy}, \citenamefont {Kundu}, \citenamefont {Chand}, \citenamefont {Hazra}, \citenamefont {Nehra}, \citenamefont {Cosmic}, \citenamefont {Ranadive}, \citenamefont {Patankar}, \citenamefont {Damle},\ and\ \citenamefont {Vijay}}]{roy_jrm_2017}%
  \BibitemOpen
  \bibfield  {author} {\bibinfo {author} {\bibfnamefont {T.}~\bibnamefont {Roy}}, \bibinfo {author} {\bibfnamefont {S.}~\bibnamefont {Kundu}}, \bibinfo {author} {\bibfnamefont {M.}~\bibnamefont {Chand}}, \bibinfo {author} {\bibfnamefont {S.}~\bibnamefont {Hazra}}, \bibinfo {author} {\bibfnamefont {N.}~\bibnamefont {Nehra}}, \bibinfo {author} {\bibfnamefont {R.}~\bibnamefont {Cosmic}}, \bibinfo {author} {\bibfnamefont {A.}~\bibnamefont {Ranadive}}, \bibinfo {author} {\bibfnamefont {M.~P.}\ \bibnamefont {Patankar}}, \bibinfo {author} {\bibfnamefont {K.}~\bibnamefont {Damle}},\ and\ \bibinfo {author} {\bibfnamefont {R.}~\bibnamefont {Vijay}},\ }\bibfield  {title} {\bibinfo {title} {Implementation of pairwise longitudinal coupling in a three-qubit superconducting circuit},\ }\href {https://doi.org/10.1103/PhysRevApplied.7.054025} {\bibfield  {journal} {\bibinfo  {journal} {Phys. Rev. Appl.}\ }\textbf {\bibinfo {volume} {7}},\ \bibinfo {pages} {054025} (\bibinfo {year} {2017})}\BibitemShut {NoStop}%
\bibitem [{\citenamefont {Pfeiffer}\ \emph {et~al.}(2024)\citenamefont {Pfeiffer}, \citenamefont {Werninghaus}, \citenamefont {Schweizer}, \citenamefont {Bruckmoser}, \citenamefont {Koch}, \citenamefont {Glaser}, \citenamefont {Huber}, \citenamefont {Bunch}, \citenamefont {Haslbeck}, \citenamefont {Knudsen}, \citenamefont {Krylov}, \citenamefont {Liegener}, \citenamefont {Marx}, \citenamefont {Richard}, \citenamefont {Romeiro}, \citenamefont {Roy}, \citenamefont {Schirk}, \citenamefont {Schneider}, \citenamefont {Singh}, \citenamefont {S\"odergren}, \citenamefont {Tsitsilin}, \citenamefont {Wallner}, \citenamefont {Riofr\'{\i}o},\ and\ \citenamefont {Filipp}}]{pfeiffer_pmon_2024}%
  \BibitemOpen
  \bibfield  {author} {\bibinfo {author} {\bibfnamefont {F.}~\bibnamefont {Pfeiffer}}, \bibinfo {author} {\bibfnamefont {M.}~\bibnamefont {Werninghaus}}, \bibinfo {author} {\bibfnamefont {C.}~\bibnamefont {Schweizer}}, \bibinfo {author} {\bibfnamefont {N.}~\bibnamefont {Bruckmoser}}, \bibinfo {author} {\bibfnamefont {L.}~\bibnamefont {Koch}}, \bibinfo {author} {\bibfnamefont {N.~J.}\ \bibnamefont {Glaser}}, \bibinfo {author} {\bibfnamefont {G.~B.~P.}\ \bibnamefont {Huber}}, \bibinfo {author} {\bibfnamefont {D.}~\bibnamefont {Bunch}}, \bibinfo {author} {\bibfnamefont {F.~X.}\ \bibnamefont {Haslbeck}}, \bibinfo {author} {\bibfnamefont {M.}~\bibnamefont {Knudsen}}, \bibinfo {author} {\bibfnamefont {G.}~\bibnamefont {Krylov}}, \bibinfo {author} {\bibfnamefont {K.}~\bibnamefont {Liegener}}, \bibinfo {author} {\bibfnamefont {A.}~\bibnamefont {Marx}}, \bibinfo {author} {\bibfnamefont {L.}~\bibnamefont {Richard}}, \bibinfo {author} {\bibfnamefont {J.~H.}\ \bibnamefont {Romeiro}}, \bibinfo {author} {\bibfnamefont
  {F.~A.}\ \bibnamefont {Roy}}, \bibinfo {author} {\bibfnamefont {J.}~\bibnamefont {Schirk}}, \bibinfo {author} {\bibfnamefont {C.}~\bibnamefont {Schneider}}, \bibinfo {author} {\bibfnamefont {M.}~\bibnamefont {Singh}}, \bibinfo {author} {\bibfnamefont {L.}~\bibnamefont {S\"odergren}}, \bibinfo {author} {\bibfnamefont {I.}~\bibnamefont {Tsitsilin}}, \bibinfo {author} {\bibfnamefont {F.}~\bibnamefont {Wallner}}, \bibinfo {author} {\bibfnamefont {C.~A.}\ \bibnamefont {Riofr\'{\i}o}},\ and\ \bibinfo {author} {\bibfnamefont {S.}~\bibnamefont {Filipp}},\ }\bibfield  {title} {\bibinfo {title} {Efficient decoupling of a nonlinear qubit mode from its environment},\ }\href {https://doi.org/10.1103/PhysRevX.14.041007} {\bibfield  {journal} {\bibinfo  {journal} {Phys. Rev. X}\ }\textbf {\bibinfo {volume} {14}},\ \bibinfo {pages} {041007} (\bibinfo {year} {2024})}\BibitemShut {NoStop}%
\bibitem [{\citenamefont {Finck}\ \emph {et~al.}(2021)\citenamefont {Finck}, \citenamefont {Carnevale}, \citenamefont {Klaus}, \citenamefont {Scerbo}, \citenamefont {Blair}, \citenamefont {McConkey}, \citenamefont {Kurter}, \citenamefont {Carniol}, \citenamefont {Keefe}, \citenamefont {Kumph},\ and\ \citenamefont {Dial}}]{finck_tcq-gates_2021}%
  \BibitemOpen
  \bibfield  {author} {\bibinfo {author} {\bibfnamefont {A.}~\bibnamefont {Finck}}, \bibinfo {author} {\bibfnamefont {S.}~\bibnamefont {Carnevale}}, \bibinfo {author} {\bibfnamefont {D.}~\bibnamefont {Klaus}}, \bibinfo {author} {\bibfnamefont {C.}~\bibnamefont {Scerbo}}, \bibinfo {author} {\bibfnamefont {J.}~\bibnamefont {Blair}}, \bibinfo {author} {\bibfnamefont {T.}~\bibnamefont {McConkey}}, \bibinfo {author} {\bibfnamefont {C.}~\bibnamefont {Kurter}}, \bibinfo {author} {\bibfnamefont {A.}~\bibnamefont {Carniol}}, \bibinfo {author} {\bibfnamefont {G.}~\bibnamefont {Keefe}}, \bibinfo {author} {\bibfnamefont {M.}~\bibnamefont {Kumph}},\ and\ \bibinfo {author} {\bibfnamefont {O.}~\bibnamefont {Dial}},\ }\bibfield  {title} {\bibinfo {title} {Suppressed crosstalk between two-junction superconducting qubits with mode-selective exchange coupling},\ }\bibfield  {journal} {\bibinfo  {journal} {Physical Review Applied}\ }\textbf {\bibinfo {volume} {16}},\ \href {https://doi.org/10.1103/physrevapplied.16.054041}
  {10.1103/physrevapplied.16.054041} (\bibinfo {year} {2021})\BibitemShut {NoStop}%
\bibitem [{\citenamefont {Dassonneville}\ \emph {et~al.}(2020)\citenamefont {Dassonneville}, \citenamefont {Ramos}, \citenamefont {Milchakov}, \citenamefont {Planat}, \citenamefont {Dumur}, \citenamefont {Foroughi}, \citenamefont {Puertas}, \citenamefont {Leger}, \citenamefont {Bharadwaj}, \citenamefont {Delaforce}, \citenamefont {Naud}, \citenamefont {Hasch-Guichard}, \citenamefont {Garc\'{\i}a-Ripoll}, \citenamefont {Roch},\ and\ \citenamefont {Buisson}}]{dassonneville_transmon-molecule_2020}%
  \BibitemOpen
  \bibfield  {author} {\bibinfo {author} {\bibfnamefont {R.}~\bibnamefont {Dassonneville}}, \bibinfo {author} {\bibfnamefont {T.}~\bibnamefont {Ramos}}, \bibinfo {author} {\bibfnamefont {V.}~\bibnamefont {Milchakov}}, \bibinfo {author} {\bibfnamefont {L.}~\bibnamefont {Planat}}, \bibinfo {author} {\bibfnamefont {E.}~\bibnamefont {Dumur}}, \bibinfo {author} {\bibfnamefont {F.}~\bibnamefont {Foroughi}}, \bibinfo {author} {\bibfnamefont {J.}~\bibnamefont {Puertas}}, \bibinfo {author} {\bibfnamefont {S.}~\bibnamefont {Leger}}, \bibinfo {author} {\bibfnamefont {K.}~\bibnamefont {Bharadwaj}}, \bibinfo {author} {\bibfnamefont {J.}~\bibnamefont {Delaforce}}, \bibinfo {author} {\bibfnamefont {C.}~\bibnamefont {Naud}}, \bibinfo {author} {\bibfnamefont {W.}~\bibnamefont {Hasch-Guichard}}, \bibinfo {author} {\bibfnamefont {J.~J.}\ \bibnamefont {Garc\'{\i}a-Ripoll}}, \bibinfo {author} {\bibfnamefont {N.}~\bibnamefont {Roch}},\ and\ \bibinfo {author} {\bibfnamefont {O.}~\bibnamefont {Buisson}},\ }\bibfield  {title}
  {\bibinfo {title} {Fast high-fidelity quantum nondemolition qubit readout via a nonperturbative cross-kerr coupling},\ }\href {https://doi.org/10.1103/PhysRevX.10.011045} {\bibfield  {journal} {\bibinfo  {journal} {Phys. Rev. X}\ }\textbf {\bibinfo {volume} {10}},\ \bibinfo {pages} {011045} (\bibinfo {year} {2020})}\BibitemShut {NoStop}%
\bibitem [{\citenamefont {Dassonneville}\ \emph {et~al.}(2023)\citenamefont {Dassonneville}, \citenamefont {Ramos}, \citenamefont {Milchakov}, \citenamefont {Mori}, \citenamefont {Planat}, \citenamefont {Foroughi}, \citenamefont {Naud}, \citenamefont {Hasch-Guichard}, \citenamefont {Garc\'{\i}a-Ripoll}, \citenamefont {Roch},\ and\ \citenamefont {Buisson}}]{dassonneville_transmon-molecule-bifurcation_2023}%
  \BibitemOpen
  \bibfield  {author} {\bibinfo {author} {\bibfnamefont {R.}~\bibnamefont {Dassonneville}}, \bibinfo {author} {\bibfnamefont {T.}~\bibnamefont {Ramos}}, \bibinfo {author} {\bibfnamefont {V.}~\bibnamefont {Milchakov}}, \bibinfo {author} {\bibfnamefont {C.}~\bibnamefont {Mori}}, \bibinfo {author} {\bibfnamefont {L.}~\bibnamefont {Planat}}, \bibinfo {author} {\bibfnamefont {F.}~\bibnamefont {Foroughi}}, \bibinfo {author} {\bibfnamefont {C.}~\bibnamefont {Naud}}, \bibinfo {author} {\bibfnamefont {W.}~\bibnamefont {Hasch-Guichard}}, \bibinfo {author} {\bibfnamefont {J.}~\bibnamefont {Garc\'{\i}a-Ripoll}}, \bibinfo {author} {\bibfnamefont {N.}~\bibnamefont {Roch}},\ and\ \bibinfo {author} {\bibfnamefont {O.}~\bibnamefont {Buisson}},\ }\bibfield  {title} {\bibinfo {title} {Transmon-qubit readout using an in situ bifurcation amplification in the mesoscopic regime},\ }\href {https://doi.org/10.1103/PhysRevApplied.20.044050} {\bibfield  {journal} {\bibinfo  {journal} {Phys. Rev. Appl.}\ }\textbf {\bibinfo {volume}
  {20}},\ \bibinfo {pages} {044050} (\bibinfo {year} {2023})}\BibitemShut {NoStop}%
\bibitem [{\citenamefont {Roy}\ \emph {et~al.}(2018)\citenamefont {Roy}, \citenamefont {Chand}, \citenamefont {Bhattacharjee}, \citenamefont {Hazra}, \citenamefont {Kundu}, \citenamefont {Damle},\ and\ \citenamefont {Vijay}}]{roy_generalized-jrm_2018}%
  \BibitemOpen
  \bibfield  {author} {\bibinfo {author} {\bibfnamefont {T.}~\bibnamefont {Roy}}, \bibinfo {author} {\bibfnamefont {M.}~\bibnamefont {Chand}}, \bibinfo {author} {\bibfnamefont {A.}~\bibnamefont {Bhattacharjee}}, \bibinfo {author} {\bibfnamefont {S.}~\bibnamefont {Hazra}}, \bibinfo {author} {\bibfnamefont {S.}~\bibnamefont {Kundu}}, \bibinfo {author} {\bibfnamefont {K.}~\bibnamefont {Damle}},\ and\ \bibinfo {author} {\bibfnamefont {R.}~\bibnamefont {Vijay}},\ }\bibfield  {title} {\bibinfo {title} {Multimode superconducting circuits for realizing strongly coupled multiqubit processor units},\ }\href {https://doi.org/10.1103/PhysRevA.98.052318} {\bibfield  {journal} {\bibinfo  {journal} {Phys. Rev. A}\ }\textbf {\bibinfo {volume} {98}},\ \bibinfo {pages} {052318} (\bibinfo {year} {2018})}\BibitemShut {NoStop}%
\bibitem [{\citenamefont {Salunkhe}\ \emph {et~al.}(2025)\citenamefont {Salunkhe}, \citenamefont {Kundu}, \citenamefont {Das}, \citenamefont {Deshmukh}, \citenamefont {Patankar},\ and\ \citenamefont {Vijay}}]{salunkhe_quantromon_2025}%
  \BibitemOpen
  \bibfield  {author} {\bibinfo {author} {\bibfnamefont {K.~V.}\ \bibnamefont {Salunkhe}}, \bibinfo {author} {\bibfnamefont {S.}~\bibnamefont {Kundu}}, \bibinfo {author} {\bibfnamefont {S.}~\bibnamefont {Das}}, \bibinfo {author} {\bibfnamefont {J.}~\bibnamefont {Deshmukh}}, \bibinfo {author} {\bibfnamefont {M.~P.}\ \bibnamefont {Patankar}},\ and\ \bibinfo {author} {\bibfnamefont {R.}~\bibnamefont {Vijay}},\ }\href {https://arxiv.org/abs/2501.17439} {\bibinfo {title} {The quantromon: A qubit-resonator system with orthogonal qubit and readout modes}} (\bibinfo {year} {2025}),\ \Eprint {https://arxiv.org/abs/2501.17439} {arXiv:2501.17439 [quant-ph]} \BibitemShut {NoStop}%
\bibitem [{\citenamefont {Brooks}\ \emph {et~al.}(2013)\citenamefont {Brooks}, \citenamefont {Kitaev},\ and\ \citenamefont {Preskill}}]{brooks_0pi_2013}%
  \BibitemOpen
  \bibfield  {author} {\bibinfo {author} {\bibfnamefont {P.}~\bibnamefont {Brooks}}, \bibinfo {author} {\bibfnamefont {A.}~\bibnamefont {Kitaev}},\ and\ \bibinfo {author} {\bibfnamefont {J.}~\bibnamefont {Preskill}},\ }\bibfield  {title} {\bibinfo {title} {Protected gates for superconducting qubits},\ }\href {https://doi.org/10.1103/PhysRevA.87.052306} {\bibfield  {journal} {\bibinfo  {journal} {Phys. Rev. A}\ }\textbf {\bibinfo {volume} {87}},\ \bibinfo {pages} {052306} (\bibinfo {year} {2013})}\BibitemShut {NoStop}%
\bibitem [{\citenamefont {Smith}\ \emph {et~al.}(2020)\citenamefont {Smith}, \citenamefont {Kou}, \citenamefont {Xiao}, \citenamefont {Vool},\ and\ \citenamefont {Devoret}}]{smith_cos2phi_2020}%
  \BibitemOpen
  \bibfield  {author} {\bibinfo {author} {\bibfnamefont {W.~C.}\ \bibnamefont {Smith}}, \bibinfo {author} {\bibfnamefont {A.}~\bibnamefont {Kou}}, \bibinfo {author} {\bibfnamefont {X.}~\bibnamefont {Xiao}}, \bibinfo {author} {\bibfnamefont {U.}~\bibnamefont {Vool}},\ and\ \bibinfo {author} {\bibfnamefont {M.~H.}\ \bibnamefont {Devoret}},\ }\bibfield  {title} {\bibinfo {title} {Superconducting circuit protected by two-cooper-pair tunneling},\ }\href {https://doi.org/10.1038/s41534-019-0231-2} {\bibfield  {journal} {\bibinfo  {journal} {npj Quantum Information}\ }\textbf {\bibinfo {volume} {6}},\ \bibinfo {pages} {8} (\bibinfo {year} {2020})}\BibitemShut {NoStop}%
\bibitem [{\citenamefont {Larsen}\ \emph {et~al.}(2020)\citenamefont {Larsen}, \citenamefont {Gershenson}, \citenamefont {Casparis}, \citenamefont {Kringh\o{}j}, \citenamefont {Pearson}, \citenamefont {McNeil}, \citenamefont {Kuemmeth}, \citenamefont {Krogstrup}, \citenamefont {Petersson},\ and\ \citenamefont {Marcus}}]{larsen_cos2phi_2020}%
  \BibitemOpen
  \bibfield  {author} {\bibinfo {author} {\bibfnamefont {T.~W.}\ \bibnamefont {Larsen}}, \bibinfo {author} {\bibfnamefont {M.~E.}\ \bibnamefont {Gershenson}}, \bibinfo {author} {\bibfnamefont {L.}~\bibnamefont {Casparis}}, \bibinfo {author} {\bibfnamefont {A.}~\bibnamefont {Kringh\o{}j}}, \bibinfo {author} {\bibfnamefont {N.~J.}\ \bibnamefont {Pearson}}, \bibinfo {author} {\bibfnamefont {R.~P.~G.}\ \bibnamefont {McNeil}}, \bibinfo {author} {\bibfnamefont {F.}~\bibnamefont {Kuemmeth}}, \bibinfo {author} {\bibfnamefont {P.}~\bibnamefont {Krogstrup}}, \bibinfo {author} {\bibfnamefont {K.~D.}\ \bibnamefont {Petersson}},\ and\ \bibinfo {author} {\bibfnamefont {C.~M.}\ \bibnamefont {Marcus}},\ }\bibfield  {title} {\bibinfo {title} {Parity-protected superconductor-semiconductor qubit},\ }\href {https://doi.org/10.1103/PhysRevLett.125.056801} {\bibfield  {journal} {\bibinfo  {journal} {Phys. Rev. Lett.}\ }\textbf {\bibinfo {volume} {125}},\ \bibinfo {pages} {056801} (\bibinfo {year} {2020})}\BibitemShut {NoStop}%
\bibitem [{\citenamefont {Hays}\ \emph {et~al.}(2025)\citenamefont {Hays}, \citenamefont {Kim},\ and\ \citenamefont {Oliver}}]{hays_harmonium_2025}%
  \BibitemOpen
  \bibfield  {author} {\bibinfo {author} {\bibfnamefont {M.}~\bibnamefont {Hays}}, \bibinfo {author} {\bibfnamefont {J.}~\bibnamefont {Kim}},\ and\ \bibinfo {author} {\bibfnamefont {W.~D.}\ \bibnamefont {Oliver}},\ }\href {https://arxiv.org/abs/2502.15459} {\bibinfo {title} {Non-degenerate noise-resilient superconducting qubit}} (\bibinfo {year} {2025}),\ \Eprint {https://arxiv.org/abs/2502.15459} {arXiv:2502.15459 [quant-ph]} \BibitemShut {NoStop}%
\bibitem [{\citenamefont {Kalashnikov}\ \emph {et~al.}(2020)\citenamefont {Kalashnikov}, \citenamefont {Hsieh}, \citenamefont {Zhang}, \citenamefont {Lu}, \citenamefont {Kamenov}, \citenamefont {Di~Paolo}, \citenamefont {Blais}, \citenamefont {Gershenson},\ and\ \citenamefont {Bell}}]{kalashnikov_bifluxon_2020}%
  \BibitemOpen
  \bibfield  {author} {\bibinfo {author} {\bibfnamefont {K.}~\bibnamefont {Kalashnikov}}, \bibinfo {author} {\bibfnamefont {W.~T.}\ \bibnamefont {Hsieh}}, \bibinfo {author} {\bibfnamefont {W.}~\bibnamefont {Zhang}}, \bibinfo {author} {\bibfnamefont {W.-S.}\ \bibnamefont {Lu}}, \bibinfo {author} {\bibfnamefont {P.}~\bibnamefont {Kamenov}}, \bibinfo {author} {\bibfnamefont {A.}~\bibnamefont {Di~Paolo}}, \bibinfo {author} {\bibfnamefont {A.}~\bibnamefont {Blais}}, \bibinfo {author} {\bibfnamefont {M.~E.}\ \bibnamefont {Gershenson}},\ and\ \bibinfo {author} {\bibfnamefont {M.}~\bibnamefont {Bell}},\ }\bibfield  {title} {\bibinfo {title} {Bifluxon: Fluxon-parity-protected superconducting qubit},\ }\href {https://doi.org/10.1103/PRXQuantum.1.010307} {\bibfield  {journal} {\bibinfo  {journal} {PRX Quantum}\ }\textbf {\bibinfo {volume} {1}},\ \bibinfo {pages} {010307} (\bibinfo {year} {2020})}\BibitemShut {NoStop}%
\bibitem [{\citenamefont {Schrade}\ \emph {et~al.}(2022)\citenamefont {Schrade}, \citenamefont {Marcus},\ and\ \citenamefont {Gyenis}}]{schrade_cos2phi_2022}%
  \BibitemOpen
  \bibfield  {author} {\bibinfo {author} {\bibfnamefont {C.}~\bibnamefont {Schrade}}, \bibinfo {author} {\bibfnamefont {C.~M.}\ \bibnamefont {Marcus}},\ and\ \bibinfo {author} {\bibfnamefont {A.}~\bibnamefont {Gyenis}},\ }\bibfield  {title} {\bibinfo {title} {Protected hybrid superconducting qubit in an array of gate-tunable josephson interferometers},\ }\href {https://doi.org/10.1103/PRXQuantum.3.030303} {\bibfield  {journal} {\bibinfo  {journal} {PRX Quantum}\ }\textbf {\bibinfo {volume} {3}},\ \bibinfo {pages} {030303} (\bibinfo {year} {2022})}\BibitemShut {NoStop}%
\bibitem [{\citenamefont {Levine}\ \emph {et~al.}(2024)\citenamefont {Levine}, \citenamefont {Haim}, \citenamefont {Hung}, \citenamefont {Alidoust}, \citenamefont {Kalaee}, \citenamefont {DeLorenzo}, \citenamefont {Wollack}, \citenamefont {Arrangoiz-Arriola}, \citenamefont {Khalajhedayati}, \citenamefont {Sanil}, \citenamefont {Moradinejad}, \citenamefont {Vaknin}, \citenamefont {Kubica}, \citenamefont {Hover}, \citenamefont {Aghaeimeibodi}, \citenamefont {Alcid}, \citenamefont {Baek}, \citenamefont {Barnett}, \citenamefont {Bawdekar}, \citenamefont {Bienias}, \citenamefont {Carson}, \citenamefont {Chen}, \citenamefont {Chen}, \citenamefont {Chinkezian}, \citenamefont {Chisholm}, \citenamefont {Clifford}, \citenamefont {Cosmic}, \citenamefont {Crisosto}, \citenamefont {Dalzell}, \citenamefont {Davis}, \citenamefont {D'Ewart}, \citenamefont {Diez}, \citenamefont {D'Souza}, \citenamefont {Dumitrescu}, \citenamefont {Elkhouly}, \citenamefont {Fang}, \citenamefont {Fang}, \citenamefont {Flammia}, \citenamefont
  {Fling}, \citenamefont {Garcia}, \citenamefont {Gharzai}, \citenamefont {Gorshkov}, \citenamefont {Gray}, \citenamefont {Grimberg}, \citenamefont {Grimsmo}, \citenamefont {Hann}, \citenamefont {He}, \citenamefont {Heidel}, \citenamefont {Howell}, \citenamefont {Hunt}, \citenamefont {Iverson}, \citenamefont {Jarrige}, \citenamefont {Jiang}, \citenamefont {Jones}, \citenamefont {Karabalin}, \citenamefont {Karalekas}, \citenamefont {Keller}, \citenamefont {Lasi}, \citenamefont {Lee}, \citenamefont {Ly}, \citenamefont {MacCabe}, \citenamefont {Mahuli}, \citenamefont {Marcaud}, \citenamefont {Matheny}, \citenamefont {McArdle}, \citenamefont {McCabe}, \citenamefont {Merton}, \citenamefont {Miles}, \citenamefont {Milsted}, \citenamefont {Mishra}, \citenamefont {Moncelsi}, \citenamefont {Naghiloo}, \citenamefont {Noh}, \citenamefont {Oblepias}, \citenamefont {Ortuno}, \citenamefont {Owens}, \citenamefont {Pagdilao}, \citenamefont {Panduro}, \citenamefont {Paquette}, \citenamefont {Patel}, \citenamefont {Peairs},
  \citenamefont {Perello}, \citenamefont {Peterson}, \citenamefont {Ponte}, \citenamefont {Putterman}, \citenamefont {Refael}, \citenamefont {Reinhold}, \citenamefont {Resnick}, \citenamefont {Reyna}, \citenamefont {Rodriguez}, \citenamefont {Rose}, \citenamefont {Rubin}, \citenamefont {Runyan}, \citenamefont {Ryan}, \citenamefont {Sahmoud}, \citenamefont {Scaffidi}, \citenamefont {Shah}, \citenamefont {Siavoshi}, \citenamefont {Sivarajah}, \citenamefont {Skogland}, \citenamefont {Su}, \citenamefont {Swenson}, \citenamefont {Sylvia}, \citenamefont {Teo}, \citenamefont {Tomada}, \citenamefont {Torlai}, \citenamefont {Wistrom}, \citenamefont {Zhang}, \citenamefont {Zuk}, \citenamefont {Clerk}, \citenamefont {Brand\~ao}, \citenamefont {Retzker},\ and\ \citenamefont {Painter}}]{levine_transmon-dual-rail_2024}%
  \BibitemOpen
  \bibfield  {author} {\bibinfo {author} {\bibfnamefont {H.}~\bibnamefont {Levine}}, \bibinfo {author} {\bibfnamefont {A.}~\bibnamefont {Haim}}, \bibinfo {author} {\bibfnamefont {J.~S.~C.}\ \bibnamefont {Hung}}, \bibinfo {author} {\bibfnamefont {N.}~\bibnamefont {Alidoust}}, \bibinfo {author} {\bibfnamefont {M.}~\bibnamefont {Kalaee}}, \bibinfo {author} {\bibfnamefont {L.}~\bibnamefont {DeLorenzo}}, \bibinfo {author} {\bibfnamefont {E.~A.}\ \bibnamefont {Wollack}}, \bibinfo {author} {\bibfnamefont {P.}~\bibnamefont {Arrangoiz-Arriola}}, \bibinfo {author} {\bibfnamefont {A.}~\bibnamefont {Khalajhedayati}}, \bibinfo {author} {\bibfnamefont {R.}~\bibnamefont {Sanil}}, \bibinfo {author} {\bibfnamefont {H.}~\bibnamefont {Moradinejad}}, \bibinfo {author} {\bibfnamefont {Y.}~\bibnamefont {Vaknin}}, \bibinfo {author} {\bibfnamefont {A.}~\bibnamefont {Kubica}}, \bibinfo {author} {\bibfnamefont {D.}~\bibnamefont {Hover}}, \bibinfo {author} {\bibfnamefont {S.}~\bibnamefont {Aghaeimeibodi}}, \bibinfo {author}
  {\bibfnamefont {J.~A.}\ \bibnamefont {Alcid}}, \bibinfo {author} {\bibfnamefont {C.}~\bibnamefont {Baek}}, \bibinfo {author} {\bibfnamefont {J.}~\bibnamefont {Barnett}}, \bibinfo {author} {\bibfnamefont {K.}~\bibnamefont {Bawdekar}}, \bibinfo {author} {\bibfnamefont {P.}~\bibnamefont {Bienias}}, \bibinfo {author} {\bibfnamefont {H.~A.}\ \bibnamefont {Carson}}, \bibinfo {author} {\bibfnamefont {C.}~\bibnamefont {Chen}}, \bibinfo {author} {\bibfnamefont {L.}~\bibnamefont {Chen}}, \bibinfo {author} {\bibfnamefont {H.}~\bibnamefont {Chinkezian}}, \bibinfo {author} {\bibfnamefont {E.~M.}\ \bibnamefont {Chisholm}}, \bibinfo {author} {\bibfnamefont {A.}~\bibnamefont {Clifford}}, \bibinfo {author} {\bibfnamefont {R.}~\bibnamefont {Cosmic}}, \bibinfo {author} {\bibfnamefont {N.}~\bibnamefont {Crisosto}}, \bibinfo {author} {\bibfnamefont {A.~M.}\ \bibnamefont {Dalzell}}, \bibinfo {author} {\bibfnamefont {E.}~\bibnamefont {Davis}}, \bibinfo {author} {\bibfnamefont {J.~M.}\ \bibnamefont {D'Ewart}}, \bibinfo {author}
  {\bibfnamefont {S.}~\bibnamefont {Diez}}, \bibinfo {author} {\bibfnamefont {N.}~\bibnamefont {D'Souza}}, \bibinfo {author} {\bibfnamefont {P.~T.}\ \bibnamefont {Dumitrescu}}, \bibinfo {author} {\bibfnamefont {E.}~\bibnamefont {Elkhouly}}, \bibinfo {author} {\bibfnamefont {M.~T.}\ \bibnamefont {Fang}}, \bibinfo {author} {\bibfnamefont {Y.}~\bibnamefont {Fang}}, \bibinfo {author} {\bibfnamefont {S.}~\bibnamefont {Flammia}}, \bibinfo {author} {\bibfnamefont {M.~J.}\ \bibnamefont {Fling}}, \bibinfo {author} {\bibfnamefont {G.}~\bibnamefont {Garcia}}, \bibinfo {author} {\bibfnamefont {M.~K.}\ \bibnamefont {Gharzai}}, \bibinfo {author} {\bibfnamefont {A.~V.}\ \bibnamefont {Gorshkov}}, \bibinfo {author} {\bibfnamefont {M.~J.}\ \bibnamefont {Gray}}, \bibinfo {author} {\bibfnamefont {S.}~\bibnamefont {Grimberg}}, \bibinfo {author} {\bibfnamefont {A.~L.}\ \bibnamefont {Grimsmo}}, \bibinfo {author} {\bibfnamefont {C.~T.}\ \bibnamefont {Hann}}, \bibinfo {author} {\bibfnamefont {Y.}~\bibnamefont {He}}, \bibinfo {author}
  {\bibfnamefont {S.}~\bibnamefont {Heidel}}, \bibinfo {author} {\bibfnamefont {S.}~\bibnamefont {Howell}}, \bibinfo {author} {\bibfnamefont {M.}~\bibnamefont {Hunt}}, \bibinfo {author} {\bibfnamefont {J.}~\bibnamefont {Iverson}}, \bibinfo {author} {\bibfnamefont {I.}~\bibnamefont {Jarrige}}, \bibinfo {author} {\bibfnamefont {L.}~\bibnamefont {Jiang}}, \bibinfo {author} {\bibfnamefont {W.~M.}\ \bibnamefont {Jones}}, \bibinfo {author} {\bibfnamefont {R.}~\bibnamefont {Karabalin}}, \bibinfo {author} {\bibfnamefont {P.~J.}\ \bibnamefont {Karalekas}}, \bibinfo {author} {\bibfnamefont {A.~J.}\ \bibnamefont {Keller}}, \bibinfo {author} {\bibfnamefont {D.}~\bibnamefont {Lasi}}, \bibinfo {author} {\bibfnamefont {M.}~\bibnamefont {Lee}}, \bibinfo {author} {\bibfnamefont {V.}~\bibnamefont {Ly}}, \bibinfo {author} {\bibfnamefont {G.}~\bibnamefont {MacCabe}}, \bibinfo {author} {\bibfnamefont {N.}~\bibnamefont {Mahuli}}, \bibinfo {author} {\bibfnamefont {G.}~\bibnamefont {Marcaud}}, \bibinfo {author} {\bibfnamefont
  {M.~H.}\ \bibnamefont {Matheny}}, \bibinfo {author} {\bibfnamefont {S.}~\bibnamefont {McArdle}}, \bibinfo {author} {\bibfnamefont {G.}~\bibnamefont {McCabe}}, \bibinfo {author} {\bibfnamefont {G.}~\bibnamefont {Merton}}, \bibinfo {author} {\bibfnamefont {C.}~\bibnamefont {Miles}}, \bibinfo {author} {\bibfnamefont {A.}~\bibnamefont {Milsted}}, \bibinfo {author} {\bibfnamefont {A.}~\bibnamefont {Mishra}}, \bibinfo {author} {\bibfnamefont {L.}~\bibnamefont {Moncelsi}}, \bibinfo {author} {\bibfnamefont {M.}~\bibnamefont {Naghiloo}}, \bibinfo {author} {\bibfnamefont {K.}~\bibnamefont {Noh}}, \bibinfo {author} {\bibfnamefont {E.}~\bibnamefont {Oblepias}}, \bibinfo {author} {\bibfnamefont {G.}~\bibnamefont {Ortuno}}, \bibinfo {author} {\bibfnamefont {J.~C.}\ \bibnamefont {Owens}}, \bibinfo {author} {\bibfnamefont {J.}~\bibnamefont {Pagdilao}}, \bibinfo {author} {\bibfnamefont {A.}~\bibnamefont {Panduro}}, \bibinfo {author} {\bibfnamefont {J.-P.}\ \bibnamefont {Paquette}}, \bibinfo {author} {\bibfnamefont {R.~N.}\
  \bibnamefont {Patel}}, \bibinfo {author} {\bibfnamefont {G.}~\bibnamefont {Peairs}}, \bibinfo {author} {\bibfnamefont {D.~J.}\ \bibnamefont {Perello}}, \bibinfo {author} {\bibfnamefont {E.~C.}\ \bibnamefont {Peterson}}, \bibinfo {author} {\bibfnamefont {S.}~\bibnamefont {Ponte}}, \bibinfo {author} {\bibfnamefont {H.}~\bibnamefont {Putterman}}, \bibinfo {author} {\bibfnamefont {G.}~\bibnamefont {Refael}}, \bibinfo {author} {\bibfnamefont {P.}~\bibnamefont {Reinhold}}, \bibinfo {author} {\bibfnamefont {R.}~\bibnamefont {Resnick}}, \bibinfo {author} {\bibfnamefont {O.~A.}\ \bibnamefont {Reyna}}, \bibinfo {author} {\bibfnamefont {R.}~\bibnamefont {Rodriguez}}, \bibinfo {author} {\bibfnamefont {J.}~\bibnamefont {Rose}}, \bibinfo {author} {\bibfnamefont {A.~H.}\ \bibnamefont {Rubin}}, \bibinfo {author} {\bibfnamefont {M.}~\bibnamefont {Runyan}}, \bibinfo {author} {\bibfnamefont {C.~A.}\ \bibnamefont {Ryan}}, \bibinfo {author} {\bibfnamefont {A.}~\bibnamefont {Sahmoud}}, \bibinfo {author} {\bibfnamefont
  {T.}~\bibnamefont {Scaffidi}}, \bibinfo {author} {\bibfnamefont {B.}~\bibnamefont {Shah}}, \bibinfo {author} {\bibfnamefont {S.}~\bibnamefont {Siavoshi}}, \bibinfo {author} {\bibfnamefont {P.}~\bibnamefont {Sivarajah}}, \bibinfo {author} {\bibfnamefont {T.}~\bibnamefont {Skogland}}, \bibinfo {author} {\bibfnamefont {C.-J.}\ \bibnamefont {Su}}, \bibinfo {author} {\bibfnamefont {L.~J.}\ \bibnamefont {Swenson}}, \bibinfo {author} {\bibfnamefont {J.}~\bibnamefont {Sylvia}}, \bibinfo {author} {\bibfnamefont {S.~M.}\ \bibnamefont {Teo}}, \bibinfo {author} {\bibfnamefont {A.}~\bibnamefont {Tomada}}, \bibinfo {author} {\bibfnamefont {G.}~\bibnamefont {Torlai}}, \bibinfo {author} {\bibfnamefont {M.}~\bibnamefont {Wistrom}}, \bibinfo {author} {\bibfnamefont {K.}~\bibnamefont {Zhang}}, \bibinfo {author} {\bibfnamefont {I.}~\bibnamefont {Zuk}}, \bibinfo {author} {\bibfnamefont {A.~A.}\ \bibnamefont {Clerk}}, \bibinfo {author} {\bibfnamefont {F.~G. S.~L.}\ \bibnamefont {Brand\~ao}}, \bibinfo {author} {\bibfnamefont
  {A.}~\bibnamefont {Retzker}},\ and\ \bibinfo {author} {\bibfnamefont {O.}~\bibnamefont {Painter}},\ }\bibfield  {title} {\bibinfo {title} {Demonstrating a long-coherence dual-rail erasure qubit using tunable transmons},\ }\href {https://doi.org/10.1103/PhysRevX.14.011051} {\bibfield  {journal} {\bibinfo  {journal} {Phys. Rev. X}\ }\textbf {\bibinfo {volume} {14}},\ \bibinfo {pages} {011051} (\bibinfo {year} {2024})}\BibitemShut {NoStop}%
\bibitem [{\citenamefont {Huang}\ \emph {et~al.}(2025)\citenamefont {Huang}, \citenamefont {Sun}, \citenamefont {Zhang}, \citenamefont {Guo}, \citenamefont {Huang}, \citenamefont {Liang}, \citenamefont {Liu}, \citenamefont {Sun}, \citenamefont {Wang}, \citenamefont {Xiong}, \citenamefont {Yang}, \citenamefont {Zhang}, \citenamefont {Zhang}, \citenamefont {Chu}, \citenamefont {Guo}, \citenamefont {Jiang}, \citenamefont {Liu}, \citenamefont {Niu}, \citenamefont {Qiu}, \citenamefont {Tao}, \citenamefont {Zhou}, \citenamefont {Linpeng}, \citenamefont {Zhong},\ and\ \citenamefont {Yu}}]{huang_dual-rail-entanglement_2025}%
  \BibitemOpen
  \bibfield  {author} {\bibinfo {author} {\bibfnamefont {W.}~\bibnamefont {Huang}}, \bibinfo {author} {\bibfnamefont {X.}~\bibnamefont {Sun}}, \bibinfo {author} {\bibfnamefont {J.}~\bibnamefont {Zhang}}, \bibinfo {author} {\bibfnamefont {Z.}~\bibnamefont {Guo}}, \bibinfo {author} {\bibfnamefont {P.}~\bibnamefont {Huang}}, \bibinfo {author} {\bibfnamefont {Y.}~\bibnamefont {Liang}}, \bibinfo {author} {\bibfnamefont {Y.}~\bibnamefont {Liu}}, \bibinfo {author} {\bibfnamefont {D.}~\bibnamefont {Sun}}, \bibinfo {author} {\bibfnamefont {Z.}~\bibnamefont {Wang}}, \bibinfo {author} {\bibfnamefont {Y.}~\bibnamefont {Xiong}}, \bibinfo {author} {\bibfnamefont {X.}~\bibnamefont {Yang}}, \bibinfo {author} {\bibfnamefont {J.}~\bibnamefont {Zhang}}, \bibinfo {author} {\bibfnamefont {L.}~\bibnamefont {Zhang}}, \bibinfo {author} {\bibfnamefont {J.}~\bibnamefont {Chu}}, \bibinfo {author} {\bibfnamefont {W.}~\bibnamefont {Guo}}, \bibinfo {author} {\bibfnamefont {J.}~\bibnamefont {Jiang}}, \bibinfo {author} {\bibfnamefont
  {S.}~\bibnamefont {Liu}}, \bibinfo {author} {\bibfnamefont {J.}~\bibnamefont {Niu}}, \bibinfo {author} {\bibfnamefont {J.}~\bibnamefont {Qiu}}, \bibinfo {author} {\bibfnamefont {Z.}~\bibnamefont {Tao}}, \bibinfo {author} {\bibfnamefont {Y.}~\bibnamefont {Zhou}}, \bibinfo {author} {\bibfnamefont {X.}~\bibnamefont {Linpeng}}, \bibinfo {author} {\bibfnamefont {Y.}~\bibnamefont {Zhong}},\ and\ \bibinfo {author} {\bibfnamefont {D.}~\bibnamefont {Yu}},\ }\href {https://arxiv.org/abs/2504.12099} {\bibinfo {title} {Logical multi-qubit entanglement with dual-rail superconducting qubits}} (\bibinfo {year} {2025}),\ \Eprint {https://arxiv.org/abs/2504.12099} {arXiv:2504.12099 [quant-ph]} \BibitemShut {NoStop}%
\bibitem [{\citenamefont {Chou}\ \emph {et~al.}(2024)\citenamefont {Chou}, \citenamefont {Shemma}, \citenamefont {McCarrick}, \citenamefont {Chien}, \citenamefont {Teoh}, \citenamefont {Winkel}, \citenamefont {Anderson}, \citenamefont {Chen}, \citenamefont {Curtis}, \citenamefont {de~Graaf}, \citenamefont {Garmon}, \citenamefont {Gudlewski}, \citenamefont {Kalfus}, \citenamefont {Keen}, \citenamefont {Khedkar}, \citenamefont {Lei}, \citenamefont {Liu}, \citenamefont {Lu}, \citenamefont {Lu}, \citenamefont {Maiti}, \citenamefont {Mastalli-Kelly}, \citenamefont {Mehta}, \citenamefont {Mundhada}, \citenamefont {Narla}, \citenamefont {Noh}, \citenamefont {Tsunoda}, \citenamefont {Xue}, \citenamefont {Yuan}, \citenamefont {Frunzio}, \citenamefont {Aumentado}, \citenamefont {Puri}, \citenamefont {Girvin}, \citenamefont {Moseley},\ and\ \citenamefont {Schoelkopf}}]{chou_dual-rail-measurement_2024}%
  \BibitemOpen
  \bibfield  {author} {\bibinfo {author} {\bibfnamefont {K.~S.}\ \bibnamefont {Chou}}, \bibinfo {author} {\bibfnamefont {T.}~\bibnamefont {Shemma}}, \bibinfo {author} {\bibfnamefont {H.}~\bibnamefont {McCarrick}}, \bibinfo {author} {\bibfnamefont {T.-C.}\ \bibnamefont {Chien}}, \bibinfo {author} {\bibfnamefont {J.~D.}\ \bibnamefont {Teoh}}, \bibinfo {author} {\bibfnamefont {P.}~\bibnamefont {Winkel}}, \bibinfo {author} {\bibfnamefont {A.}~\bibnamefont {Anderson}}, \bibinfo {author} {\bibfnamefont {J.}~\bibnamefont {Chen}}, \bibinfo {author} {\bibfnamefont {J.~C.}\ \bibnamefont {Curtis}}, \bibinfo {author} {\bibfnamefont {S.~J.}\ \bibnamefont {de~Graaf}}, \bibinfo {author} {\bibfnamefont {J.~W.~O.}\ \bibnamefont {Garmon}}, \bibinfo {author} {\bibfnamefont {B.}~\bibnamefont {Gudlewski}}, \bibinfo {author} {\bibfnamefont {W.~D.}\ \bibnamefont {Kalfus}}, \bibinfo {author} {\bibfnamefont {T.}~\bibnamefont {Keen}}, \bibinfo {author} {\bibfnamefont {N.}~\bibnamefont {Khedkar}}, \bibinfo {author} {\bibfnamefont
  {C.~U.}\ \bibnamefont {Lei}}, \bibinfo {author} {\bibfnamefont {G.}~\bibnamefont {Liu}}, \bibinfo {author} {\bibfnamefont {P.}~\bibnamefont {Lu}}, \bibinfo {author} {\bibfnamefont {Y.}~\bibnamefont {Lu}}, \bibinfo {author} {\bibfnamefont {A.}~\bibnamefont {Maiti}}, \bibinfo {author} {\bibfnamefont {L.}~\bibnamefont {Mastalli-Kelly}}, \bibinfo {author} {\bibfnamefont {N.}~\bibnamefont {Mehta}}, \bibinfo {author} {\bibfnamefont {S.~O.}\ \bibnamefont {Mundhada}}, \bibinfo {author} {\bibfnamefont {A.}~\bibnamefont {Narla}}, \bibinfo {author} {\bibfnamefont {T.}~\bibnamefont {Noh}}, \bibinfo {author} {\bibfnamefont {T.}~\bibnamefont {Tsunoda}}, \bibinfo {author} {\bibfnamefont {S.~H.}\ \bibnamefont {Xue}}, \bibinfo {author} {\bibfnamefont {J.~O.}\ \bibnamefont {Yuan}}, \bibinfo {author} {\bibfnamefont {L.}~\bibnamefont {Frunzio}}, \bibinfo {author} {\bibfnamefont {J.}~\bibnamefont {Aumentado}}, \bibinfo {author} {\bibfnamefont {S.}~\bibnamefont {Puri}}, \bibinfo {author} {\bibfnamefont {S.~M.}\ \bibnamefont
  {Girvin}}, \bibinfo {author} {\bibfnamefont {S.~H.}\ \bibnamefont {Moseley}},\ and\ \bibinfo {author} {\bibfnamefont {R.~J.}\ \bibnamefont {Schoelkopf}},\ }\bibfield  {title} {\bibinfo {title} {A superconducting dual-rail cavity qubit with erasure-detected logical measurements},\ }\href {https://doi.org/10.1038/s41567-024-02539-4} {\bibfield  {journal} {\bibinfo  {journal} {Nature Physics}\ }\textbf {\bibinfo {volume} {20}},\ \bibinfo {pages} {1454–1460} (\bibinfo {year} {2024})}\BibitemShut {NoStop}%
\bibitem [{\citenamefont {Mehta}\ \emph {et~al.}(2025)\citenamefont {Mehta}, \citenamefont {Teoh}, \citenamefont {Noh}, \citenamefont {Agrawal}, \citenamefont {Anderson}, \citenamefont {Birdsall}, \citenamefont {Brahmbhatt}, \citenamefont {Byrd}, \citenamefont {Cacioppo}, \citenamefont {Cabrera}, \citenamefont {Carroll}, \citenamefont {Chen}, \citenamefont {Chien}, \citenamefont {Chamberlain}, \citenamefont {Curtis}, \citenamefont {Danso}, \citenamefont {Desigan}, \citenamefont {D'Acounto}, \citenamefont {Elfeky}, \citenamefont {Farzaneh}, \citenamefont {Foley}, \citenamefont {Gudlewski}, \citenamefont {Hastings}, \citenamefont {Johnson}, \citenamefont {Khedkar}, \citenamefont {Keen}, \citenamefont {Kumar}, \citenamefont {Kurter}, \citenamefont {Krawczuk}, \citenamefont {Langstengel}, \citenamefont {Li}, \citenamefont {Liu}, \citenamefont {Lu}, \citenamefont {Lu}, \citenamefont {Mastalli-Kelly}, \citenamefont {Maines}, \citenamefont {Maxwell}, \citenamefont {McCarrick}, \citenamefont {Mirzaei}, \citenamefont
  {Narla}, \citenamefont {Rashad}, \citenamefont {Reikes}, \citenamefont {Rahman}, \citenamefont {Primiani}, \citenamefont {Schwaller}, \citenamefont {Sabbah}, \citenamefont {Shemma}, \citenamefont {Shi}, \citenamefont {Satapathy}, \citenamefont {Stolpe}, \citenamefont {Strenczewilk}, \citenamefont {Szperka}, \citenamefont {Sze}, \citenamefont {Sweeney}, \citenamefont {Tikkireddi}, \citenamefont {Tsung}, \citenamefont {Sam}, \citenamefont {Weiss}, \citenamefont {Yang}, \citenamefont {Yu}, \citenamefont {Zhang}, \citenamefont {Boireau}, \citenamefont {Horton}, \citenamefont {Weinberg}, \citenamefont {Aumentado}, \citenamefont {Cord}, \citenamefont {Lei}, \citenamefont {Yuan}, \citenamefont {Mundhada}, \citenamefont {Chou}, \citenamefont {S.~Harvey~Moseleley},\ and\ \citenamefont {Schoelkopf}}]{mehta_bias-preserving_2025}%
  \BibitemOpen
  \bibfield  {author} {\bibinfo {author} {\bibfnamefont {N.}~\bibnamefont {Mehta}}, \bibinfo {author} {\bibfnamefont {J.~D.}\ \bibnamefont {Teoh}}, \bibinfo {author} {\bibfnamefont {T.}~\bibnamefont {Noh}}, \bibinfo {author} {\bibfnamefont {A.}~\bibnamefont {Agrawal}}, \bibinfo {author} {\bibfnamefont {A.}~\bibnamefont {Anderson}}, \bibinfo {author} {\bibfnamefont {B.}~\bibnamefont {Birdsall}}, \bibinfo {author} {\bibfnamefont {A.}~\bibnamefont {Brahmbhatt}}, \bibinfo {author} {\bibfnamefont {W.}~\bibnamefont {Byrd}}, \bibinfo {author} {\bibfnamefont {M.}~\bibnamefont {Cacioppo}}, \bibinfo {author} {\bibfnamefont {A.}~\bibnamefont {Cabrera}}, \bibinfo {author} {\bibfnamefont {L.}~\bibnamefont {Carroll}}, \bibinfo {author} {\bibfnamefont {J.}~\bibnamefont {Chen}}, \bibinfo {author} {\bibfnamefont {T.-C.}\ \bibnamefont {Chien}}, \bibinfo {author} {\bibfnamefont {R.}~\bibnamefont {Chamberlain}}, \bibinfo {author} {\bibfnamefont {J.~C.}\ \bibnamefont {Curtis}}, \bibinfo {author} {\bibfnamefont {D.}~\bibnamefont
  {Danso}}, \bibinfo {author} {\bibfnamefont {S.~R.}\ \bibnamefont {Desigan}}, \bibinfo {author} {\bibfnamefont {F.}~\bibnamefont {D'Acounto}}, \bibinfo {author} {\bibfnamefont {B.~H.}\ \bibnamefont {Elfeky}}, \bibinfo {author} {\bibfnamefont {S.~M.}\ \bibnamefont {Farzaneh}}, \bibinfo {author} {\bibfnamefont {C.}~\bibnamefont {Foley}}, \bibinfo {author} {\bibfnamefont {B.}~\bibnamefont {Gudlewski}}, \bibinfo {author} {\bibfnamefont {H.}~\bibnamefont {Hastings}}, \bibinfo {author} {\bibfnamefont {R.}~\bibnamefont {Johnson}}, \bibinfo {author} {\bibfnamefont {N.}~\bibnamefont {Khedkar}}, \bibinfo {author} {\bibfnamefont {T.}~\bibnamefont {Keen}}, \bibinfo {author} {\bibfnamefont {A.}~\bibnamefont {Kumar}}, \bibinfo {author} {\bibfnamefont {C.}~\bibnamefont {Kurter}}, \bibinfo {author} {\bibfnamefont {K.}~\bibnamefont {Krawczuk}}, \bibinfo {author} {\bibfnamefont {E.}~\bibnamefont {Langstengel}}, \bibinfo {author} {\bibfnamefont {R.~D.}\ \bibnamefont {Li}}, \bibinfo {author} {\bibfnamefont {G.}~\bibnamefont
  {Liu}}, \bibinfo {author} {\bibfnamefont {H.}~\bibnamefont {Lu}}, \bibinfo {author} {\bibfnamefont {P.}~\bibnamefont {Lu}}, \bibinfo {author} {\bibfnamefont {L.}~\bibnamefont {Mastalli-Kelly}}, \bibinfo {author} {\bibfnamefont {A.}~\bibnamefont {Maines}}, \bibinfo {author} {\bibfnamefont {M.}~\bibnamefont {Maxwell}}, \bibinfo {author} {\bibfnamefont {H.}~\bibnamefont {McCarrick}}, \bibinfo {author} {\bibfnamefont {M.}~\bibnamefont {Mirzaei}}, \bibinfo {author} {\bibfnamefont {A.}~\bibnamefont {Narla}}, \bibinfo {author} {\bibfnamefont {O.}~\bibnamefont {Rashad}}, \bibinfo {author} {\bibfnamefont {E.}~\bibnamefont {Reikes}}, \bibinfo {author} {\bibfnamefont {M.}~\bibnamefont {Rahman}}, \bibinfo {author} {\bibfnamefont {R.}~\bibnamefont {Primiani}}, \bibinfo {author} {\bibfnamefont {M.}~\bibnamefont {Schwaller}}, \bibinfo {author} {\bibfnamefont {A.}~\bibnamefont {Sabbah}}, \bibinfo {author} {\bibfnamefont {T.}~\bibnamefont {Shemma}}, \bibinfo {author} {\bibfnamefont {R.~A.}\ \bibnamefont {Shi}}, \bibinfo
  {author} {\bibfnamefont {S.}~\bibnamefont {Satapathy}}, \bibinfo {author} {\bibfnamefont {D.}~\bibnamefont {Stolpe}}, \bibinfo {author} {\bibfnamefont {J.}~\bibnamefont {Strenczewilk}}, \bibinfo {author} {\bibfnamefont {D.}~\bibnamefont {Szperka}}, \bibinfo {author} {\bibfnamefont {I.-W.}\ \bibnamefont {Sze}}, \bibinfo {author} {\bibfnamefont {D.}~\bibnamefont {Sweeney}}, \bibinfo {author} {\bibfnamefont {P.}~\bibnamefont {Tikkireddi}}, \bibinfo {author} {\bibfnamefont {C.-L.}\ \bibnamefont {Tsung}}, \bibinfo {author} {\bibfnamefont {D.~V.}\ \bibnamefont {Sam}}, \bibinfo {author} {\bibfnamefont {D.~K.}\ \bibnamefont {Weiss}}, \bibinfo {author} {\bibfnamefont {Z.}~\bibnamefont {Yang}}, \bibinfo {author} {\bibfnamefont {L.}~\bibnamefont {Yu}}, \bibinfo {author} {\bibfnamefont {T.}~\bibnamefont {Zhang}}, \bibinfo {author} {\bibfnamefont {O.}~\bibnamefont {Boireau}}, \bibinfo {author} {\bibfnamefont {S.}~\bibnamefont {Horton}}, \bibinfo {author} {\bibfnamefont {S.}~\bibnamefont {Weinberg}}, \bibinfo {author}
  {\bibfnamefont {J.}~\bibnamefont {Aumentado}}, \bibinfo {author} {\bibfnamefont {B.}~\bibnamefont {Cord}}, \bibinfo {author} {\bibfnamefont {C.~U.}\ \bibnamefont {Lei}}, \bibinfo {author} {\bibfnamefont {J.~O.}\ \bibnamefont {Yuan}}, \bibinfo {author} {\bibfnamefont {S.~O.}\ \bibnamefont {Mundhada}}, \bibinfo {author} {\bibfnamefont {K.~S.}\ \bibnamefont {Chou}}, \bibinfo {author} {\bibfnamefont {J.}~\bibnamefont {S.~Harvey~Moseleley}},\ and\ \bibinfo {author} {\bibfnamefont {R.~J.}\ \bibnamefont {Schoelkopf}},\ }\href {https://arxiv.org/abs/2503.10935} {\bibinfo {title} {Bias-preserving and error-detectable entangling operations in a superconducting dual-rail system}} (\bibinfo {year} {2025}),\ \Eprint {https://arxiv.org/abs/2503.10935} {arXiv:2503.10935 [quant-ph]} \BibitemShut {NoStop}%
\bibitem [{\citenamefont {Ye}\ \emph {et~al.}(2021)\citenamefont {Ye}, \citenamefont {Peng}, \citenamefont {Naghiloo}, \citenamefont {Cunningham},\ and\ \citenamefont {O'Brien}}]{ye_engineering_2021}%
  \BibitemOpen
  \bibfield  {author} {\bibinfo {author} {\bibfnamefont {Y.}~\bibnamefont {Ye}}, \bibinfo {author} {\bibfnamefont {K.}~\bibnamefont {Peng}}, \bibinfo {author} {\bibfnamefont {M.}~\bibnamefont {Naghiloo}}, \bibinfo {author} {\bibfnamefont {G.}~\bibnamefont {Cunningham}},\ and\ \bibinfo {author} {\bibfnamefont {K.~P.}\ \bibnamefont {O'Brien}},\ }\bibfield  {title} {\bibinfo {title} {Engineering purely nonlinear coupling between superconducting qubits using a quarton},\ }\href {https://doi.org/10.1103/PhysRevLett.127.050502} {\bibfield  {journal} {\bibinfo  {journal} {Phys. Rev. Lett.}\ }\textbf {\bibinfo {volume} {127}},\ \bibinfo {pages} {050502} (\bibinfo {year} {2021})}\BibitemShut {NoStop}%
\bibitem [{\citenamefont {Rosenfeld}\ \emph {et~al.}(2024)\citenamefont {Rosenfeld}, \citenamefont {Hann}, \citenamefont {Schuster}, \citenamefont {Matheny},\ and\ \citenamefont {Clerk}}]{rosenfeld_fluxonium-resonator_2024}%
  \BibitemOpen
  \bibfield  {author} {\bibinfo {author} {\bibfnamefont {E.~L.}\ \bibnamefont {Rosenfeld}}, \bibinfo {author} {\bibfnamefont {C.~T.}\ \bibnamefont {Hann}}, \bibinfo {author} {\bibfnamefont {D.~I.}\ \bibnamefont {Schuster}}, \bibinfo {author} {\bibfnamefont {M.~H.}\ \bibnamefont {Matheny}},\ and\ \bibinfo {author} {\bibfnamefont {A.~A.}\ \bibnamefont {Clerk}},\ }\bibfield  {title} {\bibinfo {title} {High-fidelity two-qubit gates between fluxonium qubits with a resonator coupler},\ }\href {https://doi.org/10.1103/PRXQuantum.5.040317} {\bibfield  {journal} {\bibinfo  {journal} {PRX Quantum}\ }\textbf {\bibinfo {volume} {5}},\ \bibinfo {pages} {040317} (\bibinfo {year} {2024})}\BibitemShut {NoStop}%
\bibitem [{\citenamefont {Lambert}\ \emph {et~al.}(2024)\citenamefont {Lambert}, \citenamefont {Giguère}, \citenamefont {Menczel}, \citenamefont {Li}, \citenamefont {Hopf}, \citenamefont {Suárez}, \citenamefont {Gali}, \citenamefont {Lishman}, \citenamefont {Gadhvi}, \citenamefont {Agarwal}, \citenamefont {Galicia}, \citenamefont {Shammah}, \citenamefont {Nation}, \citenamefont {Johansson}, \citenamefont {Ahmed}, \citenamefont {Cross}, \citenamefont {Pitchford},\ and\ \citenamefont {Nori}}]{lambert_qutip5_2024}%
  \BibitemOpen
  \bibfield  {author} {\bibinfo {author} {\bibfnamefont {N.}~\bibnamefont {Lambert}}, \bibinfo {author} {\bibfnamefont {E.}~\bibnamefont {Giguère}}, \bibinfo {author} {\bibfnamefont {P.}~\bibnamefont {Menczel}}, \bibinfo {author} {\bibfnamefont {B.}~\bibnamefont {Li}}, \bibinfo {author} {\bibfnamefont {P.}~\bibnamefont {Hopf}}, \bibinfo {author} {\bibfnamefont {G.}~\bibnamefont {Suárez}}, \bibinfo {author} {\bibfnamefont {M.}~\bibnamefont {Gali}}, \bibinfo {author} {\bibfnamefont {J.}~\bibnamefont {Lishman}}, \bibinfo {author} {\bibfnamefont {R.}~\bibnamefont {Gadhvi}}, \bibinfo {author} {\bibfnamefont {R.}~\bibnamefont {Agarwal}}, \bibinfo {author} {\bibfnamefont {A.}~\bibnamefont {Galicia}}, \bibinfo {author} {\bibfnamefont {N.}~\bibnamefont {Shammah}}, \bibinfo {author} {\bibfnamefont {P.~D.}\ \bibnamefont {Nation}}, \bibinfo {author} {\bibfnamefont {J.~R.}\ \bibnamefont {Johansson}}, \bibinfo {author} {\bibfnamefont {S.}~\bibnamefont {Ahmed}}, \bibinfo {author} {\bibfnamefont {S.}~\bibnamefont {Cross}},
  \bibinfo {author} {\bibfnamefont {A.}~\bibnamefont {Pitchford}},\ and\ \bibinfo {author} {\bibfnamefont {F.}~\bibnamefont {Nori}},\ }\href {https://doi.org/10.48550/arXiv.2412.04705} {\bibinfo {title} {{QuTiP} 5: The quantum toolbox in {Python}}} (\bibinfo {year} {2024}),\ \Eprint {https://arxiv.org/abs/2412.04705} {arXiv:2412.04705 [quant-ph]} \BibitemShut {NoStop}%
\bibitem [{\citenamefont {Pedersen}\ \emph {et~al.}(2007)\citenamefont {Pedersen}, \citenamefont {Møller},\ and\ \citenamefont {Mølmer}}]{pedersen_fidelity_2007}%
  \BibitemOpen
  \bibfield  {author} {\bibinfo {author} {\bibfnamefont {L.~H.}\ \bibnamefont {Pedersen}}, \bibinfo {author} {\bibfnamefont {N.~M.}\ \bibnamefont {Møller}},\ and\ \bibinfo {author} {\bibfnamefont {K.}~\bibnamefont {Mølmer}},\ }\bibfield  {title} {\bibinfo {title} {Fidelity of quantum operations},\ }\href {https://doi.org/https://doi.org/10.1016/j.physleta.2007.02.069} {\bibfield  {journal} {\bibinfo  {journal} {Physics Letters A}\ }\textbf {\bibinfo {volume} {367}},\ \bibinfo {pages} {47} (\bibinfo {year} {2007})}\BibitemShut {NoStop}%
\bibitem [{\citenamefont {Nielsen}\ and\ \citenamefont {Chuang}(2010)}]{nielsen_quantum-computation_2010}%
  \BibitemOpen
  \bibfield  {author} {\bibinfo {author} {\bibfnamefont {M.~A.}\ \bibnamefont {Nielsen}}\ and\ \bibinfo {author} {\bibfnamefont {I.~L.}\ \bibnamefont {Chuang}},\ }\href@noop {} {\emph {\bibinfo {title} {Quantum Computation and Quantum Information: 10th Anniversary Edition}}}\ (\bibinfo  {publisher} {Cambridge University Press},\ \bibinfo {year} {2010})\BibitemShut {NoStop}%
\bibitem [{\citenamefont {Wang}\ \emph {et~al.}(2022)\citenamefont {Wang}, \citenamefont {Li}, \citenamefont {Xu}, \citenamefont {Li}, \citenamefont {Wang}, \citenamefont {Yang}, \citenamefont {Mi}, \citenamefont {Liang}, \citenamefont {Su}, \citenamefont {Yang}, \citenamefont {Wang}, \citenamefont {Wang}, \citenamefont {Li}, \citenamefont {Chen}, \citenamefont {Li}, \citenamefont {Linghu}, \citenamefont {Han}, \citenamefont {Zhang}, \citenamefont {Feng}, \citenamefont {Song}, \citenamefont {Ma}, \citenamefont {Zhang}, \citenamefont {Wang}, \citenamefont {Zhao}, \citenamefont {Liu}, \citenamefont {Xue}, \citenamefont {Jin},\ and\ \citenamefont {Yu}}]{wang_towards-practical_2022}%
  \BibitemOpen
  \bibfield  {author} {\bibinfo {author} {\bibfnamefont {C.}~\bibnamefont {Wang}}, \bibinfo {author} {\bibfnamefont {X.}~\bibnamefont {Li}}, \bibinfo {author} {\bibfnamefont {H.}~\bibnamefont {Xu}}, \bibinfo {author} {\bibfnamefont {Z.}~\bibnamefont {Li}}, \bibinfo {author} {\bibfnamefont {J.}~\bibnamefont {Wang}}, \bibinfo {author} {\bibfnamefont {Z.}~\bibnamefont {Yang}}, \bibinfo {author} {\bibfnamefont {Z.}~\bibnamefont {Mi}}, \bibinfo {author} {\bibfnamefont {X.}~\bibnamefont {Liang}}, \bibinfo {author} {\bibfnamefont {T.}~\bibnamefont {Su}}, \bibinfo {author} {\bibfnamefont {C.}~\bibnamefont {Yang}}, \bibinfo {author} {\bibfnamefont {G.}~\bibnamefont {Wang}}, \bibinfo {author} {\bibfnamefont {W.}~\bibnamefont {Wang}}, \bibinfo {author} {\bibfnamefont {Y.}~\bibnamefont {Li}}, \bibinfo {author} {\bibfnamefont {M.}~\bibnamefont {Chen}}, \bibinfo {author} {\bibfnamefont {C.}~\bibnamefont {Li}}, \bibinfo {author} {\bibfnamefont {K.}~\bibnamefont {Linghu}}, \bibinfo {author} {\bibfnamefont {J.}~\bibnamefont
  {Han}}, \bibinfo {author} {\bibfnamefont {Y.}~\bibnamefont {Zhang}}, \bibinfo {author} {\bibfnamefont {Y.}~\bibnamefont {Feng}}, \bibinfo {author} {\bibfnamefont {Y.}~\bibnamefont {Song}}, \bibinfo {author} {\bibfnamefont {T.}~\bibnamefont {Ma}}, \bibinfo {author} {\bibfnamefont {J.}~\bibnamefont {Zhang}}, \bibinfo {author} {\bibfnamefont {R.}~\bibnamefont {Wang}}, \bibinfo {author} {\bibfnamefont {P.}~\bibnamefont {Zhao}}, \bibinfo {author} {\bibfnamefont {W.}~\bibnamefont {Liu}}, \bibinfo {author} {\bibfnamefont {G.}~\bibnamefont {Xue}}, \bibinfo {author} {\bibfnamefont {Y.}~\bibnamefont {Jin}},\ and\ \bibinfo {author} {\bibfnamefont {H.}~\bibnamefont {Yu}},\ }\bibfield  {title} {\bibinfo {title} {Towards practical quantum computers: transmon qubit with a lifetime approaching 0.5 milliseconds},\ }\href {https://doi.org/10.1038/s41534-021-00510-2} {\bibfield  {journal} {\bibinfo  {journal} {npj Quantum Information}\ }\textbf {\bibinfo {volume} {8}},\ \bibinfo {pages} {3} (\bibinfo {year}
  {2022})}\BibitemShut {NoStop}%
\bibitem [{\citenamefont {Bland}\ \emph {et~al.}(2025)\citenamefont {Bland}, \citenamefont {Bahrami}, \citenamefont {Martinez}, \citenamefont {Prestegaard}, \citenamefont {Smitham}, \citenamefont {Joshi}, \citenamefont {Hedrick}, \citenamefont {Pakpour-Tabrizi}, \citenamefont {Kumar}, \citenamefont {Jindal}, \citenamefont {Chang}, \citenamefont {Yang}, \citenamefont {Cheng}, \citenamefont {Yao}, \citenamefont {Cava}, \citenamefont {de~Leon},\ and\ \citenamefont {Houck}}]{bland_2d-transmons_2025}%
  \BibitemOpen
  \bibfield  {author} {\bibinfo {author} {\bibfnamefont {M.~P.}\ \bibnamefont {Bland}}, \bibinfo {author} {\bibfnamefont {F.}~\bibnamefont {Bahrami}}, \bibinfo {author} {\bibfnamefont {J.~G.~C.}\ \bibnamefont {Martinez}}, \bibinfo {author} {\bibfnamefont {P.~H.}\ \bibnamefont {Prestegaard}}, \bibinfo {author} {\bibfnamefont {B.~M.}\ \bibnamefont {Smitham}}, \bibinfo {author} {\bibfnamefont {A.}~\bibnamefont {Joshi}}, \bibinfo {author} {\bibfnamefont {E.}~\bibnamefont {Hedrick}}, \bibinfo {author} {\bibfnamefont {A.}~\bibnamefont {Pakpour-Tabrizi}}, \bibinfo {author} {\bibfnamefont {S.}~\bibnamefont {Kumar}}, \bibinfo {author} {\bibfnamefont {A.}~\bibnamefont {Jindal}}, \bibinfo {author} {\bibfnamefont {R.~D.}\ \bibnamefont {Chang}}, \bibinfo {author} {\bibfnamefont {A.}~\bibnamefont {Yang}}, \bibinfo {author} {\bibfnamefont {G.}~\bibnamefont {Cheng}}, \bibinfo {author} {\bibfnamefont {N.}~\bibnamefont {Yao}}, \bibinfo {author} {\bibfnamefont {R.~J.}\ \bibnamefont {Cava}}, \bibinfo {author} {\bibfnamefont
  {N.~P.}\ \bibnamefont {de~Leon}},\ and\ \bibinfo {author} {\bibfnamefont {A.~A.}\ \bibnamefont {Houck}},\ }\href {https://arxiv.org/abs/2503.14798} {\bibinfo {title} {2d transmons with lifetimes and coherence times exceeding 1 millisecond}} (\bibinfo {year} {2025}),\ \Eprint {https://arxiv.org/abs/2503.14798} {arXiv:2503.14798 [quant-ph]} \BibitemShut {NoStop}%
\bibitem [{\citenamefont {Jones}\ \emph {et~al.}(1993)\citenamefont {Jones}, \citenamefont {Perttunen},\ and\ \citenamefont {Stuckman}}]{jones_direct_1993}%
  \BibitemOpen
  \bibfield  {author} {\bibinfo {author} {\bibfnamefont {D.~R.}\ \bibnamefont {Jones}}, \bibinfo {author} {\bibfnamefont {C.~D.}\ \bibnamefont {Perttunen}},\ and\ \bibinfo {author} {\bibfnamefont {B.~E.}\ \bibnamefont {Stuckman}},\ }\bibfield  {title} {\bibinfo {title} {Lipschitzian optimization without the lipschitz constant},\ }\href {https://doi.org/10.1007/BF00941892} {\bibfield  {journal} {\bibinfo  {journal} {Journal of Optimization Theory and Applications}\ }\textbf {\bibinfo {volume} {79}},\ \bibinfo {pages} {157–181} (\bibinfo {year} {1993})}\BibitemShut {NoStop}%
\bibitem [{\citenamefont {Gablonsky}\ and\ \citenamefont {Kelley}(2001)}]{gablonsky_direct-L_2001}%
  \BibitemOpen
  \bibfield  {author} {\bibinfo {author} {\bibfnamefont {J.}~\bibnamefont {Gablonsky}}\ and\ \bibinfo {author} {\bibfnamefont {C.}~\bibnamefont {Kelley}},\ }\bibfield  {title} {\bibinfo {title} {A locally-biased form of the direct algorithm},\ }\href {https://doi.org/10.1023/A:1017930332101} {\bibfield  {journal} {\bibinfo  {journal} {Journal of Global Optimization}\ }\textbf {\bibinfo {volume} {21}},\ \bibinfo {pages} {27–37} (\bibinfo {year} {2001})}\BibitemShut {NoStop}%
\bibitem [{\citenamefont {Clerk}\ and\ \citenamefont {Utami}(2007)}]{clerk_shot-noise_2007}%
  \BibitemOpen
  \bibfield  {author} {\bibinfo {author} {\bibfnamefont {A.~A.}\ \bibnamefont {Clerk}}\ and\ \bibinfo {author} {\bibfnamefont {D.~W.}\ \bibnamefont {Utami}},\ }\bibfield  {title} {\bibinfo {title} {Using a qubit to measure photon-number statistics of a driven thermal oscillator},\ }\href {https://doi.org/10.1103/PhysRevA.75.042302} {\bibfield  {journal} {\bibinfo  {journal} {Phys. Rev. A}\ }\textbf {\bibinfo {volume} {75}},\ \bibinfo {pages} {042302} (\bibinfo {year} {2007})}\BibitemShut {NoStop}%
\bibitem [{\citenamefont {Gambetta}\ \emph {et~al.}(2007)\citenamefont {Gambetta}, \citenamefont {Braff}, \citenamefont {Wallraff}, \citenamefont {Girvin},\ and\ \citenamefont {Schoelkopf}}]{gambetta_readout_2007}%
  \BibitemOpen
  \bibfield  {author} {\bibinfo {author} {\bibfnamefont {J.}~\bibnamefont {Gambetta}}, \bibinfo {author} {\bibfnamefont {W.~A.}\ \bibnamefont {Braff}}, \bibinfo {author} {\bibfnamefont {A.}~\bibnamefont {Wallraff}}, \bibinfo {author} {\bibfnamefont {S.~M.}\ \bibnamefont {Girvin}},\ and\ \bibinfo {author} {\bibfnamefont {R.~J.}\ \bibnamefont {Schoelkopf}},\ }\bibfield  {title} {\bibinfo {title} {Protocols for optimal readout of qubits using a continuous quantum nondemolition measurement},\ }\href {https://doi.org/10.1103/PhysRevA.76.012325} {\bibfield  {journal} {\bibinfo  {journal} {Phys. Rev. A}\ }\textbf {\bibinfo {volume} {76}},\ \bibinfo {pages} {012325} (\bibinfo {year} {2007})}\BibitemShut {NoStop}%
\bibitem [{\citenamefont {Swiadek}\ \emph {et~al.}(2024)\citenamefont {Swiadek}, \citenamefont {Shillito}, \citenamefont {Magnard}, \citenamefont {Remm}, \citenamefont {Hellings}, \citenamefont {Lacroix}, \citenamefont {Ficheux}, \citenamefont {Zanuz}, \citenamefont {Norris}, \citenamefont {Blais}, \citenamefont {Krinner},\ and\ \citenamefont {Wallraff}}]{swiadek_readout_2024}%
  \BibitemOpen
  \bibfield  {author} {\bibinfo {author} {\bibfnamefont {F.~m.~c.}\ \bibnamefont {Swiadek}}, \bibinfo {author} {\bibfnamefont {R.}~\bibnamefont {Shillito}}, \bibinfo {author} {\bibfnamefont {P.}~\bibnamefont {Magnard}}, \bibinfo {author} {\bibfnamefont {A.}~\bibnamefont {Remm}}, \bibinfo {author} {\bibfnamefont {C.}~\bibnamefont {Hellings}}, \bibinfo {author} {\bibfnamefont {N.}~\bibnamefont {Lacroix}}, \bibinfo {author} {\bibfnamefont {Q.}~\bibnamefont {Ficheux}}, \bibinfo {author} {\bibfnamefont {D.~C.}\ \bibnamefont {Zanuz}}, \bibinfo {author} {\bibfnamefont {G.~J.}\ \bibnamefont {Norris}}, \bibinfo {author} {\bibfnamefont {A.}~\bibnamefont {Blais}}, \bibinfo {author} {\bibfnamefont {S.}~\bibnamefont {Krinner}},\ and\ \bibinfo {author} {\bibfnamefont {A.}~\bibnamefont {Wallraff}},\ }\bibfield  {title} {\bibinfo {title} {Enhancing dispersive readout of superconducting qubits through dynamic control of the dispersive shift: Experiment and theory},\ }\href {https://doi.org/10.1103/PRXQuantum.5.040326}
  {\bibfield  {journal} {\bibinfo  {journal} {PRX Quantum}\ }\textbf {\bibinfo {volume} {5}},\ \bibinfo {pages} {040326} (\bibinfo {year} {2024})}\BibitemShut {NoStop}%
\bibitem [{\citenamefont {Rigetti}\ \emph {et~al.}(2012)\citenamefont {Rigetti}, \citenamefont {Gambetta}, \citenamefont {Poletto}, \citenamefont {Plourde}, \citenamefont {Chow}, \citenamefont {C\'orcoles}, \citenamefont {Smolin}, \citenamefont {Merkel}, \citenamefont {Rozen}, \citenamefont {Keefe}, \citenamefont {Rothwell}, \citenamefont {Ketchen},\ and\ \citenamefont {Steffen}}]{rigetti_shot-noise_2012}%
  \BibitemOpen
  \bibfield  {author} {\bibinfo {author} {\bibfnamefont {C.}~\bibnamefont {Rigetti}}, \bibinfo {author} {\bibfnamefont {J.~M.}\ \bibnamefont {Gambetta}}, \bibinfo {author} {\bibfnamefont {S.}~\bibnamefont {Poletto}}, \bibinfo {author} {\bibfnamefont {B.~L.~T.}\ \bibnamefont {Plourde}}, \bibinfo {author} {\bibfnamefont {J.~M.}\ \bibnamefont {Chow}}, \bibinfo {author} {\bibfnamefont {A.~D.}\ \bibnamefont {C\'orcoles}}, \bibinfo {author} {\bibfnamefont {J.~A.}\ \bibnamefont {Smolin}}, \bibinfo {author} {\bibfnamefont {S.~T.}\ \bibnamefont {Merkel}}, \bibinfo {author} {\bibfnamefont {J.~R.}\ \bibnamefont {Rozen}}, \bibinfo {author} {\bibfnamefont {G.~A.}\ \bibnamefont {Keefe}}, \bibinfo {author} {\bibfnamefont {M.~B.}\ \bibnamefont {Rothwell}}, \bibinfo {author} {\bibfnamefont {M.~B.}\ \bibnamefont {Ketchen}},\ and\ \bibinfo {author} {\bibfnamefont {M.}~\bibnamefont {Steffen}},\ }\bibfield  {title} {\bibinfo {title} {Superconducting qubit in a waveguide cavity with a coherence time approaching 0.1 ms},\ }\href
  {https://doi.org/10.1103/PhysRevB.86.100506} {\bibfield  {journal} {\bibinfo  {journal} {Phys. Rev. B}\ }\textbf {\bibinfo {volume} {86}},\ \bibinfo {pages} {100506} (\bibinfo {year} {2012})}\BibitemShut {NoStop}%
\bibitem [{\citenamefont {Wang}\ \emph {et~al.}(2019)\citenamefont {Wang}, \citenamefont {Shankar}, \citenamefont {Minev}, \citenamefont {Campagne-Ibarcq}, \citenamefont {Narla},\ and\ \citenamefont {Devoret}}]{wang_cavity-attenuators_2019}%
  \BibitemOpen
  \bibfield  {author} {\bibinfo {author} {\bibfnamefont {Z.}~\bibnamefont {Wang}}, \bibinfo {author} {\bibfnamefont {S.}~\bibnamefont {Shankar}}, \bibinfo {author} {\bibfnamefont {Z.}~\bibnamefont {Minev}}, \bibinfo {author} {\bibfnamefont {P.}~\bibnamefont {Campagne-Ibarcq}}, \bibinfo {author} {\bibfnamefont {A.}~\bibnamefont {Narla}},\ and\ \bibinfo {author} {\bibfnamefont {M.}~\bibnamefont {Devoret}},\ }\bibfield  {title} {\bibinfo {title} {Cavity attenuators for superconducting qubits},\ }\href {https://doi.org/10.1103/PhysRevApplied.11.014031} {\bibfield  {journal} {\bibinfo  {journal} {Phys. Rev. Appl.}\ }\textbf {\bibinfo {volume} {11}},\ \bibinfo {pages} {014031} (\bibinfo {year} {2019})}\BibitemShut {NoStop}%
\bibitem [{\citenamefont {Opremcak}\ \emph {et~al.}(2018)\citenamefont {Opremcak}, \citenamefont {Pechenezhskiy}, \citenamefont {Howington}, \citenamefont {Christensen}, \citenamefont {Beck}, \citenamefont {Leonard}, \citenamefont {Suttle}, \citenamefont {Wilen}, \citenamefont {Nesterov}, \citenamefont {Ribeill}, \citenamefont {Thorbeck}, \citenamefont {Schlenker}, \citenamefont {Vavilov}, \citenamefont {Plourde},\ and\ \citenamefont {McDermott}}]{opremcak_JPM-readout_2018}%
  \BibitemOpen
  \bibfield  {author} {\bibinfo {author} {\bibfnamefont {A.}~\bibnamefont {Opremcak}}, \bibinfo {author} {\bibfnamefont {I.~V.}\ \bibnamefont {Pechenezhskiy}}, \bibinfo {author} {\bibfnamefont {C.}~\bibnamefont {Howington}}, \bibinfo {author} {\bibfnamefont {B.~G.}\ \bibnamefont {Christensen}}, \bibinfo {author} {\bibfnamefont {M.~A.}\ \bibnamefont {Beck}}, \bibinfo {author} {\bibfnamefont {E.}~\bibnamefont {Leonard}}, \bibinfo {author} {\bibfnamefont {J.}~\bibnamefont {Suttle}}, \bibinfo {author} {\bibfnamefont {C.}~\bibnamefont {Wilen}}, \bibinfo {author} {\bibfnamefont {K.~N.}\ \bibnamefont {Nesterov}}, \bibinfo {author} {\bibfnamefont {G.~J.}\ \bibnamefont {Ribeill}}, \bibinfo {author} {\bibfnamefont {T.}~\bibnamefont {Thorbeck}}, \bibinfo {author} {\bibfnamefont {F.}~\bibnamefont {Schlenker}}, \bibinfo {author} {\bibfnamefont {M.~G.}\ \bibnamefont {Vavilov}}, \bibinfo {author} {\bibfnamefont {B.~L.~T.}\ \bibnamefont {Plourde}},\ and\ \bibinfo {author} {\bibfnamefont {R.}~\bibnamefont {McDermott}},\
  }\bibfield  {title} {\bibinfo {title} {Measurement of a superconducting qubit with a microwave photon counter},\ }\href {https://doi.org/10.1126/science.aat4625} {\bibfield  {journal} {\bibinfo  {journal} {Science}\ }\textbf {\bibinfo {volume} {361}},\ \bibinfo {pages} {1239} (\bibinfo {year} {2018})},\ \Eprint {https://arxiv.org/abs/https://www.science.org/doi/pdf/10.1126/science.aat4625} {https://www.science.org/doi/pdf/10.1126/science.aat4625} \BibitemShut {NoStop}%
\bibitem [{\citenamefont {Opremcak}\ \emph {et~al.}(2021)\citenamefont {Opremcak}, \citenamefont {Liu}, \citenamefont {Wilen}, \citenamefont {Okubo}, \citenamefont {Christensen}, \citenamefont {Sank}, \citenamefont {White}, \citenamefont {Vainsencher}, \citenamefont {Giustina}, \citenamefont {Megrant}, \citenamefont {Burkett}, \citenamefont {Plourde},\ and\ \citenamefont {McDermott}}]{opremcak_JPM-readout_2021}%
  \BibitemOpen
  \bibfield  {author} {\bibinfo {author} {\bibfnamefont {A.}~\bibnamefont {Opremcak}}, \bibinfo {author} {\bibfnamefont {C.~H.}\ \bibnamefont {Liu}}, \bibinfo {author} {\bibfnamefont {C.}~\bibnamefont {Wilen}}, \bibinfo {author} {\bibfnamefont {K.}~\bibnamefont {Okubo}}, \bibinfo {author} {\bibfnamefont {B.~G.}\ \bibnamefont {Christensen}}, \bibinfo {author} {\bibfnamefont {D.}~\bibnamefont {Sank}}, \bibinfo {author} {\bibfnamefont {T.~C.}\ \bibnamefont {White}}, \bibinfo {author} {\bibfnamefont {A.}~\bibnamefont {Vainsencher}}, \bibinfo {author} {\bibfnamefont {M.}~\bibnamefont {Giustina}}, \bibinfo {author} {\bibfnamefont {A.}~\bibnamefont {Megrant}}, \bibinfo {author} {\bibfnamefont {B.}~\bibnamefont {Burkett}}, \bibinfo {author} {\bibfnamefont {B.~L.~T.}\ \bibnamefont {Plourde}},\ and\ \bibinfo {author} {\bibfnamefont {R.}~\bibnamefont {McDermott}},\ }\bibfield  {title} {\bibinfo {title} {High-fidelity measurement of a superconducting qubit using an on-chip microwave photon counter},\ }\href
  {https://doi.org/10.1103/PhysRevX.11.011027} {\bibfield  {journal} {\bibinfo  {journal} {Phys. Rev. X}\ }\textbf {\bibinfo {volume} {11}},\ \bibinfo {pages} {011027} (\bibinfo {year} {2021})}\BibitemShut {NoStop}%
\bibitem [{\citenamefont {Kerman}(2020)}]{kerman_efficient-simulation_2020}%
  \BibitemOpen
  \bibfield  {author} {\bibinfo {author} {\bibfnamefont {A.~J.}\ \bibnamefont {Kerman}},\ }\href {https://arxiv.org/abs/2010.14929} {\bibinfo {title} {Efficient numerical simulation of complex josephson quantum circuits}} (\bibinfo {year} {2020}),\ \Eprint {https://arxiv.org/abs/2010.14929} {arXiv:2010.14929 [quant-ph]} \BibitemShut {NoStop}%
\bibitem [{\citenamefont {Chitta}\ \emph {et~al.}(2022)\citenamefont {Chitta}, \citenamefont {Zhao}, \citenamefont {Huang}, \citenamefont {Mondragon-Shem},\ and\ \citenamefont {Koch}}]{chitta_scqubits_2022}%
  \BibitemOpen
  \bibfield  {author} {\bibinfo {author} {\bibfnamefont {S.~P.}\ \bibnamefont {Chitta}}, \bibinfo {author} {\bibfnamefont {T.}~\bibnamefont {Zhao}}, \bibinfo {author} {\bibfnamefont {Z.}~\bibnamefont {Huang}}, \bibinfo {author} {\bibfnamefont {I.}~\bibnamefont {Mondragon-Shem}},\ and\ \bibinfo {author} {\bibfnamefont {J.}~\bibnamefont {Koch}},\ }\bibfield  {title} {\bibinfo {title} {Computer-aided quantization and numerical analysis of superconducting circuits},\ }\href {https://doi.org/10.1088/1367-2630/ac94f2} {\bibfield  {journal} {\bibinfo  {journal} {New Journal of Physics}\ }\textbf {\bibinfo {volume} {24}},\ \bibinfo {pages} {103020} (\bibinfo {year} {2022})}\BibitemShut {NoStop}%
\bibitem [{\citenamefont {Groszkowski}\ and\ \citenamefont {Koch}(2021)}]{groszkowski_scqubits_2021}%
  \BibitemOpen
  \bibfield  {author} {\bibinfo {author} {\bibfnamefont {P.}~\bibnamefont {Groszkowski}}\ and\ \bibinfo {author} {\bibfnamefont {J.}~\bibnamefont {Koch}},\ }\bibfield  {title} {\bibinfo {title} {Scqubits: a {P}ython package for superconducting qubits},\ }\href {https://doi.org/10.22331/q-2021-11-17-583} {\bibfield  {journal} {\bibinfo  {journal} {{Quantum}}\ }\textbf {\bibinfo {volume} {5}},\ \bibinfo {pages} {583} (\bibinfo {year} {2021})}\BibitemShut {NoStop}%
\bibitem [{\citenamefont {Ding}\ \emph {et~al.}(2021)\citenamefont {Ding}, \citenamefont {Ku}, \citenamefont {Shi},\ and\ \citenamefont {Zhao}}]{ding_mode-removal_2021}%
  \BibitemOpen
  \bibfield  {author} {\bibinfo {author} {\bibfnamefont {D.}~\bibnamefont {Ding}}, \bibinfo {author} {\bibfnamefont {H.-S.}\ \bibnamefont {Ku}}, \bibinfo {author} {\bibfnamefont {Y.}~\bibnamefont {Shi}},\ and\ \bibinfo {author} {\bibfnamefont {H.-H.}\ \bibnamefont {Zhao}},\ }\bibfield  {title} {\bibinfo {title} {Free-mode removal and mode decoupling for simulating general superconducting quantum circuits},\ }\href {https://doi.org/10.1103/PhysRevB.103.174501} {\bibfield  {journal} {\bibinfo  {journal} {Phys. Rev. B}\ }\textbf {\bibinfo {volume} {103}},\ \bibinfo {pages} {174501} (\bibinfo {year} {2021})}\BibitemShut {NoStop}%
\bibitem [{\citenamefont {Beaudoin}\ \emph {et~al.}(2011)\citenamefont {Beaudoin}, \citenamefont {Gambetta},\ and\ \citenamefont {Blais}}]{beaudoin_dissipation_2011}%
  \BibitemOpen
  \bibfield  {author} {\bibinfo {author} {\bibfnamefont {F.}~\bibnamefont {Beaudoin}}, \bibinfo {author} {\bibfnamefont {J.~M.}\ \bibnamefont {Gambetta}},\ and\ \bibinfo {author} {\bibfnamefont {A.}~\bibnamefont {Blais}},\ }\bibfield  {title} {\bibinfo {title} {Dissipation and ultrastrong coupling in circuit {{QED}}},\ }\bibfield  {journal} {\bibinfo  {journal} {Physical Review A}\ }\textbf {\bibinfo {volume} {84}},\ \href {https://doi.org/10.1103/physreva.84.043832} {10.1103/physreva.84.043832} (\bibinfo {year} {2011})\BibitemShut {NoStop}%
\bibitem [{\citenamefont {Schoelkopf}\ \emph {et~al.}(2003)\citenamefont {Schoelkopf}, \citenamefont {Clerk}, \citenamefont {Girvin}, \citenamefont {Lehnert},\ and\ \citenamefont {Devoret}}]{schoelkopf_qubits-as-spectrometers_2003}%
  \BibitemOpen
  \bibfield  {author} {\bibinfo {author} {\bibfnamefont {R.~J.}\ \bibnamefont {Schoelkopf}}, \bibinfo {author} {\bibfnamefont {A.~A.}\ \bibnamefont {Clerk}}, \bibinfo {author} {\bibfnamefont {S.~M.}\ \bibnamefont {Girvin}}, \bibinfo {author} {\bibfnamefont {K.~W.}\ \bibnamefont {Lehnert}},\ and\ \bibinfo {author} {\bibfnamefont {M.~H.}\ \bibnamefont {Devoret}},\ }\bibinfo {title} {Qubits as spectrometers of quantum noise},\ in\ \href {https://doi.org/10.1007/978-94-010-0089-5_9} {\emph {\bibinfo {booktitle} {Quantum Noise in Mesoscopic Physics}}},\ \bibinfo {editor} {edited by\ \bibinfo {editor} {\bibfnamefont {Y.~V.}\ \bibnamefont {Nazarov}}}\ (\bibinfo  {publisher} {Springer Netherlands},\ \bibinfo {address} {Dordrecht},\ \bibinfo {year} {2003})\ pp.\ \bibinfo {pages} {175--203}\BibitemShut {NoStop}%
\bibitem [{\citenamefont {Nguyen}\ \emph {et~al.}(2019)\citenamefont {Nguyen}, \citenamefont {Lin}, \citenamefont {Somoroff}, \citenamefont {Mencia}, \citenamefont {Grabon},\ and\ \citenamefont {Manucharyan}}]{nguyen_high-coherence-fluxonium_2019}%
  \BibitemOpen
  \bibfield  {author} {\bibinfo {author} {\bibfnamefont {L.~B.}\ \bibnamefont {Nguyen}}, \bibinfo {author} {\bibfnamefont {Y.-H.}\ \bibnamefont {Lin}}, \bibinfo {author} {\bibfnamefont {A.}~\bibnamefont {Somoroff}}, \bibinfo {author} {\bibfnamefont {R.}~\bibnamefont {Mencia}}, \bibinfo {author} {\bibfnamefont {N.}~\bibnamefont {Grabon}},\ and\ \bibinfo {author} {\bibfnamefont {V.~E.}\ \bibnamefont {Manucharyan}},\ }\bibfield  {title} {\bibinfo {title} {High-coherence fluxonium qubit},\ }\href {https://doi.org/10.1103/PhysRevX.9.041041} {\bibfield  {journal} {\bibinfo  {journal} {Phys. Rev. X}\ }\textbf {\bibinfo {volume} {9}},\ \bibinfo {pages} {041041} (\bibinfo {year} {2019})}\BibitemShut {NoStop}%
\bibitem [{\citenamefont {You}\ \emph {et~al.}(2019)\citenamefont {You}, \citenamefont {Sauls},\ and\ \citenamefont {Koch}}]{you_circuit-quantization-flux_2019}%
  \BibitemOpen
  \bibfield  {author} {\bibinfo {author} {\bibfnamefont {X.}~\bibnamefont {You}}, \bibinfo {author} {\bibfnamefont {J.~A.}\ \bibnamefont {Sauls}},\ and\ \bibinfo {author} {\bibfnamefont {J.}~\bibnamefont {Koch}},\ }\bibfield  {title} {\bibinfo {title} {Circuit quantization in the presence of time-dependent external flux},\ }\href {https://doi.org/10.1103/PhysRevB.99.174512} {\bibfield  {journal} {\bibinfo  {journal} {Phys. Rev. B}\ }\textbf {\bibinfo {volume} {99}},\ \bibinfo {pages} {174512} (\bibinfo {year} {2019})}\BibitemShut {NoStop}%
\bibitem [{\citenamefont {Koch}\ \emph {et~al.}(2007)\citenamefont {Koch}, \citenamefont {Yu}, \citenamefont {Gambetta}, \citenamefont {Houck}, \citenamefont {Schuster}, \citenamefont {Majer}, \citenamefont {Blais}, \citenamefont {Devoret}, \citenamefont {Girvin},\ and\ \citenamefont {Schoelkopf}}]{koch_charge-insensitive_2007}%
  \BibitemOpen
  \bibfield  {author} {\bibinfo {author} {\bibfnamefont {J.}~\bibnamefont {Koch}}, \bibinfo {author} {\bibfnamefont {T.~M.}\ \bibnamefont {Yu}}, \bibinfo {author} {\bibfnamefont {J.}~\bibnamefont {Gambetta}}, \bibinfo {author} {\bibfnamefont {A.~A.}\ \bibnamefont {Houck}}, \bibinfo {author} {\bibfnamefont {D.~I.}\ \bibnamefont {Schuster}}, \bibinfo {author} {\bibfnamefont {J.}~\bibnamefont {Majer}}, \bibinfo {author} {\bibfnamefont {A.}~\bibnamefont {Blais}}, \bibinfo {author} {\bibfnamefont {M.~H.}\ \bibnamefont {Devoret}}, \bibinfo {author} {\bibfnamefont {S.~M.}\ \bibnamefont {Girvin}},\ and\ \bibinfo {author} {\bibfnamefont {R.~J.}\ \bibnamefont {Schoelkopf}},\ }\bibfield  {title} {\bibinfo {title} {Charge-insensitive qubit design derived from the {{Cooper}} pair box},\ }\href {https://doi.org/10.1103/PhysRevA.76.042319} {\bibfield  {journal} {\bibinfo  {journal} {Physical Review A}\ }\textbf {\bibinfo {volume} {76}},\ \bibinfo {pages} {042319} (\bibinfo {year} {2007})}\BibitemShut {NoStop}%
\bibitem [{\citenamefont {Nguyen}\ \emph {et~al.}(2022)\citenamefont {Nguyen}, \citenamefont {Koolstra}, \citenamefont {Kim}, \citenamefont {Morvan}, \citenamefont {Chistolini}, \citenamefont {Singh}, \citenamefont {Nesterov}, \citenamefont {J{\"u}nger}, \citenamefont {Chen}, \citenamefont {Pedramrazi} \emph {et~al.}}]{nguyen_fluxonium_blueprint}%
  \BibitemOpen
  \bibfield  {author} {\bibinfo {author} {\bibfnamefont {L.~B.}\ \bibnamefont {Nguyen}}, \bibinfo {author} {\bibfnamefont {G.}~\bibnamefont {Koolstra}}, \bibinfo {author} {\bibfnamefont {Y.}~\bibnamefont {Kim}}, \bibinfo {author} {\bibfnamefont {A.}~\bibnamefont {Morvan}}, \bibinfo {author} {\bibfnamefont {T.}~\bibnamefont {Chistolini}}, \bibinfo {author} {\bibfnamefont {S.}~\bibnamefont {Singh}}, \bibinfo {author} {\bibfnamefont {K.~N.}\ \bibnamefont {Nesterov}}, \bibinfo {author} {\bibfnamefont {C.}~\bibnamefont {J{\"u}nger}}, \bibinfo {author} {\bibfnamefont {L.}~\bibnamefont {Chen}}, \bibinfo {author} {\bibfnamefont {Z.}~\bibnamefont {Pedramrazi}}, \emph {et~al.},\ }\bibfield  {title} {\bibinfo {title} {Blueprint for a high-performance fluxonium quantum processor},\ }\href@noop {} {\bibfield  {journal} {\bibinfo  {journal} {PRX Quantum}\ }\textbf {\bibinfo {volume} {3}},\ \bibinfo {pages} {037001} (\bibinfo {year} {2022})}\BibitemShut {NoStop}%
\bibitem [{\citenamefont {Pop}\ \emph {et~al.}(2014)\citenamefont {Pop}, \citenamefont {Geerlings}, \citenamefont {Catelani}, \citenamefont {Schoelkopf}, \citenamefont {Glazman},\ and\ \citenamefont {Devoret}}]{pop_coherent-suppression_2014}%
  \BibitemOpen
  \bibfield  {author} {\bibinfo {author} {\bibfnamefont {I.~M.}\ \bibnamefont {Pop}}, \bibinfo {author} {\bibfnamefont {K.}~\bibnamefont {Geerlings}}, \bibinfo {author} {\bibfnamefont {G.}~\bibnamefont {Catelani}}, \bibinfo {author} {\bibfnamefont {R.~J.}\ \bibnamefont {Schoelkopf}}, \bibinfo {author} {\bibfnamefont {L.~I.}\ \bibnamefont {Glazman}},\ and\ \bibinfo {author} {\bibfnamefont {M.~H.}\ \bibnamefont {Devoret}},\ }\bibfield  {title} {\bibinfo {title} {Coherent suppression of electromagnetic dissipation due to superconducting quasiparticles},\ }\href {https://doi.org/10.1038/nature13017} {\bibfield  {journal} {\bibinfo  {journal} {Nature}\ }\textbf {\bibinfo {volume} {508}},\ \bibinfo {pages} {369–372} (\bibinfo {year} {2014})}\BibitemShut {NoStop}%
\bibitem [{\citenamefont {Somoroff}\ \emph {et~al.}(2023)\citenamefont {Somoroff}, \citenamefont {Ficheux}, \citenamefont {Mencia}, \citenamefont {Xiong}, \citenamefont {Kuzmin},\ and\ \citenamefont {Manucharyan}}]{somoroff_millisecond_2023}%
  \BibitemOpen
  \bibfield  {author} {\bibinfo {author} {\bibfnamefont {A.}~\bibnamefont {Somoroff}}, \bibinfo {author} {\bibfnamefont {Q.}~\bibnamefont {Ficheux}}, \bibinfo {author} {\bibfnamefont {R.~A.}\ \bibnamefont {Mencia}}, \bibinfo {author} {\bibfnamefont {H.}~\bibnamefont {Xiong}}, \bibinfo {author} {\bibfnamefont {R.}~\bibnamefont {Kuzmin}},\ and\ \bibinfo {author} {\bibfnamefont {V.~E.}\ \bibnamefont {Manucharyan}},\ }\bibfield  {title} {\bibinfo {title} {Millisecond {{Coherence}} in a {{Superconducting Qubit}}},\ }\href {https://doi.org/10.1103/PhysRevLett.130.267001} {\bibfield  {journal} {\bibinfo  {journal} {Physical Review Letters}\ }\textbf {\bibinfo {volume} {130}},\ \bibinfo {pages} {267001} (\bibinfo {year} {2023})}\BibitemShut {NoStop}%
\bibitem [{\citenamefont {McEwen}\ \emph {et~al.}(2024)\citenamefont {McEwen}, \citenamefont {Miao}, \citenamefont {Atalaya}, \citenamefont {Bilmes}, \citenamefont {Crook}, \citenamefont {Bovaird}, \citenamefont {Kreikebaum}, \citenamefont {Zobrist}, \citenamefont {Jeffrey}, \citenamefont {Ying}, \citenamefont {Bengtsson}, \citenamefont {Chang}, \citenamefont {Dunsworth}, \citenamefont {Kelly}, \citenamefont {Zhang}, \citenamefont {Forati}, \citenamefont {Acharya}, \citenamefont {Iveland}, \citenamefont {Liu}, \citenamefont {Kim}, \citenamefont {Burkett}, \citenamefont {Megrant}, \citenamefont {Chen}, \citenamefont {Neill}, \citenamefont {Sank}, \citenamefont {Devoret},\ and\ \citenamefont {Opremcak}}]{mcewen_resisting_2024}%
  \BibitemOpen
  \bibfield  {author} {\bibinfo {author} {\bibfnamefont {M.}~\bibnamefont {McEwen}}, \bibinfo {author} {\bibfnamefont {K.~C.}\ \bibnamefont {Miao}}, \bibinfo {author} {\bibfnamefont {J.}~\bibnamefont {Atalaya}}, \bibinfo {author} {\bibfnamefont {A.}~\bibnamefont {Bilmes}}, \bibinfo {author} {\bibfnamefont {A.}~\bibnamefont {Crook}}, \bibinfo {author} {\bibfnamefont {J.}~\bibnamefont {Bovaird}}, \bibinfo {author} {\bibfnamefont {J.~M.}\ \bibnamefont {Kreikebaum}}, \bibinfo {author} {\bibfnamefont {N.}~\bibnamefont {Zobrist}}, \bibinfo {author} {\bibfnamefont {E.}~\bibnamefont {Jeffrey}}, \bibinfo {author} {\bibfnamefont {B.}~\bibnamefont {Ying}}, \bibinfo {author} {\bibfnamefont {A.}~\bibnamefont {Bengtsson}}, \bibinfo {author} {\bibfnamefont {H.-S.}\ \bibnamefont {Chang}}, \bibinfo {author} {\bibfnamefont {A.}~\bibnamefont {Dunsworth}}, \bibinfo {author} {\bibfnamefont {J.}~\bibnamefont {Kelly}}, \bibinfo {author} {\bibfnamefont {Y.}~\bibnamefont {Zhang}}, \bibinfo {author} {\bibfnamefont {E.}~\bibnamefont
  {Forati}}, \bibinfo {author} {\bibfnamefont {R.}~\bibnamefont {Acharya}}, \bibinfo {author} {\bibfnamefont {J.}~\bibnamefont {Iveland}}, \bibinfo {author} {\bibfnamefont {W.}~\bibnamefont {Liu}}, \bibinfo {author} {\bibfnamefont {S.}~\bibnamefont {Kim}}, \bibinfo {author} {\bibfnamefont {B.}~\bibnamefont {Burkett}}, \bibinfo {author} {\bibfnamefont {A.}~\bibnamefont {Megrant}}, \bibinfo {author} {\bibfnamefont {Y.}~\bibnamefont {Chen}}, \bibinfo {author} {\bibfnamefont {C.}~\bibnamefont {Neill}}, \bibinfo {author} {\bibfnamefont {D.}~\bibnamefont {Sank}}, \bibinfo {author} {\bibfnamefont {M.}~\bibnamefont {Devoret}},\ and\ \bibinfo {author} {\bibfnamefont {A.}~\bibnamefont {Opremcak}},\ }\bibfield  {title} {\bibinfo {title} {Resisting high-energy impact events through gap engineering in superconducting qubit arrays},\ }\href {https://doi.org/10.1103/PhysRevLett.133.240601} {\bibfield  {journal} {\bibinfo  {journal} {Phys. Rev. Lett.}\ }\textbf {\bibinfo {volume} {133}},\ \bibinfo {pages} {240601} (\bibinfo
  {year} {2024})}\BibitemShut {NoStop}%
\end{thebibliography}%

\end{document}